\documentclass[acmsmall]{acmart}
\usepackage{cite}
\usepackage{algorithmic}
\usepackage{graphicx}
\usepackage{textcomp}
\usepackage{xcolor}
\usepackage{xspace}
\usepackage{booktabs}
\usepackage{hyperref}
\usepackage{cleveref}
\usepackage{colortbl}
\usepackage{balance}
\usepackage{subfig}
\usepackage{multirow}
\usepackage{diagbox}
\usepackage{caption}
\usepackage{pifont}
\newcommand{\cmark}{\ding{51}}%
\newcommand{\xmark}{\ding{55}}%

\usepackage{enumitem}
\setlist[itemize]{leftmargin=10pt}

\def\BibTeX{{\rm B\kern-.05em{\sc i\kern-.025em b}\kern-.08em
    T\kern-.1667em\lower.7ex\hbox{E}\kern-.125emX}}

\newcommand{\cmt}[1]{\textcolor{red}{\textbf{comment}: {#1}}}
\newcommand{\sihan}[1][\textcolor{purple}]{#1}
\newcommand{\ling}[1]{\textcolor{orange}{#1}}
\newcommand{\gy}[1]{\textcolor{green}{#1}}
\newcommand{\revise}[1]{\textcolor{black}{#1}}
\newcommand{\tool}{\textsc{LiDetector}\xspace}
\graphicspath{{ase2021/}}
\usepackage{url}  
\usepackage[ruled,lined,linesnumbered,vlined,algo2e]{algorithm2e}
\definecolor{codegray}{rgb}{0.5,0.5,0.5}
\SetCommentSty{mycommfont}
    
\usepackage[skip=3pt]{caption} %font=small,
\newcommand{\distance}{3pt}
\usepackage{makecell}
\usepackage{threeparttable}
\makeatletter
\newif\if@restonecol
\renewcommand\footnoterule{%
 \kern-3\p@
 \hrule\@width\columnwidth
 \kern2.6\p@}
\begin{document}
\setcopyright{acmcopyright}
\acmJournal{TOSEM}
\acmYear{2022} \acmVolume{1} \acmNumber{1} \acmArticle{1} \acmMonth{1} \acmPrice{15.00}\acmDOI{10.1145/3518994}

\title{
\tool: License Incompatibility Detection for Open Source Software
}

\author{Sihan Xu}
\email{xusihan@nankai.edu.cn}
\affiliation{%
  \institution{TKLNDST, College of Cyber Science, Nankai University}
  \city{Tianjin}
  \country{China}
}

\author{Ya Gao}
\email{gaoya_cs@mail.nankai.edu.cn}
\affiliation{
  \institution{TKLNDST, College of Computer Science, Nankai University}
  \city{Tianjin}
  \country{China}
}

\author{Lingling Fan}
\email{linglingfan@nankai.edu.cn}
\affiliation{
  \institution{TKLNDST, College of Cyber Science, Nankai University}
  \city{Tianjin}
  \country{China}
}
\authornote{Lingling Fan is the corresponding author. Email: linglingfan@nankai.edu.cn}

\author{Zheli Liu}
\email{liuzheli@nankai.edu.cn}
\affiliation{
  \institution{TKLNDST, College of Cyber Science, Nankai University}
  \city{Tianjin}
  \country{China}
}

\author{Yang Liu}
\email{yangliu@ntu.edu.sg}
\affiliation{
  \institution{Zhejiang Sci-tech University, China and Nanyang Technological University}
  \country{Singapore}
}

\author{Hua Ji}
\email{hua.ji@nankai.edu.cn}
\affiliation{
  \institution{TKLNDST, College of Cyber Science, Nankai University}
  \city{Tianjin}
  \country{China}
}

\begin{abstract}
    Open-source software (OSS) licenses dictate the conditions which should be followed to reuse, distribute, and modify software. Apart from widely-used licenses such as {the} {MIT License}, developers are also allowed to customize their own licenses \revise{(called custom license)}, %({24.56\%} custom licenses according to our preliminary study), 
    whose descriptions are more flexible. The presence of such various licenses imposes challenges to understand licenses and their compatibility. To avoid financial and legal risks, it is essential to ensure license compatibility when integrating third-party packages \revise{or reusing code accompanied with licenses}. In this work, we propose \tool, 
    an effective tool that extracts and interprets OSS licenses \revise{(including both official licenses and custom licenses)}, and detects license incompatibility \revise{among these licenses}. 
    Specifically, \tool introduces a learning-based method to automatically identify meaningful license terms from an arbitrary license, and employs Probabilistic Context-Free Grammar (PCFG) to infer rights and obligations for incompatibility detection. Experiments demonstrate that \tool outperforms existing methods with 93.28\% precision for term identification, and 91.09\% accuracy for right and obligation inference, and can effectively detect incompatibility with {10.06\%} FP rate and {2.56\%} FN rate. Furthermore, with \tool, our large-scale empirical study on 1,846 projects reveals that {72.91\%} of the projects are suffering from license incompatibility, including popular ones such as the {MIT} License and the {Apache License}. We highlighted lessons learned from perspectives of different stakeholders and \color{black}{made all related data and the replication package publicly available} to facilitate follow-up research.
\end{abstract}




\maketitle

\section{Introduction}

%%%%Background%%%%%
Open source software (OSS) \citep{url-oss} is a type of software where source code is publicly available under certain licenses. 
The licenses dictate the conditions under which OSS can be reused, distributed, and modified \textit{legally}.
To facilitate software development, a common practice is to integrate OSS code so that developers do not need to reinvent the wheel. 
%In recent years, the availability of source code hosting services such as SourceForge~\citep{sourceforge} and GitHub~\citep{url-github} significantly increases the diffusion of OSS.
%so as to avoid financial and legal risks.
While it brings convenience for software development, it also induces security issues \citep{zhan2021atvhunter,zhan2021research,chen2018mobile,chen2020empirical} and legal risks \citep{chinacase,licensing2004software} such as copyright infringement \citep{infringement-case}, caused by license incompatibility when integrating third-party components.
As previous studies~\citep{gangadharan2012managing,jss`17-Kapitsaki-SPDX}, \textit{license incompatibility} occurs when there exists no such a license that satisfies all the rights and obligations of all the integrated third-party components, e.g, ``MUST disclose source'' declared by one license but ``CANNOT disclose source'' declared by another license within the same project.

According to our preliminary study on 1,846 GitHub projects, {48.86\%} projects are suffering from license incompatibility.
%and 37.10\% license texts are customized by developers. 
Note that, in the preliminary study, due to the lack of effective tools for incompatibility detection, we only investigate the incompatibility among some popular licenses \revise{that can be identified by an existing tool Ninka \citep{ase`10-German-Ninka}}.%, the actual situation may be even worse than what it shows when considering custom licences.

To address this issue, there have been several studies that investigate license compatibility, mainly focused on license identification and compatibility analysis~\citep{Wheeler`07-Wheeler,ase`10-German-Ninka,jss`17-Kapitsaki-SPDX,APSEC`17-Kapitsaki-termsIdentifying,tse`18-Kapitsaki-findOSSLicense,url-tldrlegal}. However, there are two problems that limit the application of previous studies. \textbf{First}, previous works can only investigate the compatibility between a predefined set of common licenses, and can hardly be adapted to other licenses automatically. 
For instance, SPDX-VT~\citep{jss`17-Kapitsaki-SPDX} predefines a dependency graph to tease apart the compatibility relationships specifically for 20 well-known licenses,
%license use that violates the rules are considered to have incompatibility issues. Although strict and precise, 
however, other licenses 
%outside its application scope 
cannot be addressed by SPDX-VT. \textbf{Second}, 
{the rules to detect incompatibility need to be manually defined and specified for each license}, which is a major obstacle to automatically detect incompatibility when licenses are changed, updated, or customized by developers. Among previous studies, only FOSS-LTE~\citep{APSEC`17-Kapitsaki-termsIdentifying} can be adapted to interpret an arbitrary license. Nevertheless, it can only detect 38 regulations in licenses, and developers need to manually analyze the compatibility for a given project.  
%manually analyzed 25 common licenses and defined 38 phrases to express the conditions of software use, which limits its flexibility in interpreting other licenses.    
%, while few works has attempted to propose a general method that is suited to an arbitrary license
Actually, popular licenses \revise{or official licenses in SPDX~\citep{url-spdx}} often have various versions and exceptions, apart from which, developers are also allowed to create their own licenses (i.e., \textit{custom licenses}). According to our preliminary study, {24.56\%} license texts are customized by developers.
Different licenses regulate different rights and obligations. As a result, it is impractical to identify and manually define the relationships for all licenses in the community. Instead, it is crucial to propose an effective method that automatically interprets licenses and detects license incompatibility issues throughout all kinds of licenses including custom licenses. 

%%%%Our proposed work and experiment results%%%%
To this end, in this paper, we proposed \tool, an automated tool for interpreting licenses to detect license incompatibility for open source software. It first constructs a probabilistic model to identify meaningful license terms, and then performs sentiment analysis based on grammar parsing to infer rights and obligations from licenses. Based on the identified terms and the attitudes implied by licenses, \tool can identify license incompatibility for arbitrary licenses. Comparative experiments demonstrate the effectiveness of \tool, with 93.28\% precision and 75.70\% recall for license term identification, 91.09\% accuracy for right and obligation inference, and {169} incompatible projects identified from 200 Github projects {(with 10.06\% false positive rate and 2.56\% false negative rate).}
To further investigate license incompatibility in real-word OSS, we leverage \tool to conduct an empirical study on 1,846 Github projects and find that {72.91\%} projects are suffering from license incompatibility, involving some very popular licenses such as the MIT License~\citep{url-MIT} and the Apache License~\citep{url-Apache-2.0}. In addition, {\textit{Disclose Source}} induces the most number of conflicts {(7,186)}, which deserves more attention to avoid serious legal risks. 
Finally, lessons learned are summarized based on our study from the perspectives of different stakeholders (e.g., developers) to shed light on the importance of license compatibility and the usefulness of \tool. 

%%%%Contributions%%%%
In summary, we made the following novel contributions:
\begin{itemize}
    \item 
    We proposed \tool, a hybrid and effective method that automatically understands license texts and infers rights and obligations to detect license incompatibility in open source software with arbitrary licenses, including the widely-used ones and the custom ones. %It can identify incompatibility of not only popular licenses but also custom ones.
    %We propose a two-stage method to interpret OSS licenses automatically. Specifically, we first employ semi-supervised learning to identify license terms from texts. Then, based on the identified terms, we infer the permissive or restrictive attitudes towards these terms by sentiment analysis. Experimental results demonstrate that the proposed tool, \tool, outperforms the baselines in the two independent tasks, as well as the joint task. 
    \item Extensive and comparative experiments demonstrate the effectiveness of \tool over existing methods, {with {10.06\%} false positive rate and {2.56\%} false negative rate} in license incompatibility detection for open source projects.
    %We propose a \tool to automatically detect license incompatibility for an arbitrary project. Extensive experiments on 200 GitHub projects with 11,721 licenses demonstrate the superiority of the proposed tool, achieving 95.24\% precision and 96.39\% recall, respectively. 
    \item We further conduct a large-scale empirical study on 1,846 GitHub projects by leveraging \tool, and find that {72.91\%} of the projects are suffering from license incompatibility, involving popular licenses such as the MIT and Apache License, which deserve more attention from developers and software companies.
    \color{black}{{We released all the datasets~\citep{lidetector} and the replication package~\citep{lidetector-github} on Github for the community.}}
    %To further investigate the incompatibility issues, we conduct a large-scale empirical study on 1,846 GitHub projects using \tool, and reported 5,112 conflicts in 877 projects. We reveal the top licenses and terms involved in compatibility issues, which should arouse the attention of developers and software companies. All the datasets\footnote{https://sites.google.com/view/ase2021-lidetector} are released publicly available for the community.
\end{itemize}

% \begin{figure*}[htp]
%     \centering
%     \includegraphics[width=16cm]{figs/motivation example.pdf}
%     \caption{A Motivation Example}
%     \label{fig_motivation}
% \end{figure*}

\section{Background}

In this section, we first introduce OSS licences and the compliance issues. Then, we present a motivating study, which shows the importance of detecting license incompatibility for OSS.

\subsection{OSS License}
Licenses applied in open source software (OSS) regulate the rights, obligations, and prohibitions of OSS use. 
An OSS license is represented in the form of text description, where the copyright holders specify the conditions under which users can freely use, modify, and distribute software~\citep{AISEW`12-Mathur-Empirical}.
{The Software Package Data Exchange specification (SPDX)~\citep{mancinelli2006managing,url-spdx} maintains over
% 200
400 licenses including very popular ones such as the Apache License, the Academic Free License (AFL), and the GNU General Public License (GPL).}
When users use OSS to facilitate software development, they are expected to comply with the rights and obligations implicated by the licenses, e.g., \textit{can use}, \textit{cannot redistribute}, and \textit{refuse commercial use}.
%and \textit{must include copyright notice}.  

\noindent \textbf{License term \textit{vs.} license term entity.}
 In this paper, \revise{a \textit{license term} refers to a formal and unified description of the conditions of software use} (e.g., \textit{commercial use}), while a \textit{license term entity} refers to a specific expression of a license term in real-world license texts (e.g., \textit{sell} or \textit{offer for sale}). As previous studies~\citep{url-tldrlegal}, there are 23 license terms as displayed in Table~\ref{table-23terms}, each of which represents a type of action that users may do. 
 \revise{To better understand these terms and facilitate incompatibility detection, following by the previous studies~\citep{APSEC`17-Kapitsaki-termsIdentifying,tse`18-Kapitsaki-findOSSLicense}, we further classify the 23 terms into \textit{Rights} and \textit{Obligations}, and consider the conditions of license terms.}
 \revise{Nevertheless, there are a variety of licenses (such as official ones and custom ones), leading the expressions of a license term to be various, which impose challenges in identifying license terms from an arbitrary license~\citep{jss`17-Kapitsaki-SPDX,tse`18-Kapitsaki-findOSSLicense}}.

\begin{table*}[t]
	\centering
	%\color{blue}
	\small
	\caption{License Terms and the Descriptions}
	\label{table-23terms}
	\scalebox{0.83}{\begin{tabular}{c|cll}
		\toprule
		\textbf{Category}& \textbf{No.} & \textbf{Term}    & \textbf{Description} \\
		\midrule
		
		\multirow{11}{*}{\textbf{Rights}} &\cellcolor{gray!20} 0 & \cellcolor{gray!20}Distribute & \cellcolor{gray!20}Distribute original or modified derivative works \\
		& 1 & Modify	& Modify the software and create derivatives\\
		 & \cellcolor{gray!20}2 & \cellcolor{gray!20}Commercial Use 	& \cellcolor{gray!20} Use the software for commercial purposes\\
		& 3 & Relicense	& Add other licenses with the software\\
		&\cellcolor{gray!20} 4 & \cellcolor{gray!20}Hold Liable &\cellcolor{gray!20}	Hold the author responsible for subsequent impacts\\
		& 5 & Use Patent Claims &	Practice patent claims of contributors to the code\\
		&\cellcolor{gray!20} 6 & \cellcolor{gray!20}Sublicense	& \cellcolor{gray!20}Incorporate the work into something that has a more restrictive license\\
		& 7 & Statically Link &     {\begin{tabular}[c]{@{}l@{}} The library can be compiled into the program linked at compile time rather \\\ than runtime\end{tabular}} \\ 
		&\cellcolor{gray!20} 8 & \cellcolor{gray!20}Private Use &\cellcolor{gray!20}	Use or modify software freely without distributing it\\
		& 9 & Use Trademark & Use contributors' names, trademarks or logos\\
		&\cellcolor{gray!20} 10 & \cellcolor{gray!20}Place Warranty &\cellcolor{gray!20}	Place warranty on the software licensed\\
		\hline
		
		\multirow{12}{*}{\textbf{Obligations}}
         & 11 & Include Copyright &	Retain the copyright notice in all copies or substantial uses of the work.\\
       &\cellcolor{gray!20} 12 & \cellcolor{gray!20}Include License &\cellcolor{gray!20}	Include the full text of license in modified software\\
       & 13 & Include Notice &
        {\begin{tabular}[c]{@{}l@{}} Include that NOTICE when you distribute if the library has a NOTICE file \\\ with  attribution notes\end{tabular}}\\
        & \cellcolor{gray!20}14 & \cellcolor{gray!20}Disclose Source &	\cellcolor{gray!20}
      {\begin{tabular}[c]{@{}l@{}} Disclose your source code when you distribute the software and make the \\\ source for the library available\end{tabular}} \\
         & 15 & State Changes &	State significant changes made to software\\
       &\cellcolor{gray!20} 16 & \cellcolor{gray!20}Include Original & \cellcolor{gray!20}{\begin{tabular}[c]{@{}l@{}} Distribute copies of the original software or instructions to obtain copies \\\ with the  software\end{tabular}}	 \\
         & 17 & Give Credit &	Give explicit credit or acknowledgement to the author with the software\\
      &\cellcolor{gray!20} 18 & \cellcolor{gray!20}Rename &	\cellcolor{gray!20}Change software name as to not misrepresent them as the original software\\
       & 19 & Contact Author &{\begin{tabular}[c]{@{}l@{}}	Get permission from author or contact the author about the module you \\\ are using\end{tabular}}\\
      &\cellcolor{gray!20} 20 & \cellcolor{gray!20}Include Install Instructions &\cellcolor{gray!20}{\begin{tabular}[c]{@{}l@{}}	Include the installation information necessary to modify and reinstall \\\ the software\end{tabular}}\\
         & 21 & Compensate for Damages &	Compensate the author for any damages cased by your work\\
       &\cellcolor{gray!20} 22 & \cellcolor{gray!20}Pay Above Use Threshold &\cellcolor{gray!20}	Pay the licensor after a certain amount of use
 
         \\
		\bottomrule
	\end{tabular}}
	\label{table-terms}
\end{table*}

\noindent \textbf{Project licenses (PL) \textit{vs.} component licenses (CL).} Typically, a software product may contain a \textit{project license}, usually in the form of \revise{a LICENSE file in the main directory of the software}, which states the conditions of software use. 
In addition, when incorporating third-party software packages or reusing code \citep{zhan2020automated}, \revise{licenses that accompany each third-party package or file should also be conformed to.
In this paper, to distinguish with the project license, licenses in software other than \textit{project licenses} are called \textit{component licenses}. }
%}which we denote as \textit{component licenses}.

\noindent \textbf{Declared licenses \textit{vs.} referenced licenses {\textit{vs.} inline licenses}.} Licenses in OSS products are presented mainly in %two
{three forms, i.e., declaration, reference, and text in source code.} 
%The former 
{The \textit{declared licenses}} dictate rights and obligations in one or more license files (e.g., LICENSE.txt). Users can capture license information from these files directly without any external resource. %The latter 
The \textit{referenced licenses} indicate licenses referenced by direct or indirect links, where direct links refer to the license name, version, or the website of the license, {\color{black}and indirect links refer to imported software packages according to which licenses can be found. The detailed information needs to be obtained from external sources, such as pypi~\citep{url-pypi} (the Python Package Index), SPDX~\citep{mancinelli2006managing,url-spdx}, and the hosted pages for OSS licenses.}
The \textit{inline licenses} refer to license text in the same file of source code, which usually appear on the top of source code files. The \textit{inline licenses} are considered as the most fine-grained licenses, since their scope only cover the source code in the same file with license text.

\subsection{License Incompatibility}

\textit{License incompatibility} %(interchangeably used with “license conflicts”) %\footnote{We use the terms “license incompatibility” and “license conflicts” interchangeably throughout the paper.}
refers to the conflicts of multiple licenses within the same projects. %when they are combined into the same project. 
A license consists of permissive and restrictive statements that specify the requirements for a derivative work. Given a license term, rules associated with it span a range from very permissive ones to highly restrictive ones (i.e., \textit{strong copyleft}). 
To facilitate software development, developers often need to integrate multiple third-party OSS within one project. For this reason, an open source project may need to comply with more than one licenses. {However, there are a variety of licenses and exceptions that regulate different rights and obligations. Moreover, developers are also allowed to create their own licenses (denoted by \textbf{custom licenses}), which are more flexible in expressions. The combination of such a variety of licenses} often carries out incompatibility issues that prevent correct incorporation of third-party software packages.

Compatibility analysis aims to integrate software components with multiple licenses in a newly-developed software~\citep{gangadharan2012managing}. 
%Therefore, we define \textit{license incompatibility} as the presence of conflict rules indicated by licenses. 
\revise{Therefore, we define \textit{license incompatibility} as follows: two licenses ($l_1$ and $l_2$) are incompatible if there exists no such a license that can integrate $l_1$ and $l_2$ in a newly-defined software without right/obligation conflicts.
For instance, if one license declares that \textit{``must contact author for software use''} and another license states \textit{``do not contact authors''}, it indicates that there exists a conflict between the two licenses upon the term \textit{``contact author''}}. 
%Developers need to address such a conflict, for instance, developing a restrictive license to avoid legal risks.
\revise{Developers need to address such a conflict, for instance, choosing another restrictive license to avoid legal risks.}

%Before integrating software packages accompanied with their own licenses, developers need to perform compatibility analysis on the component licenses. 
%The goal of compatibility analysis is to combine multiple licenses in a newly-developed software~\citep{gangadharan2012managing}. For this reason, we define license incompatibility as the presence of conflict rules indicated by licenses. For instance, if one license declares that \textit{commercial use is allowed} and another license \textit{refuses commercial use}, it indicates that there exist different rights and obligations towards the term \textit{commercial use}, and developers need to figure out a restrictive license to conform.

%A motivation example is shown in Figure~\ref{fig_motivation}. \ling{briefly describe it using the figure.}

%\textit{Referenced license}, refers to the license not shown as whole texts but linked directly or indirectly. Directly linked ones, are represented in its name, version, or just a url of the license. Indirectly linked ones, refer that there may only be the name of the third-party library in the project code, and the license name does not appear, which can only be found on the official website of these OSS packages.We use the available license terms tag information to search for its terms and polarity.\textit{Included license}, refers to the license shown as its whole text in the code. In this paper uses our proposed term extraction model to obtain its terms and polarity. 

\subsection{Motivating Study}
\label{sec:motivatingstudy}

To better motivate our work on license incompatibility detection, we conduct an empirical study on real-world open source projects to investigate the prevalence of custom licenses and the incompatibility issues.

\begin{figure}[t]
    \centering
    \includegraphics[width=0.7\textwidth]{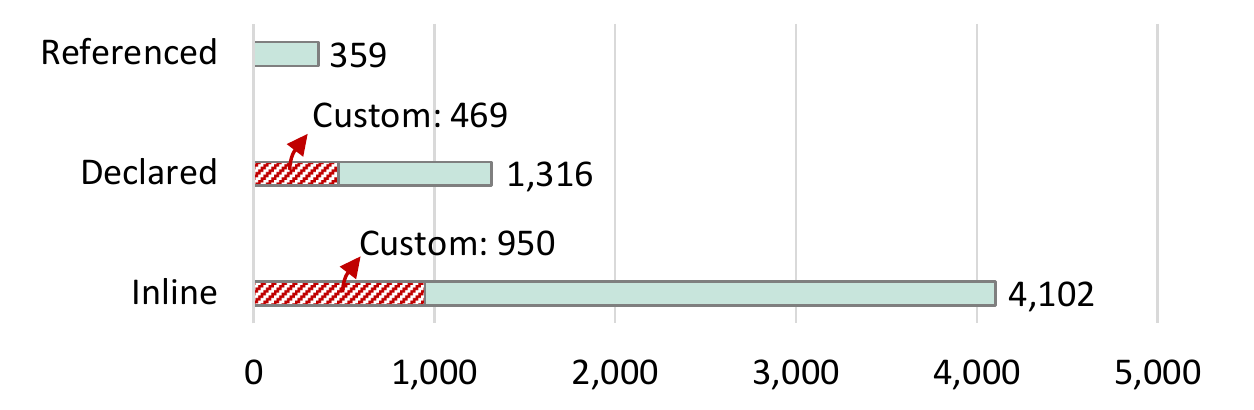}
    \caption{The Prevalence of Custom Licenses}
    \label{fig-prevalence}
\end{figure}

\smallskip
\noindent \textbf{The prevalence of custom licenses.}
%We crawled 1,846 popular Python projects ordered by the number of stars from GitHub. For each project, we only extracted declared licenses to investigate the prevalence of custom licenses. The rational behind is that most referenced licenses are widely-used ones that can be found from external sources by users. We use Ninka~\citep{ase`10-German-Ninka}, a notable license identification tool, to identify well-known licenses by the license texts. Licenses that cannot be recognized by Ninka are considered as custom licenses. As shown in Figure~\ref{fig-prevalence}, we obtained 11,578 declared licenses in total, only 62.9\% of which are popular ones that can be identified by Ninka. The preliminary study shows that 37.1\% of declared licenses are customized by the authors of software products. \textcolor{brown}{Additionally, we obtained 96,606 referenced licenses and after de-duplication, 10258 were left within all declared and referenced licenses, in which the custom licenses occupied 41.88\%.} 
%\cmt{check it.....}
{We crawled 1,846 popular Python projects ordered by the number of stars from GitHub. For each project, 
%we extracted both the declared and referenced licenses to investigate the prevalence of custom licenses. 
{we extract three types of licenses (i.e., the declared, the referenced, and the inline licenses) to investigate the prevalence of custom licenses. }
%\ling{Specifically, to extract the declared licenses, we extract.... To extract the referenced licenses, we .... To extract the inline licenses, we ....}
{Specifically, to extract the declared licenses, we conducted regular matching to identify license files, such as the LICENSE.txt and the COPYING files. To extract the referenced licenses, we obtained license names and versions directly from the project. In addition, {\color{black}for indirectly linked licenses of imported third-party packages, inspired by LicenseFinder~\citep{url-LicenseFinder}, we first extracted their names using QDox~\citep{url-qdox}, 
%and then queried the website PyPi~\citep{url-pypi}, which is an official repository of software for the Python programming language, by the package names to search for the accompanied licenses. 
and then queried pypi~\citep{url-pypi} (the Python Package Index) by the package names to search for the accompanied licenses.}
To extract the inline licenses, we 
%used Ninka~\citep{ase`10-German-Ninka} to 
extracted license text from comments in the source code files (typically on the top of code files).}
For extracted license texts, we use Ninka~\citep{ase`10-German-Ninka}, a notable
license identification tool, to identify the names and versions of well-known licenses. 
%\ling{In addition, we refer to SPDX \citep{url-spdx} to filter out known ones, and the licenses that are not recognized by Ninka and not in the list of SPDX are considered as custom licenses.}
%Licenses that cannot be recognized by Ninka are considered as custom licenses. 

As shown in Fig.~\ref{fig-prevalence}, {from %507,688 licenses extracted from 
1,846 projects, we obtained {359} unique referenced licenses, {1,316} unique declared licenses, and {4,102} unique inline licenses. In total, we found {5,777} unique licenses. Then, we used Ninka to detect well-known licenses. It was observed that {75.44\%}} of licenses are popular licences that can be detected by Ninka, and {{24.56\%} of the licenses ({1,419}) are customized} by the authors of software products. Note that licenses reported by Ninka may include popular and custom exceptions that are slightly different from the original licenses. Moreover, all the custom licenses belong to the declared and inline licenses. The rational behind is that most referenced licenses are widely-used ones that can be found from external sources by users. Therefore, we further investigated the portion of custom licenses in declared and inline ones, and {found that {35.64\%} of declared licenses and {23.16\%} of inline licenses are customized licenses.}} %\gy{Besides, we consider declared licenses and inline licenses together because of they both display license content directly in texts compared with the referenced licenses obtained by links.}
Compared with widely-used licenses, custom ones are more flexible in text. For this reason, although previous works mainly focus on a set of popular licenses, the presence of such a variety of custom licenses requires adaptive methods that are capable of understanding the implications of an arbitrary license.

\smallskip \noindent \textbf{Incompatible licenses in real-world projects.} To investigate the compliance issues of licenses in open source software, {we extract the declared, referenced, and inline licenses from the 1,846 projects using the aforementioned extraction method}. 
%For declared licenses, we used Ninka to identify the license names and versions. 
%For referenced licenses, we obtained license names and versions directly from the project. Moreover, for indirectly linked licenses, we also queried pypi~\citep{url-pypi} by the package names to search for the accompanied licenses. 
Note that to observe the prevalence of licence incompatibility, in this motivating study, we only focus on widely-used licenses that can be identified by Ninka and have labels on tldrlegal~\citep{url-tldrlegal}, a
platform that provides the rights and obligations towards license terms (e.g., redistribute, modify) for well-known licenses. 

With the assistance of tldrlegal, we conduct an investigation towards the incompatibility issues in real-world OSS. The result shows that about {\textbf{48.86\%}} of the projects suffer from license incompatibility issues. To avoid involving illegal issues, 
authors of these projects need to analyze component licenses and address the incompatibility issues before distributing their software products.

\subsection{Running Example}
\revise{Fig.~\ref{fig-example} depicts two real-world running examples of license incompatibility, where PL denotes a project license, and CL denotes a component license.
The first project is Augmented Traffic Control~\citep{url-example1}, an open source project to simulate network traffic. It contains a license file for the whole project (i.e., a project license), which is an official license named BSD License~\citep{url-BSD-3-Clause}. Meanwhile, the project also contains a component of atc cookbooks, where a custom license can be found. 
%In Fig.~\ref{fig-example}, we use PL to denote a project license, and use CL to denote a component license. 
The project license states that ``\textit{Redistribution and use in source and binary forms, with or without modification, are permitted}''. However, the component license declares that ``\textit{Do Not Redistribute}''. It can be seen that the two licenses convey different attitudes towards the same license term (i.e., Distribute in Table~\ref{table-23terms}). Users who comply with the project license (CAN distribute) may still violate the component license (CANNOT distribute). For this reason, we say there exists license incompatibility between the project and component licenses.} 

\begin{figure}[t]
\centering
\subfloat[Incompatibility between PL and CL in \textit{Augmented Traffic Control}~\citep{url-example1}]{\label{fig:example1}{\includegraphics[width=0.48\textwidth]{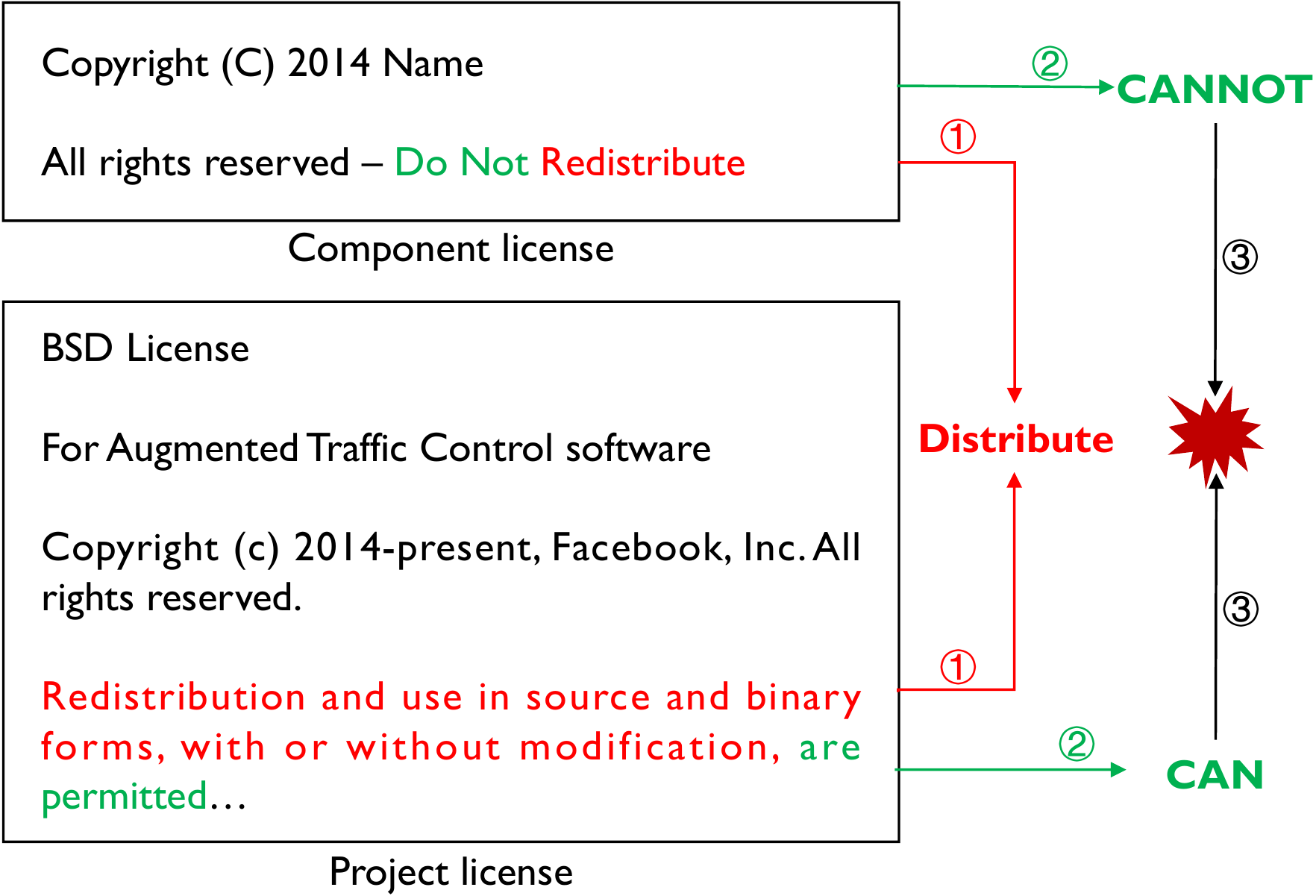}}}
\hfill
\subfloat[Incompatibility between CL and CL in \textit{Faust}~\citep{url-example2}]{\label{fig:example2}{\includegraphics[width=0.48\textwidth]{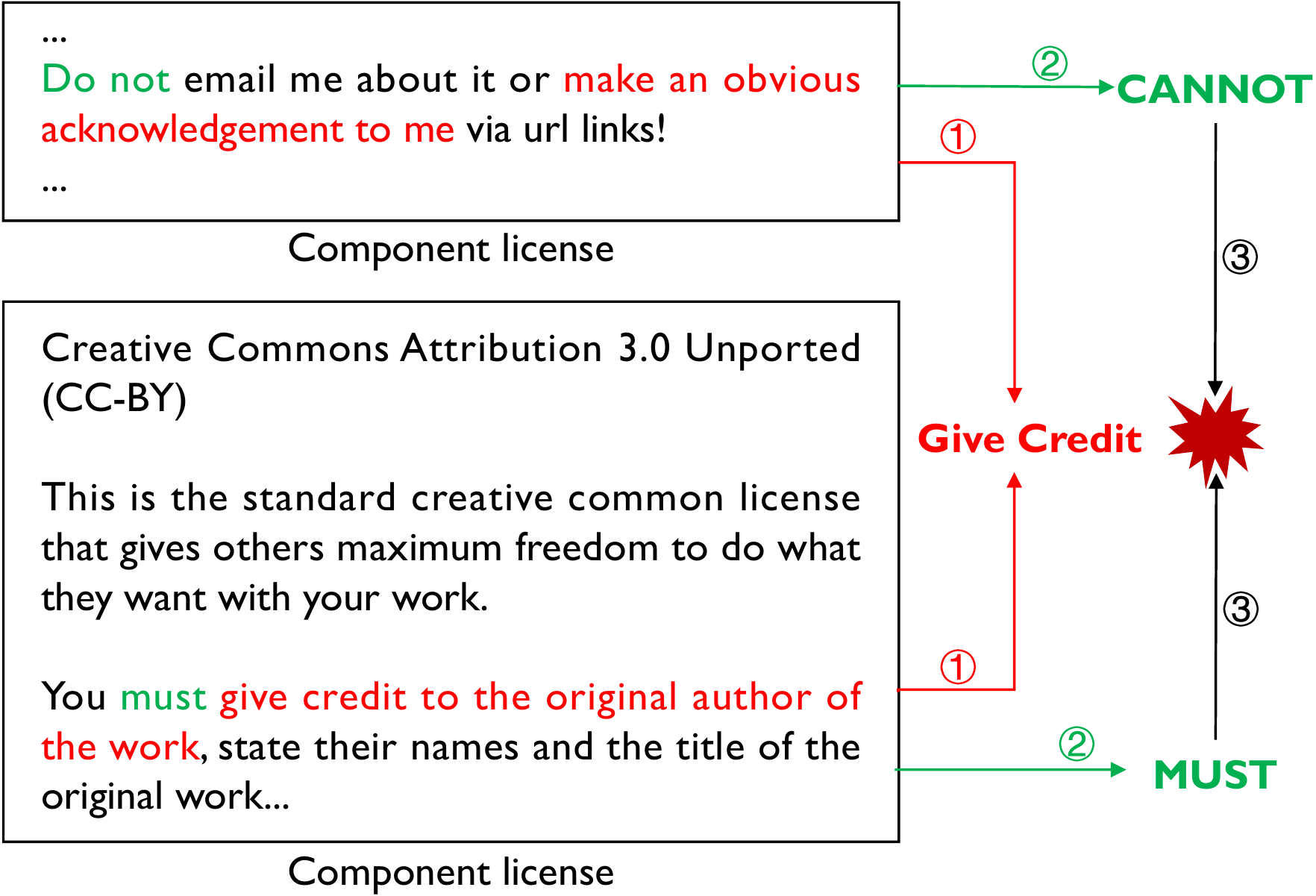}}}
\caption{Two Running Examples}
\label{fig-example}
\end{figure}

\revise{Fig.~\ref{fig:example2} illustrates another example from Faust~\citep{url-example2}, a Python stream processing library. The project contains two component licences, one of which is a custom license that states ``\textit{Do not email me about it or make an obvious acknowledgement to me via url links}''. Another component license is an official license named Creative Commons Attribution 3.0 Unported~\citep{url-CC3.0}, where the authors declare that ``\textit{You must give credit to the original author of the work}''. In this project, two component licenses convey conflict attitudes towards the same license term (i.e., CANNOT give credit \textit{vs.} MUST give credit). 
Since one can not develop a new license to satisfy the two component licenses simultaneously, %that anyone who conforms to the new license will not violate the component licenses, 
we say there exists license incompatibility between these component licenses.
Such license incompatibility can represent a serious threat to the legal use of OSS.}

\iffalse
\begin{figure}[t]
    \centering
    \includegraphics[width=0.7\textwidth]{figs/motivating.pdf}
    \caption{Running Example}
    \label{fig-example}
\end{figure}
\fi

\subsection{Problem Statement}

\revise{In this paper, given a project, we address all types of licenses involved in the project, including the inline, the declared, and the referenced licenses. The collected licenses usually include a project license $PL$ and $n$ component licences, i.e., \{$CL_1$, $CL_2$, $CL_3$, ..., $CL_n$\}. Each license declares rights and obligations that users should comply with. Our problem is to check whether there exists license incompatibility (i.e., with conflict rights or obligations) among all kinds of licenses, with consideration for (1) the differences between project licenses and component licenses, i.e., $pl$ \textit{vs.} $cl_i$ and $cl_i$ \textit{vs.} $cl_j$, where $i$ and $j$ represent the $i^{th}$ and $j^{th}$ component license, respectively; (2) the diversity of licenses; and (3) condition constraints between license terms.
%we aim to first extract all the licenses texts included or referenced by the project (including popular licenses and custom ones which are treated with no discrimination), and detect license incompatibilities among these licenses by extracting the semantics (license terms and the associated rights and obligations) from licenses and detecting license incompatibilities with consideration for (1) the difference of project licenses and component licenses, and (2) condition constraints between different terms.
}

\section{Approach}
\subsection{Overview}
%Considering that our ultimate goal is to \ling{identify license conflicts in software projects through} realizing the recognition of multiple license terms and their multiple authorization polarities, and there are large differences between term recognition's and polarity recognition's domain knowledge and needed techniques. Therefore, this paper divides the task of extracting the license terms into two phases: \textit{localization of term sentence} and \textit{identification of authorization type}, in order to further improve the overall performance on the basis of seeking better solutions separately. 

\begin{figure*}[t]
    \centering
    \includegraphics[width=1\textwidth]{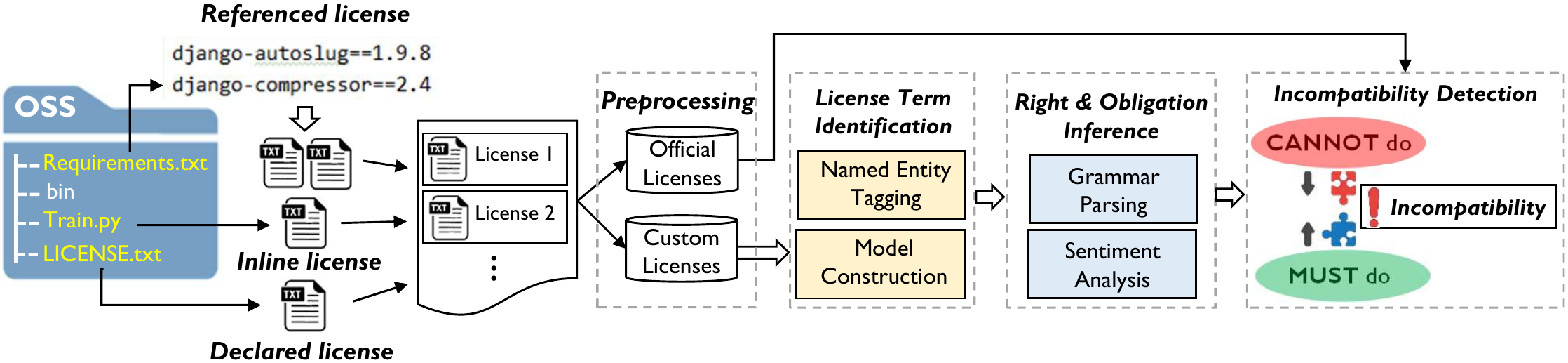}
    \vspace{-2mm}
    \caption{Overview of \tool}
    \label{fig_overview}
\end{figure*}

%Although license texts are usually long and complicated, there are only two kinds of information needed to be captured: license terms (e.g., \textit{redistribute}) and the extent to which licenses are permissive/restrictive (e.g., \textit{cannot}). 
%To identify conflicts of multiple licenses within a project, a fundamental task is to understand license texts and infer the rights and obligations under them. 
%In this paper, we propose a hybrid method that automatically understands license texts and facilitates compatibility analysis.
This section details our approach, \tool, a hybrid method that automatically understands license texts and infers rights and obligations to detect license incompatibility in open source software.
As depicted in Fig.~\ref{fig_overview}, given an open source project, we first extract three types of licenses, i.e., the referenced, the inline, and the declared licenses, for further incompatibility analysis.
After obtaining the set of licenses, the main components of \tool include: \revise{(1) \textit{Preprocessing}, which filters out official licenses and feeds custom ones into the probabilistic model for automatic understandings of license texts}; (2) \textit{License term identification}, which aims to identify the license terms (as displayed in Table~\ref{table-23terms}) relevant to rights and obligations;
(3) \textit{Right and obligation inference}, which infers the stated condition of software use defined by license terms;
(4) \textit{Incompatibility detection}, which automatically analyzes incompatibility between multiple licenses within one project based on the regulations inferred from each license. We detail each phase as follows.

\subsection{\revise{Preprocessing}}

\revise{
%Given a project, we extract three types of licenses: (1) the declared licenses that developers explicitly declare the conditions of software use in a license file; (2) the inline licenses that developers state the rights and obligations in the source code; (3) the referenced licenses that appear in the external links. 
Given a project, after obtaining all the license texts, we filter out official licenses whose rights and obligations have already been known, and feed the other licenses such as newly-defined official licenses and custom licenses into the probabilistic model for automatic understanding of license texts. 
Here, official licenses denote licenses from the Software Package Data Exchange (SPDX)~\citep{url-spdx}.
%Other licenses such as newly-defined licenses, exceptions, and custom licenses are all fed into the detection model for automatic understandings of license contents. 
To filter out official licenses, we note that only license texts that exactly match the listed official licenses are marked as official licenses. For license texts that contain or reference an official license, we extract the official license and then feed the rest texts into the machine learning model to infer additional rules. \revise{Then, given license texts, we remove non-textual parts, check spellings, perform stemming and morphological to obtain the roots of tokens. We utilize the Natural Language Toolkit~\citep{url-nltk} to preprocess license texts.} }

\subsection{License Term Identification}
\label{subsec-LicenseTermIdentification}
%Term entity recognition for phase 2: the clustering algorithm is used to preliminary filter out the sentence sets related to the license terms in each text, and then we use the neural network to extract and transform the features from natural language texts, and finally, the paper uses the discriminant probability model to predict and extract the term entities in the texts. 

In this section, we elaborate the method to identify license terms related to rights and obligations.

\subsubsection{\textbf{Named Entity Tagging}}
 \begin{figure}[t]
    \centering
    \includegraphics[width=0.65\textwidth]{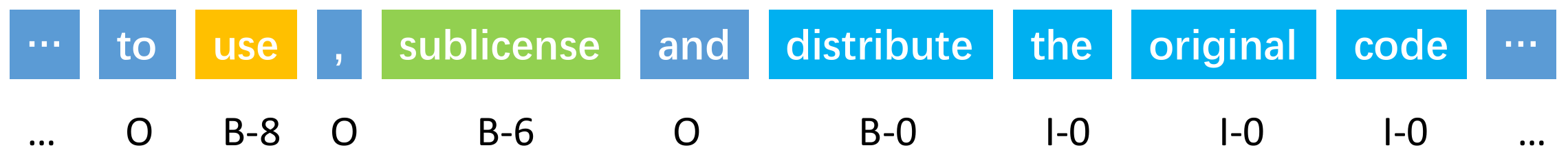}
    \caption{Illustration of the BIO Labelling Mode for Licenses}
    \label{fig_BIO}
\end{figure}
\revise{Since license texts are typically long and complicated, it is not easy to directly interpret license texts. For each license, \tool first identifies \textit{named entities} of license terms which state the conditions of software use. 
In this paper, a \textit{named entity} refers to a specific expression of a \textit{license term} shown in Table~\ref{table-23terms}, which can be a word, a phrase, or a sentence. %The license term may actually display in varieties of forms in real-world license texts, e.g., words, phrases, parts of sentences, which are so called the term entities. Besides, we labeled term entities mainly to discover the terms involved in one license and identify their localization, which is needed in \textit{Right and Obligation Inference} phase. 
Inspired by Named Entity Recognition (NER)~\citep{ISPRS`20-Runyu-NER,guo2021detecting,guo2021key} in natural language processing, this paper utilizes sequence labeling to identify and localize license term entities. Before sequence labeling, \tool first splits each license text into sentences by \textit{Stanford CoreNLP}~\citep{url-corenlp}, an integrated framework for natural language processing.} %(3) after preprocessing, {in the \textit{training phase}}, we employed a clustering technique named CURE (Clustering Using REpresentative algorithm~\citep{IS`01-Guha-CURE}) to group license sentences into clusters based on text similarity, and manually identified clusters that does not state rights and obligations of software use. {In the \textit{inference phase}, we utilize the trained model to automatically filter out sentences irrelevant to the conditions of software use. }
After that, we employ the BIO (Begin, Inside, and Outside) mode~\citep{CSCL`17-Reimers-BIO} to label each token in a sentence. As illustrated in Fig.~\ref{fig_BIO}, \textit{B-X} implies that the current token is at the beginning of a named entity of the $X^{th}$ license term in Table~\ref{table-23terms}. For instance, \textit{B-0} in Fig.~\ref{fig_BIO} implies that \textit{distribute} is the beginning token of a named entity whose license term is the first one in Table~\ref{table-23terms}. Similarly, \textit{I-X} implies that the current token is inside a named entity of the $X^{th}$ license term, and \textit{O} represents a token outside name entities. In the training phase, we manually tag each sentence by the BIO mode, and obtain a training dataset for sequence labelling of license sentences. In the inference phase, \tool automatically predicts the labels of tokens in license texts, so as to identify and localize license term entities. %Note that \textit{X} is the serial number of the term types, 0-22, as the Table~\ref{table-23terms} showed. 
%Through the figure, we can find license term entities like \textit{use}, \textit{distribute the original code}, \textit{each copy contains the above copyright notice} .etc. 

%\subsubsection{\textbf{The Probabilistic Model}}
\subsubsection{\textbf{Model Construction}} 
%\cmt{@sihan, plz slightly reorganize this part...also proof read for the whole section.}
% Then we divided the above data into training set, validation set, and test set according to the ratio of 0.64 : 0.16 : 0.20, plus additional unlabeled data to prepare for the training of the semi-supervised term entity extraction model. For details, see the next section \textit{Evaluation} please. 

\begin{figure}[t]
    \centering
    \includegraphics[width=8cm]{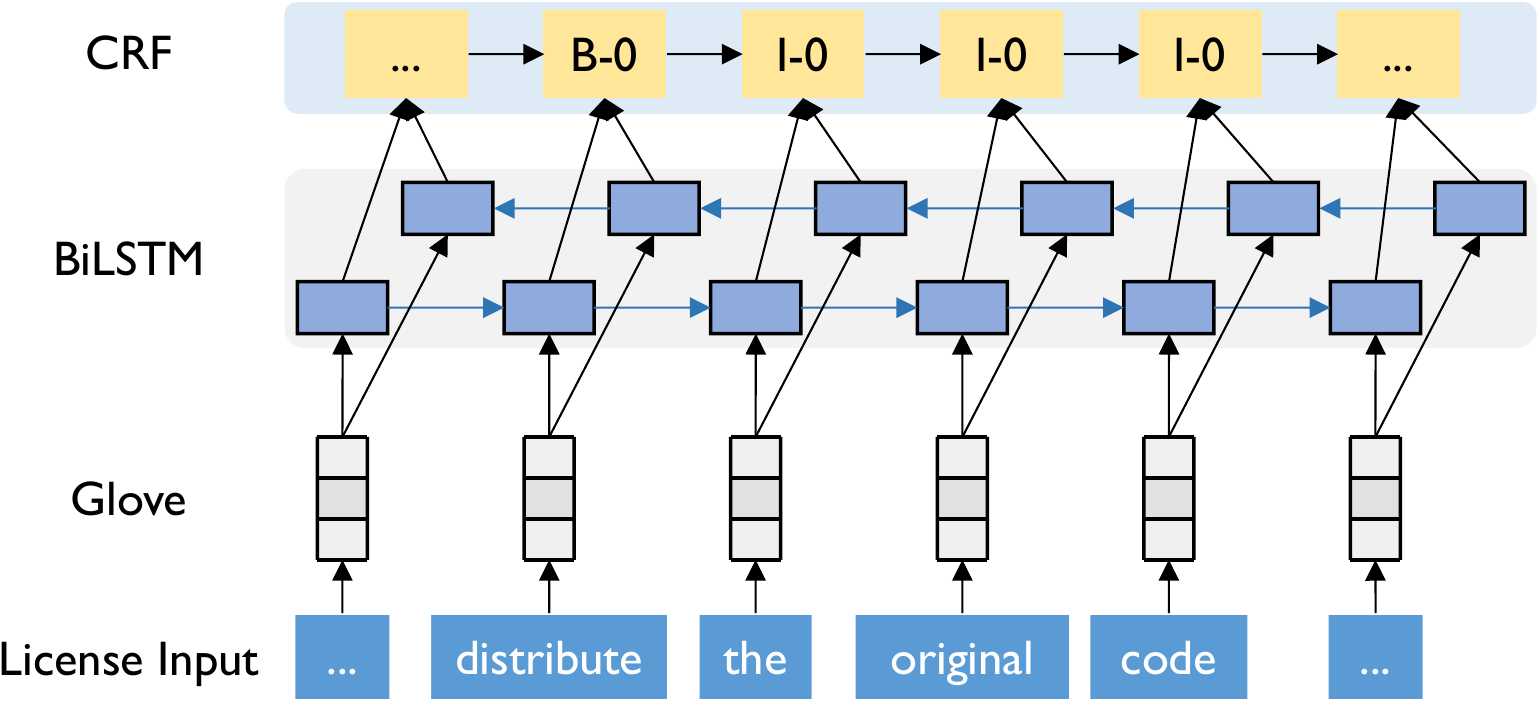}
    \caption{The Probabilistic Model}
    \label{fig_vector}
\end{figure}

Based on the tagged dataset, we train a probabilistic model to predict the label of each token in a license text. As illustrated in Fig.~\ref{fig_vector}, the model consists of three parts: word embedding, sentence representation, and probability calibration. 
(1) \textit{Word embedding}. To serve as the input of the model, we first embed words in license sentences into vectors. \revise{Since licenses are expressed by natural language, we exploit prior knowledge on word semantics and employ the pre-trained Glove model~\citep{url-glove} for word embedding. (2) \textit{Sentence representation}. To represent each sentence in the license text, we feed the results of word embeddings into a bi-Directional Long Short-Term Memory (bi-LSTM)~\citep{CSAI`19-Xia-BiLSTM} model, and learn the representation of each license sentence.} (3) \textit{Probability calibration}. Given a token and its context vector learned by bi-LSTM, the model then calibrates its probability distribution over each category (label) by Conditional Random Fields (CRF)~\citep{USENIX`19-Andow-Policylint}, so that the contextual information represented by the hidden layer state can be utilized to make a global decision. 
{To reduce the labelling effort and enhance performance, we implement the probabilistic model by semi-supervised training. Specifically, after training with labelled samples, we used the trained model to predict license terms for unlabeled samples. Then, all samples, \revise{including labelled samples and other samples with pseudo labels,} were collected to train the model. The rationale behind is that pseudo labelling of unlabelled samples are also predictive. }
Finally, the output of the probabilistic model is a sequence of labels from where license term entities can be directly inferred. 

\revise{For instance, given the first running example in Section 2.4, the label sequence of the sentence \textit{``Redistribution and use in source and binary forms, with or without modification, are permitted''} predicted by the probabilistic model is \textit{\{B-0, I-0, I-0, I-0, I-0, I-0, I-0, I-0, O, O, O, O, O, O\}}. By this means, \tool can localize the license term entity \textit{``redistribution and use in source and binary forms''}, and identify the license term ``Distribute'' from the sentence.}

\subsection{Right and Obligation Inference}
\label{subsec-RightObligationInference}

%Authorization polarity analysis for phase 3: this paper performs grammatical parsing on the sentence, which contains the term entity predicted in the phase 2, to obtain the grammatical sequence corresponding to each word. Then we extract the set of predicates that really affect the term entity polarity. Finally, according to the heuristic rules and the polarity keywords set, the authorization type is calculated and obtained. 
\revise{After localizing license terms from texts, \tool infers the attitudes of originators towards these license terms (e.g., grant/reserve certain rights), which are the rights and obligations stated by the license. \tool achieves this goal by three steps. First, it parses sentences where license terms are localized. Then, based on the grammar parsing, it identifies tokens which convey permissive or restrictive attitudes towards license terms. Finally, it infers the relationships between identified license terms, since some license terms can be the conditions of other terms.} %Based on the identified license terms, we predict the stated condition of software use defined by these terms. 
 %After localizing license term entities, \tool first parses the grammar of the sentences where license term entities are located, then it automatically analyzes the words that may influence the attitudes towards the term entity, so as to infer the rights and obligations implied by licenses.

\subsubsection{\textbf{Grammar Parsing}} 

\revise{To infer the conveyed attitudes towards identified license terms, we first conduct grammatical analysis on the sentences where term entities are localized.} 
%to obtain the combination and dependency relationship among words. 
%
%Natural language processing (NLP) technology can be roughly divided into three levels: lexical analysis, grammatical analysis, and semantic analysis. 
%Among them, grammatical analysis, a key measure in NLP, is used to capture the collocation, modification, and dependence between words within a sentence. It can help downstream semantic analysis tasks and plays an essential role in machine translation, intelligent question answering, text mining, information retrieval, etc. 
%At present, main grammar analysis algorithms are as follows : Context-Free Grammar (CFG), Probabilistic Context-Free Grammar (PCFG), training algorithm, and syntax tree prediction algorithm. 
%\ling{Note that some term-related sentences involve conditional clauses or constraints and may affect further incompatibility detection, therefore, we also extract the conditions associated with the sentences to facilitate further analysis.}
\revise{Since license texts are expressed by natural language under universe grammar rules, we exploited the pretrained  Probabilistic Context-Free Grammar (PCFG) model~\citep{FI`16-Scicluna-PCFG} from \textit{Standford CoreNLP}~\citep{url-corenlp} to assign a probability to each grammar rule}. By this means, a syntax tree can be obtained by selecting the tree with the highest probability scores. 
Fig.~\ref{fig-tree-vis} displays an example of grammar parsing {on the sentence \textit{``You can not refuse such a promise that significant changes must be declared''} from CC-BY-SA-4.0~\citep{url-CCBYSA}. In this case, \textit{``significant changes must be declared''} is an identified term entity, which refers to a license term ``State Changes'' (No.15 in Table~\ref{table-23terms})}. {In Fig.~\ref{fig-tree-vis}, non-terminal nodes such as \textit{NP} and \textit{VB} represent the part-of-speech tags~\citep{url-postagslist}}, and a leaf node {such as \textit{refuse} and \textit{changes}} represents a token in the license sentence. {Table~\ref{table-postags} shows the part-of-speech tags associated with their descriptions}. By traversing the grammar tree, we can obtain the full path of each leaf node, which is the grammatical sequence of the token in the parsed sentence. For example, from Fig.~\ref{fig-tree-vis} we can acquire the path from the \textit{ROOT} node to the leaf node \textit{refuse} as $ROOT\rightarrow S\rightarrow VP\rightarrow VP\rightarrow VB\rightarrow \textit{refuse}$. {It indicates that \textit{refuse} is a {verb} in the verb phrase \textit{``refuse such a promise that significant changes must be declared''} (denoted by $V_1$), which is dominated by \textit{can} and \textit{not}, two tokens in a larger verb phase that contains $V_1$.}
\revise{Taking Fig.~\ref{fig:example1} as an example, it can be inferred that the license term entity \textit{``redistribution and use in source and binary forms''} is a noun phrase (NP) in the sentence, and the attitude towards this license term is affected by the verb phrase (VP) \textit{``are permitted''}.} %Combined the results of license term identification and attitude inference, ``XXX'' can be inferred as one of the rights/obligations conveyed by the license.
%{This paper obtains the full path of every token in a term entity, implying its grammar roles on each grade in the syntax tree, to prepare for the next sentiment analysis.}

%\begin{equation}
%\begin{aligned}
%	ROOT->S->VP->VP->SBAR->S\\->VP->VP->VP->VBN->\textit{declared}
%\end{aligned}
%\label{eq_grammatical-sequence}
%\end{equation}

%acquire the syntax structure of sentences in license texts. obtain the tree with the highest probability score as the final result of license sentence parsing. %from the generated multiple grammar trees,. 

% may describe the implementation in the evaluation.
%In practice, we used the open source \textit{CoreNLP}~\citep{url-corenlp} tool, a basical lexical probabilistic context-free parser, to implement the PCFG algorithm. . % and it also uses the knowledge of dependency parsing.

%Through phase 2, we can obtain a number of recognized license term entities. 
%In the first step of the phase 3, 
%By leveraging the parser, we can obtain the syntax tree of the sentence containing a term entity. 
% you cannot refuse such a promise that significant changes made to software must be declared within the copies

%

%After this step, the grammatical sequences are obtained, which is the precondition to acquire the set of words that have a real influence on the attitude towards each term. 

% $P = \{VB, VBD, VBG, VBN, VBP, VBZ, MD, IN, RB, RBR, RBS\}$

\subsubsection{\textbf{Sentiment Analysis}}
Based on the results of {license term identification and grammar parsing, we perform sentiment analysis to infer the attitudes towards these terms. Generally, the attitudes towards license terms can be categorized into three types, i.e., \textit{MUST}, \textit{CANNOT}, and \textit{CAN}. Given a target sentence $l$ and a license term $k$ identified from the sentence, \tool infers the attitude towards $k$ via sentiment analysis. When the attitudes towards all identified terms are inferred, \tool obtains a summary of rights and obligations of the whole license, i.e., $T(l)=[t_0, t_1, ..., t_{22}]$, where $t_i$ represents the attitude towards the $i^{th}$ term in Table~\ref{table-terms}, $t_i\in \{\text{CAN, CANNOT, MUST, UNKNOWN}\}$, and $0\le i\leq22$. Note that absent license terms are marked with {UNKNOWN}.} Specifically, we first define a set of parts of speech that may convey permissive or restrictive attitudes of authors, i.e., Verb (VB, VBD, VBG, VBN, VBP, VBZ) and Others (MD, IN, RB, RBR, RBS). Tokens with these parts of speech are regarded as \textit{powerful tokens} (PTs).
%tokens whose parts of speech are listed in Table~\ref{table-predicates} are regarded as powerful tokens (PTs). 
\begin{figure}[t]
    \centering
    \includegraphics[width=0.95\linewidth]{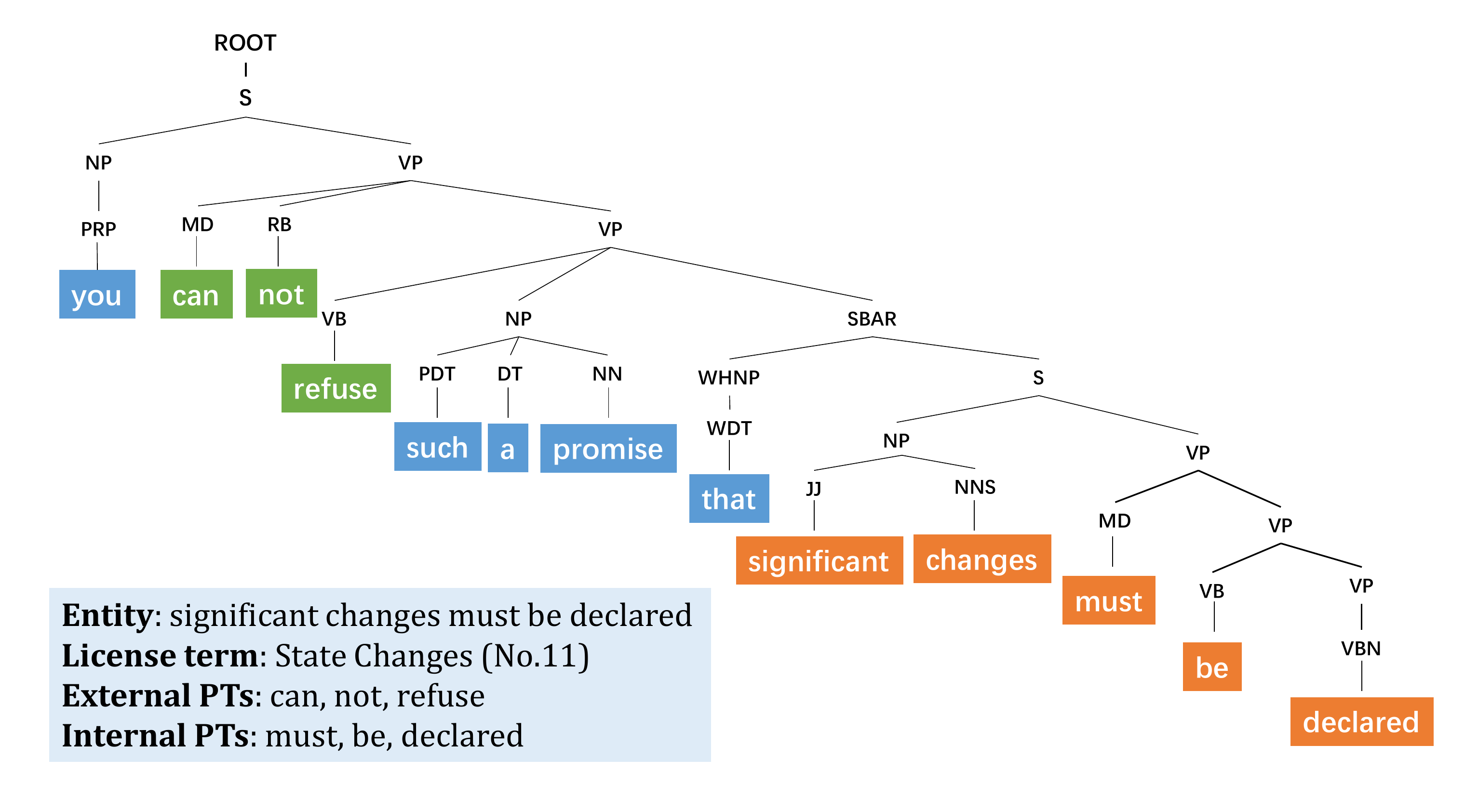}
    \caption{The Parsing Result of a Simplified Sentence in CC-BY-SA-4.0~\citep{url-CCBYSA}}
    \label{fig-tree-vis}
\end{figure}
For instance, the token \textit{you} in Fig.~\ref{fig-tree-vis} is not a powerful token since its part of speech is RPR, which is supposed to have no influence on the attitudes. After obtaining all PTs, we further divide them into two groups according to their relationships with the target entity.
\begin{itemize}
    \item \textbf{Internal PTs}:  An internal PT represents a token contained in the target entity. Typically, internal PTs directly declare the rights and obligations towards the target license term. For example, in Fig.~\ref{fig-tree-vis}, the tokens \textit{must}, \textit{be}, and \textit{declared} are three internal PTs of the entity \textit{significant changes made to software must be declared within the copies}. 
    \item \textbf{External PTs}: As a part of a sentence, the sentiment analysis towards the attitudes needs to consider external PTs which are outside the target entity but have a dominant influence on the attitude towards the target entity. For example, in Fig.~\ref{fig-tree-vis}, tokens such as \textit{can}, \textit{not}, and \textit{refuse} are three external PTs that dominates the attitude of the entire sentence.
\end{itemize}

\begin{table}[htp]
\small
\caption{Partial List of the Part-of-speech Tags and Description, more Tags are Described in~\citep{url-postagslist}}
\label{table-postags}
\scalebox{0.95}{\begin{tabular}{ll||ll}
\hline
\textbf{Tag} & \textbf{Description}                           & \textbf{Tag} & \textbf{Description} \\ \hline
ROOT & Text to process                      & VP  & Verb phrase \\
S   & Simple declarative clause             & SBAR & Clause by a subordinating conjunction \\
NP  & Noun phrase                           & WHNP & Wh-noun Phrase \\
VB  & Verb, base form                       & MD  & Modal    \\
VBD & Verb, past tense                      & IN  & Preposition or subordinating conjunction         \\
VBG & Verb, gerund or present participle    & RB  & Adverb  \\
VBP & Verb, non-3rd person singular present & RBR & Adverb, comparative            \\
VBZ & Verb, 3rd person singular present     & RBS & Adverb, superlative   \\ 
VBN	& Verb, past participle                 & PRP & Personal pronoun \\
PDT	& Predeterminer                         & NN  & Noun, singular or mass \\
DT	& Determiner                            & NNS & Noun, plural \\
JJ	& Adjective                             & WDT & Wh-determiner \\
\hline
\end{tabular}%
}
\end{table}

%The verb in the predicate is the center of the sentence and dominates other components. It is the main way of expressing the overall emotion tendency and emphasis of the sentence. 
%The \textit{predicate words} mentioned in this paper, in addition to the verbs of the predicate, are also expanded based on the characteristics of the license text language, as shown in Table~\ref{table-predicates}.

% \begin{table}[t]
% 	\centering
% 	\small
% 	\caption{Influential Parts of Speech}
% 	\label{table-predicates}
% 	\scalebox{0.92}{\begin{tabular}{c|c}
% 		\toprule
% 		%\hline
% 		\textbf{Type}    & \textbf{Part of Speech}  \\
% 		\hline
% 		Verb & VB, VBD, VBG, VBN, VBP, VBZ \\
% 		\hline
% 		Others & MD, IN, RB, RBR, RBS \\
% 		%\hline
% 		\bottomrule
% 	\end{tabular}}
% \end{table}

%As aforementioned, the term entity may appear in the form of words, phrases, sentence parts, etc., therefore, the predicate words set we are searching for consists of the following two categories:
Given a license term entity, we first collect its internal PTs that directly declare the rights and obligations. Then, based on the parsing results, we search for the external PTs that may dominate the attitudes towards the target entity. By this means, we can acquire a set of PTs that are further used to infer the rights and obligations implied by licenses. Since the expressions of attitudes are not as flexible as those of license terms, we define a set of words and phrases in Table~\ref{table-polarity keywords} as the expressions of each attitude. Finally, for each PT, we apply a heuristic strategy to infer the attitudes. Specifically, we mark a PT with CANNOT or MUST if it belongs to the corresponding expressions listed in Table~\ref{table-polarity keywords}; otherwise, we mark it with CAN. Double CANNOT is offset. 
%{Moreover, as previous studies~\citep{url-choosealicense}, absent terms (i.e., UNKNOWN) are marked with CANNOT since no rights are explicitly granted in the license}. \cmt{need to confirm... the attitude of rights and obligation terms if absent.}

\revise{We note that for right-related terms, as previous studies~\citep{url-choosealicense}, the absence of them are marked with CANNOT since no rights are explicitly granted in the license; while for obligation-related terms, the absence of them are marked with CAN.} %, since there are no mandatory obligations for users to comply with.}
\revise{In this way, we can infer the attitude towards each license term. For example, in Fig.~\ref{fig:example1}, there are no internal PTs in the license term entity \textit{``redistribution and use in source and binary forms''}, and the external PTs are \{\textit{``are'', ``permitted''}\}. According to Table~\ref{table-polarity keywords}, \tool infers the attitude towards the license term \textit{``Distribute''} as CAN, and thus the right conveyed by the project license is \textit{``CAN Distribute''}.} %Similarly, \tool infers that the component license in the running example declares \textit{``CANNOT Distribute''}.}
%Finally, for each PT, we compare it with the key words listed in Table~\ref{table-polarity keywords} it with the keywords of rights and obligations (i.e., CAN, CANNOT, MUST) to finally identify the attitude of each term. Specifically, we summarized the set of keywords through the analysis of the license text corpus, and Table~\ref{table-polarity keywords} shows the keywords for each permissive or restrictive attitude type.
%After collecting internal PTs that directly declare the rights and obligations, we search for the Based on the grammatical sequences obtained in the previous step, we finally find the internal predicate words set and external predicate words set related to the term entity, so as to obtain the final predicate words, in which the final exact attitude is acquired. 

%\noindent \textbf{Determining Authorization Type}

%In this step, the obtained predicate words set is combined with the polarity keyword set for further discrimination, and the polarity type is finally obtained corresponding to the item entity. 

%and the discrimination rules,  (in this discrimination rules, the keywords of the polarity “Can” are not required, so they are not listed here). 

\begin{table}[t]
	\small
	\centering
	\caption{Expressions of Attitudes}
	\label{table-polarity keywords}
	\begin{tabular}{c|c}
		\toprule
		%\hline
		\textbf{Attitude}    & \textbf{Expression}  \\
		\midrule
		CAN & --- \\
		\midrule
	CANNOT	& not, without, notwithstand, refuse, disallow, decline, against,\\
		 &  delete, nor, void, neither, prohibit, remove, don't, no, nothing\\
		% &  \\
		 %&  nor\\
		\midrule
		MUST & must, should, as long as, so long as, shall,\\
		 & provided that, ensure that,ask that, have to\\
		%\hline
		\bottomrule
	\end{tabular}
\end{table}

%To take into account the expression styles used in natural language, and adapt to the specific characteristics of open source software licenses at the same time, our final heuristic rules for determining the attitude type are as follows:
%
%(1) Whether it is ``Cannot'': filter the list of Cannot keywords. Double Cannot keywords is offset, and so on;
%
%(2) If it is not the type ``Cannot'', then judge whether it is ``Must'': If the ``Must'' keyword appears, it is judged as Must, otherwise it is the type ``Can''.
%
%After this step, we have obtained the authorization polarity type on various terms of the open source software license text. 

\subsubsection{\textbf{Condition Relationship}}
\revise{After identifying license terms and the attitudes towards them, rights and obligations related to each term can be inferred (e.g., CANNOT Redistribute). Although we treat each license term independently in the previous steps, license terms can also be the conditions of other terms. For instance, \textit{``you can modify if you state changes''}. In this case, there are two license terms (i.e., ``Modify'' and ``State Changes''), and the right (i.e., ``Modify'') is only granted under certain constraint (i.e., ``State Changes''). To address this issue, we analyze the condition relationships between license terms. Specifically, based on the parsing results, we identify the conditional clauses that state the conditions of software use. License terms in the conditional clauses are the conditions of terms in the main clause. Finally, we separately assume the condition is satisfied or not, and update the attitudes of license terms in both conditional and main clause. We describe incompatibility analysis when faced with conditions in Section \ref{subsec:incom}.}

%%%%%%%%%%%%%%%%%%%%%%%%%%%%%%%%%%%%%%%%%%%%%%%%%%%%%%%%%%%%%%%%%%%%%%%%%%%%%%%%%%%%%%%%%%%%%%%%%%%%%%%%%%%%%%%%%%%%%%%%%%%%%%%%%%%%%%%%%%%%%%%%%%%%%%%%%%%%%%%%%%%%%%%%%%%%%%%%%%%%%%%%%%%%%%%%%%%%%%%%%%%%%%%%%%%%%%%%%%%%%%%%%%%%%%%%%%
\subsection{Incompatibility Detection}  
\label{subsec:incom}

\SetKwInput{KwInput}{Input}
\SetKwInput{KwInput}{Input}
\SetKw{Let}{let}
\SetKw{Continue}{continue}
\SetKw{Break}{break}
\SetKw{Create}{CREATE}
\begin{algorithm2e}[t]
%\color{blue}
    \footnotesize
	\setcounter{AlgoLine}{0}
	\caption{{License Incompatibility Detection}}
	\label{algo}
	\DontPrintSemicolon
	\SetCommentSty{mycommfont}
	%\SetAlgoLined
	{
	    \KwIn{: $l_1<t_1, atti_{t1}>$ and $l_2<t_2, atti_{t2}>$: A pair of licenses with extracted terms and attitudes}
    	\KwOut{$Incompatible$ or $Compatible$}
    	\SetKwFunction{FCheck}{condiCheck}
    	\SetKwFunction{IncompCheck}{checkImcomp}
        \SetKwProg{Fn}{Function}{:}{}
                \eIf{$l_1.hasCondi()$ $\vee$ $l_2.hasCondi()$}{
                \tcp*[h]{if $l_1$ or $l_2$ has conditions, check compatibility when the conditions are satisfied or not.} \;
                    $R$ = condiCheck($l_1$, $l_2$)\;
                    \ForEach{$(r1, r2) \in R$}{
                        \If{($l_1.isCL$ $\wedge$ $l_2.isCL$) $\wedge$ $\neg$ ($r_1$ $\vee$ $r_2$)}{  \tcp*[h]{when $l_1$ and $l_2$ are both component licenses, under both conditions the licenses are incompatible.} \;
                        \Return $Incompatible$ \;
                        }
                        \If{($l_1.isCL$ $\wedge$ $l_2.isPL$) $\wedge$ $\neg$ ($r_1$ $\wedge$ $r_2$)}{
                        \tcp*[h]{when $l_1$ is a component license and $l_2$ is a project license, at least under one condition the licenses are incompatible.} \;
                        \Return $Incompatible$ \;
                        }
                    }
                    \Return $Compatible$\;
                }
                %{\ForEach{$term \in (t_1\vee t_2) \wedge term.hasNoCondi()$ }{
                {\eIf{$\neg$checkIncomp ($l_1$, $l_2$)}{
                \tcp*[h]{For terms without conditions, use Rule1\&Rule2 to check compatibility for each term.} \;
                    
                        \Return $Incompatible$ \;
                    }
                 {
                \Return $Compatible$\;}   
                }

        \SetKwProg{Pn}{Function}{:}{\KwRet}
        \Pn{\FCheck{$l_1$, $l_2$}}{
            $R<r_1, r_2> \gets \emptyset$   \tcp*[h]{Incompatibility detection result pairs when considering the path conditions} \;
            $\mathcal{T} \gets t_{condi}(l_1, l_2)$ \tcp*[h]{Terms with conditions}\;
            \ForEach{$term \in \mathcal{T}$}{
                \tcp*[h]{Assume the condition is \textit{True}, update the attitudes of the related terms, and detect incompatibility.} \;
                $condi \gets TRUE$ \;
                $L<l_{1}^{'}, l_{2}^{'}>$ $\gets$ UpdateAtti($term$, $l_1$, $l_2$, $condi$)\;
                $r_{true} \gets$ checkIncomp ($l_{1}^{'}$, $l_{2}^{'}$)\;
                
                \tcp*[h]{Assume the condition is \textit{False}} \;
                $condi \gets FALSE$ \;
                 $L<l_{1}^{'}, l_{2}^{'}>$ $\gets$ UpdateAtti($term$, $l_1$, $l_2$, $condi$)\;
                $r_{false} \gets$ checkIncomp ($l_{1}^{'}$, $l_{2}^{'}$)\;
                
                $R \gets R \bigcup <r_{true}, r_{false}>$ \;
            }
            \Return $R$
        }
     }

\end{algorithm2e}

%\textcolor{green}{We process \textit{declared licenses} and \textit{referenced licenses} differently. For declared licenses, the term extraction model already acquired in phase~\cref{subsec-LicenseTermIdentification} and phase~\cref{subsec-RightObligationInference} is applied to understand and interpret them, so as to obtain their existing rights and obligations towards various license terms. 
%For referenced licenses, we obtained their license names and versions first, directly from the project code, or indirectly linked by the names of software packages as described in~\cref{sec:motivatingstudy}. Their term labels including rights and obligations tags can be predicted by our proposed model similarly, as Figure~\ref{fig_overview} showed.  }

%\cmt{1. mention how we deal with terms with conditions.  2. check whether we can handle dual-license projects, if yes, we can add it here, otherwise discuss it in the limitation or threats. 3. use the running example to explain.}
%Based on the rights and obligations \ling{(together with the associated conditions/constraints)} implied by each license, \tool automatically analyzes incompatibility between multiple licenses within the same project. 
\revise{As previous studies~\citep{gangadharan2012managing,jss`17-Kapitsaki-SPDX}, we define license compatibility as the ability to combine multiple licenses into the same software product. Some incompatible examples are aforementioned in the running examples.
Based on the rights and obligations implied by each license, 
it is still challenging to accurately detect license incompatibility due to: (1) project licenses (PL) should be more restrictive than component licenses (CL), i.e., PL and CL should be treated differently.
(2) Some rights or obligations are only granted under certain conditions.}
%\tool automatically analyzes incompatibility between multiple licenses within the same project. 
%
%Typically, a software product may contain a project license (e.g., LICENSE.txt), which states the conditions of software use. In addition, when incorporating third-party software packages or reusing other code, licenses that accompany each component should also be conformed to. 
%\gy{It is noticing that, the definition sets \{\textit{the project license}, \textit{the component license}\} and \{\textit{the declared license}, \textit{the referenced license}, \textit{the inline license}\} are not the same. The former emphasizes the functions and influence of licenses on license compliance in one OSS project and the latter emphasizes the presentation styles of licenses.} \cmt{need to mention?}
%

\revise{To address the first challenge, }
given a target project, \tool discriminates between project and component licenses, and define license compatibility rules as follows:
\begin{itemize}
    \item \textbf{Rule 1: Compatibility between component licenses}. Two component licenses $CL_1$ and $CL_2$ are compatible if it is possible to develop a new license $L$ that anyone who conforms to license $L$ will not violate license $CL_1$ and $CL_2$. %Alternatively, there exists a license $L$ that anyone who conforms to license $L$ will not violate the rules defined by $L_1$ and $L_2$.
    \item \textbf{Rule 2: Compatibility between a project license and a component license}. A project license $PL$ is ``one-way compatible" with a component license $CL$, if anyone who conforms to license $PL$ will not violate license $CL$.   
\end{itemize}
Based on the definitions, 
%we detect license incompatibility from two aspects. Given a target project, if it contains a $PL$, we analyze the compatibility between the $PL$ and each $CL$; otherwise, we only analyze the compatibility between $CLs$.
%Specifically, 
for each pair of licenses,  we compare their attitudes towards the same license term one by one. As illustrated in \revise{Table~\ref{table:pl-vs-cl}}, a project license $PL$ is ``one-way-compatible" with a component license $CL$, if $PL$ is the same or more restrictive than $CL$. For instance, MUST and CANNOT are stricter than CAN, so that anyone who conforms to MUST or CANNOT will not violate CAN. For a pair of $CLs$, \revise{as shown in Table~\ref{table:cl-vs-cl}}, they are compatible with each other only when they can be incorporated into the same software product, so only CANNOT and MUST towards the same license term (e.g., \textit{distribute}) are regarded as incompatible attitudes in this case. Finally, we note that as declared by %\ling{Github [XXX] and} 
\textit{choosealicense}~\citep{url-choosealicense}, the absence of a license implies that nobody can copy, distribute, or modify the work. \revise{Therefore, if a project is \textbf{without a PL}, we consider all rights are reserved, so the PL is the most restrictive license that are compatible with any CL in the same project. In this case, we only check compatibility among CLs.}
\revise{Similarly, if a license term is not mentioned in the license, all right-related terms are set to CANNOT, and all obligation-related terms are set to CAN by default. For instance, if the license text does not mention anything about redistribution, then it means nobody can redistribute the work.}

\revise{%Given the first running example in Section 2.4, 
For example, in Fig.~\ref{fig:example1}, \tool infers that the project license declares {``CAN Distribute''}, while the component license declares {``CANNOT Distribute''}. Since the project license is more permissive than the component license upon the same license term, \tool detects incompatibility between two licenses. For the second example in Fig.~\ref{fig:example2}, \tool infers that the first component license conveys {``CANNOT give credit''}, while the second component license states {``MUST give credit''}. Since it is impossible to comply with both of them simultaneously, \tool infers there exists incompatibility between two component licenses.}

\revise{To address the second challenge, we consider both conditional cases separately when handling terms with conditions, i.e., separately assume the condition is True or False to detect potential incompatibility issues. 
In this way, we can eliminate the effect of conditions, and employ the above incompatibility checking rules for both cases individually.}

\revise{Algorithm~\ref{algo} details the detection process. Specifically, it takes as input a pair of licenses within a projects ($l_1$, $l_2$), with extracted terms ($t_1$, $t_2$) and the associated attitudes ($atti_{t1}$, $atti_{t2}$), and outputs whether there exist license incompatibility issues.
For $l_1$ and $l_2$, if at least one of them has terms with conditions, the method \texttt{condiCheck()} is invoked to obtain the results under both conditional cases (Lines 1-2 and Lines 14-25), otherwise, we directly invoke the \texttt{checkIncomp()} method to detect incompatibility (Lines 10-13). Here, the \texttt{checkIncomp()} method detects incompatibility using Rule1 and Rule2 described above.
As for the \texttt{condiCheck()}, we initialize a list $R<r_1,r_2>$ to store all the result pairs of both conditional cases, and extract all the terms $\mathcal{T}$ with conditions (Lines 15-16). For each $term$ in $\mathcal{T}$, we first assume the condition is $True$, update the attitudes of the corresponding terms, and check the incompatibility (result stored in $r_{true}$) (Lines 18-20); then we assume the condition is $False$, update the attitudes of the related terms again, check incompatibility in this case, and the checking result is stored in $r_{false}$ (Lines 18-20).  For instance, given a conditional term \textit{``CAN modify if you state changes''}, we split it into two cases: (1) MUST state changes \& CAN modify (2) CANNOT modify. Then, we separately check license incompatibility for both cases and obtain the detection results $<r_{true}, r_{false}>$ for both cases. Finally, the results of both cases are stored in $R$ and returned (Lines 24-25). 
For each result pair in $R$, we treat PL and CL differently. Specifically, if $l_1$ and $l_2$ are both CL, according to Rule 1, they are regarded as $incompatible$ only when both results show incompatibility (Lines 4-5).
If $l_1$ or $l_2$ is a PL, according to Rule 2, they are regarded as $incompatible$ if there exists at least one case showing 
incompatible (Lines 6-7).}

%\begin{figure}
%  \centering
  %\subfigure{\includegraphics[width=0.33\linewidth]{figs/com1.pdf}}
  %\subfigure{\includegraphics[width=0.31\linewidth]{figs/com2.pdf}}
%  \subfigure{\includegraphics[width=0.75\linewidth]{tosem/figs/compatibility-20210805.pdf}}
%  \caption{Rules of Incompatibility Detection \cmt{change l1 l2 to lc if not use table.}}
%  \label{compatibility}
%\end{figure}

\begin{table}[]
%\color{blue}
\begin{minipage}[t]{0.45\textwidth}
\centering
\scriptsize
\caption{Compatibility between PL and CL}
\begin{tabular}{|c|c|c|c|}
\hline
\diagbox{\textbf{PL}}{\textbf{CL}}     & \textbf{CAN}                   & \textbf{CANNOT}                & \textbf{MUST}                  \\ \hline
\textbf{CAN}    & \cmark & \xmark & \xmark \\\hline
\textbf{CANNOT} & \cmark & \cmark & \xmark \\\hline
\textbf{MUST}   & \cmark & \xmark & \cmark \\\hline
\end{tabular}
\label{table:pl-vs-cl}
\end{minipage}
\begin{minipage}[t]{0.45\textwidth}
\centering
\scriptsize
\caption{Compatibility between CL$_1$ and CL$_2$}
\begin{tabular}{|c|c|c|c|}
\hline
\diagbox{\textbf{CL$_1$}}{\textbf{CL$_2$}}     & \textbf{CAN}                   & \textbf{CANNOT}                & \textbf{MUST}                  \\ \hline
\textbf{CAN}    & \cmark & \cmark & \cmark \\\hline
\textbf{CANNOT} & \cmark & \cmark & \xmark \\\hline
\textbf{MUST}   & \cmark & \xmark & \cmark \\\hline
\end{tabular}
\label{table:cl-vs-cl}
\end{minipage}
\end{table}

\section{Evaluation}

In this section, we present the evaluation results of \tool to show that it can efficiently detect license incompatibility for open source software. Specifically, we first present the performance of \tool in two phases (license term identification and attitude inference), and then demonstrate the effectiveness of \tool to detect incompatibility \revise{compared with the state-of-the-art tools.}
%For comparative analysis, we also report the results of three state-of-the-art tools for license analysis, i.e., FOSS-LTE~\citep{APSEC`17-Kapitsaki-termsIdentifying}, a graph-based method~\citep{jss`17-Kapitsaki-SPDX}, and Librariesio~\citep{url-librariesio}. %\sihan{how about the github one?} 

\subsection{\textbf{Preparation}} 
\label{sec:dataset}

\subsubsection{\textbf{Data Preparation}} 

\iffalse
\begin{table}[t]
	\centering
	\small
	\color{blue}
	\caption{The Sizes of Datasets}
	\scalebox{0.9}{\begin{tabular}{cccc}
		\toprule
		%\hline
		\textbf{Phase} & \textbf{Training} & \textbf{Validation} & \textbf{Testing}\\
		\midrule
		License Term Identification  & 13,980 (XX\%) & 3,495 (XX\%) & 4,369 (XX\%) \\
		Right and Obligation Inference  & 13,980 (XX\%) & 3,495 (XX\%) & 4,369 (XX\%) \\
		Incompatibility Detection  & - & - & 200 (projects) \\
		%\hline
		\bottomrule
		\label{table-datasets}
	\end{tabular}}
\end{table}
\fi
%Table~\ref{table-datasets} shows the statistics of datasets used in the experiments. 
\revise{The evaluation was conducted in three phases (i.e., license term identification, right and obligation inference, and the overall incompatibility detection). In the first two phases, LiDetector performed tagging, training, and testing in \textbf{sentences}. %To evaluate the performance of \tool in each phase (license term identification, right and obligation inference, and incompatibility detection), we first constructed a dataset comprising labelled license texts. 
Specifically, we collected the license sentences from two sources: (1) \textbf{tldrlegal}~\citep{url-tldrlegal}, a platform where licenses can be uploaded, summarized, and managed by users. We collected license sentences accompanied with license term tags (i.e., CAN, CANNOT, and MUST), and obtained \revise{{11,973} labeled sentences from {212 labelled licenses on} \textit{tldrlegal}. }%212 labelled licenses
%To ensure the correctness of these tags, three co-authors manually verified and cross-validated the results of labelling.
%so that license terms are no omitted or incorrectly labelled. %However, not all the licenses are with tags, finally obtained 212 licenses in total. 
%classification and recommended platform for open source software licenses, where there are a large number of license texts with term tags, i.e., terms are extracted and labeled with \textit{CAN}, \textit{CANNOT}, \textit{MUST} tags for each license.
%After an in-depth analysis, we noticed that among the 212 licenses, some terms are still missing correct tags. Therefore, we invited two senior software engineering students with extensive industry experience to manually check and label them. 
%We recorded the inter-coder agreement ratio for evaluating labeling process. A Cohen’s Kappa of 0.604~\citep{F`05-kappa-statistic} and an agreement ratio (i.e., the percentage of term entities that are labeled by the two researchers at the same time) of 78.96\% are received, which indicates moderate agreement~\citep{EPM`60-agreement-ratio}. Disagreements in their tags were solved by open discussion until the researchers reach a common interpretation of the terms. 
%
(2) \textbf{Github}. We crawled 1,846 popular projects with {more than} 1,000 stars and extracted licenses from them. 
Then, we filtered out duplicated licenses %that are the same 
%with those collected from \textit{tldrlegal}, 
and finally obtained %{942} licenses 
{48,275} sentences from GitHub. %These sentences were divided into two groups: one group to construct the labeled dataset and another group for unsupervised learning.
\revise{Then, we randomly selected 9,871 sentences (i.e., 20\%) to carefully label, and the remaining 38,404 sentences from 754 unlabelled licenses were fed into the semi-supervised learning to enhance the performance of identifying license terms.}
{In total, we obtained %400 labeled licenses 
{21,844} labeled sentences (i.e., {11,973} from \textit{tldrlegal} and {9,871} from Github)}. %Note that both license term identification and sentiment analysis were trained and tested in sentences. 
\revise{We randomly split these samples into the training and testing datasets by 4:1, and further split the training dataset for training and validation by 4:1. Therefore, in the first two phases, the training, validation, and testing dataset consists of 13,980, 3,495, and 4,369 license sentences. Three authors cross-validated the labels and a lawyer from Yingke Law Firm\footnote{www.yingkeinternational.com} was involved in the validation. All the techniques and tools were evaluated on the same testing dataset.} We have made the dataset publicly available~\citep{lidetector}.} 

%each of which were manually labelled and cross-validated by two senior students

%we obtained 400 labelled licenses (i.e., 212 from \textit{tldrlegal} and 188 from Github) with cross-validated labels to form a ground-truth dataset, which is released online~\citep{lidetector}. 

%\ling{These licenses are without term tags, thus we also use the aforementioned method to manually label them for further evaluation.}
%These licenses are without term tags, to construct the ground-truth dataset and for evaluation purpose, we randomly selected 188 licenses for manual labeling by using the above method. 
%These licenses are without term tags, thus we also use the aforementioned method to manually label them for further evaluation, depending on the exact needs. 

%Specifically, we obtained 50 clusters through clustering, and then get 39 clusters that meet the conditions through manual review. 
%\ling{why delete it? see the commented sentence.}
%\textcolor{green}{GY: only 400 licenses were labeled, within 212+754, }
%\ling{@Gaoya, what is the ground-truth data (how many labeled licenses) for experiment 1 \& 2}

\begin{figure}[t]
    \centering
    \includegraphics[width=0.5\textwidth]{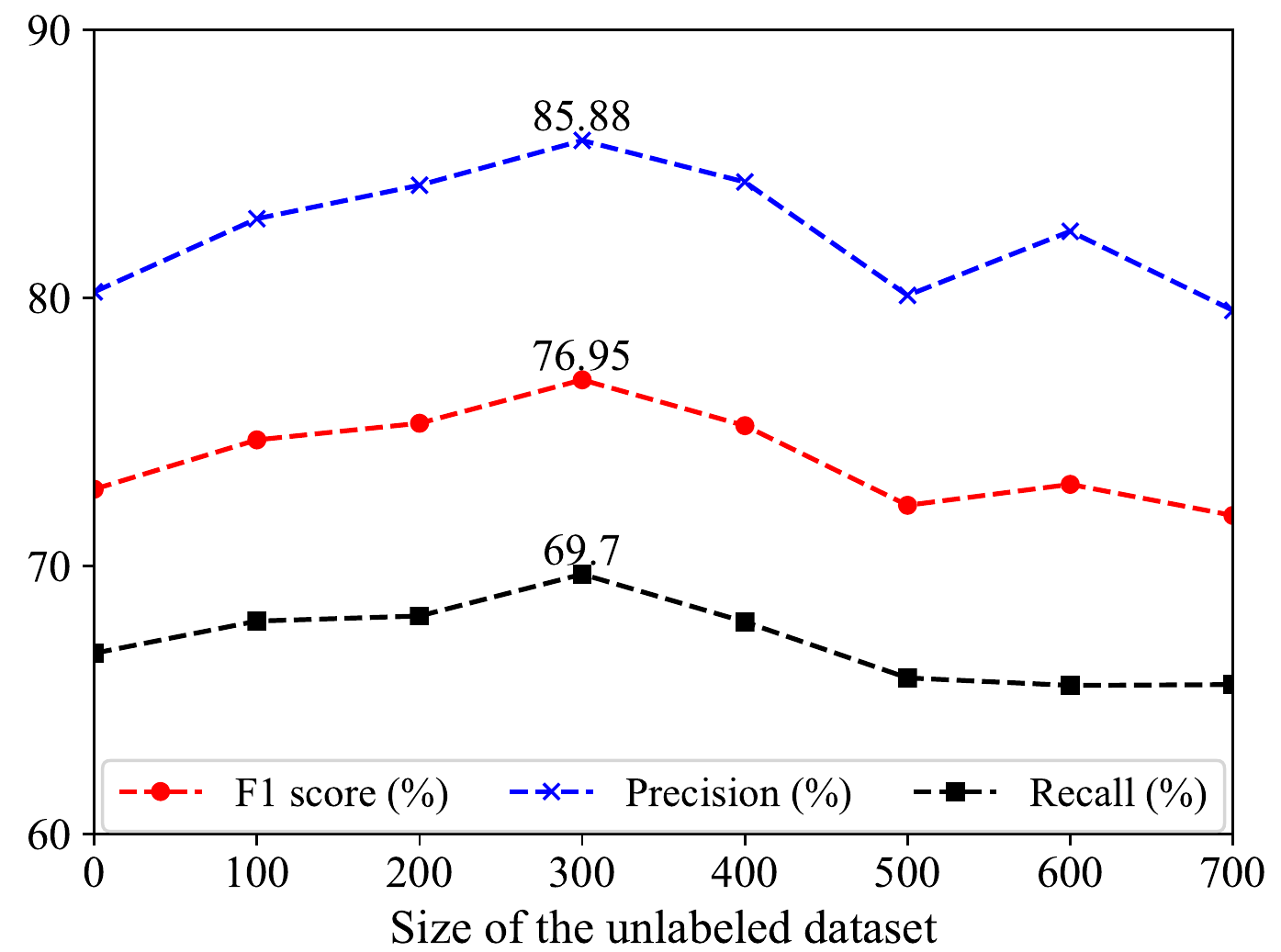}
    \caption{Unlabeled Data Size Determination for Semi-supervised Training of \tool}
    \label{fig-semi-data}
\end{figure}

%\noindent $\bullet$ \textbf{Parameter Settings for Semi-Supervised Learning.}
\subsubsection{\textbf{Parameter Settings for Semi-Supervised Learning}}
%\ling{(except for the added unlabeled ones, indicate the training set, evaluation set and the validated set, and their roles)}
To save the manual efforts of labelling licenses, we employ a semi-supervised learning method to identify license terms, so that both labelled and unlabelled licenses can be utilized to train the probabilistic model. The rationale behind is that pseudo labels predicted by the model are also predictive and may benefit the model performance when involved in training. %\sihan{(explain why we need semi-supervised learning better)}
%{Under the premise of limited labeled data, this paper puts into available unlabeled data to try to obtain a better performance for the license term identification ability of \tool}. 
Before evaluating the performance of \tool, we first investigate how the number of unlabelled samples influences the performance of \tool. 
{
Specifically, we randomly select a set of numbers of licenses (i.e., 100, 200, 300, 400, 500, 600, 700) from the {754} unlabelled licenses, \revise{extract the sentences}, and add them into the training dataset respectively. We use three metrics to evaluate the performance of \tool, i.e., precision ($P = \frac{TP}{TP+FP}$), recall ($R = \frac{TP}{TP+FN}$), and F1 score ($F1 = \frac{2*P*R}{P+R}$). 
To avoid bias, for each parameter, we randomly select the same number of unlabelled licenses for three times, and average the results.}
{Fig.~\ref{fig-semi-data} shows the performances of \tool accompanied with different sizes of unlabelled licenses. It can be observed that the performance of \tool peaked when 300 unlabeled licenses were added for semi-supervised training. %\sihan{(briefly describe Fig. 7 and give a possible explanation here. for example, increased first ..., which is consistent with the intuition that pseudo labels are also predictive and can enhance the performance. Then decreased...because...)}
{In the Fig.~\ref{fig-semi-data}, it can be observed that when more unlabeled data were involved in the semi-supervised training, the performance of \tool first increased and reached a peak when the size of unlabeled data is 300. It is consistent with the hypothesis that pseudo labels are predicted for the unlabeled data. However, when the size of unlabeled data is larger than 300, the performance of \tool decreased. A possible reason could be that pseudo labels might introduce some noisy data, which prohibited the enhancement of model performance.} 
\revise{Therefore, to achieve the best performance, we decide to add {300} unlabeled licenses ({i.e., \textbf{15,312} sentences}) into the training dataset for our semi-supervised learning phase in \tool in the following experiments.}
\revise{All experiments were conducted on a machine with Intel (R) Core (TM) i5-7200U CPU @ 2.70 GHz and 4.00 GB RAM.
}
}

%%%%%%%%%%%%%%%%%%%%%%%%%%%%%%%%%%%%%%%%%%%%%%%%%%%%%%%%%%%%%%%%%%%%%%%%%%%%%%%%%%%%%%%%%%%%%%%%%%%%%%%%%%%%%%%%%%%%%%%%%%%%%%%%%%%%%%%%%%%%%%%%%%%%%%%%%%%%%%%%%%%%%%%%%%%%%%%%%%%%%%%%%%%%%%%%%%%%%%%%%%%%%%%%%%%%%%%%%%%%%%%%%%%%%%%%%%
\subsection{\textbf{Evaluation on License Term Identification}}
\label{subsec:phase1}

\subsubsection{\textbf{Setup}}

To evaluate the ability of \tool to identify license terms, we compare it with FOSS-LTE~\citep{APSEC`17-Kapitsaki-termsIdentifying}, a state-of-the-art tool that extracts license terms from texts, and
%To conduct a comprehensive study, 
two natural language processing (NLP) techniques,
%are also adapted to identify license terms, 
i.e., regular matching~\citep{USENIX`19-Andow-Policylint} and semantic similarity~\citep{ICML'14-Quoc-doc2vec}. 
%We perform comparative experiments on the ground-truth dataset (i.e., 400 licenses). 
Moreover, to study the influence of unlabelled training samples, we also conduct an ablation study, where \tool trained without unlabeled training samples is denoted by \textsc{LiDetector$^-$}. To conduct a fair comparison, we carefully implement the following NLP techniques and adapt them for license term identification.
%, so as to conduct a fair comparison. 
%Since \ling{the terms for each method are customized in a certain context,}
%we need to implement the first three methods in our context to conduct the comparison.
%inspired by some knowledge extraction techniques involved in papers from other research tasks.  
Specifically, 
%the descriptions of three baselines in addition to FOSS-LTE are as follows:
\begin{itemize}
    \item \textbf{Regular Matching} %(Regular M.)~
    \citep{USENIX`19-Andow-Policylint}, which predefines a set of keyword patterns to guide license term identification. To implement this strategy, we manually analyzed license texts to find as more expressions of license terms as possible. Finally, 72 patterns were found for 23 license terms, as listed on our website~\citep{lidetector}.
    %Examples of multiple patterns that convey the meanings of license terms can be seen in Table~\ref{table-keywordMatching-example}. 
    We then use CoreNLP~\citep{url-corenlp} to split licenses into sentences and search for predefined expressions of license terms by regular matching.%operate on each possible keyword form of each term type to localize the sentence related to each term. Finally, the license-related terms can be obtained. 
    
    \item \textbf{Semantic Similarity} %(Semantic S.)~
    \citep{IJCNN`18-Doc2Vec-method}, which utilizes doc2vec~\citep{ICML'14-Quoc-doc2vec} to represent text as a vector and searchs for similar expressions of license terms. 
    To adapt this method, we trained the doc2vec model on the aforementioned {754} unlabeled licenses {(i.e., 38,404 sentences)}, so as to learn the representations of license sentences. After that, we manually analyzed license sentences and collected a set of representative sentences for each license term. In total, we collected 51 representative sentences for 23 license terms, as listed on our website~\citep{lidetector}.
    %Examples of representative sentences can be seen in Table~\ref{table-sematicMatching-example}. 
    Finally, given a license sentence, we use the trained doc2vec model to predict its representation. %, and calculate the cosine similarities between the representations of each representative sentence and the target sentence.  of each standard sample sentence to identify the sentence that expresses the related license term. 
    Sentences with similar representations are considered to convey similar regulations about software use. 
    %\item \textbf{Topic model (TM)}~\citep{MTA`19-Jelodar-LDA}\sihan{mark!}, which is an unsupervised machine learning technique that automatically clusters phrases and similar expressions that best characterize a set of documents. 
    %Here, we use the idea to model the topics of the sentences in license texts, and then link the topics with license terms. We first build a Latent Dirichlet Allocation model~\citep{MTA`19-Jelodar-LDA}, a commonly used topic model, \textcolor{brown}{by using the same data as the semantic matching method. }
    %The corpus used here is the same as the previous method. 
    %The output is the topics corresponding to each sentence.
    %a number of topics %(each topic is represented by a vocabulary), and it also outputs the topic corresponding to each sentence. 
    %Meanwhile, we select a vocabulary set to represent each term. 
    %Then, the cosine similarity between each term and each topic is computed through embedding their vocabulary into word vectors through a word2vec model (which is trained using the same corpus), to find the closest term for each topic. 
    %Finally, the term for each sentence is obtained by their connections to topics respectively, i.e., possible terms for each license sentence are identified.  
    
\end{itemize}

% \begin{table}[t]
% 	\centering
% 	\small
% 	\caption{Keyword Examples of Each Term for \textit{Keyword Matching} (Full list on our website \citep{lidetector})}
% 	\label{table-keywordMatching-example}
% 	\scalebox{0.92}{\begin{tabular}{c|l}
% 		%\toprule
% 		\hline
% 		\textbf{Term}    & \textbf{Word Combination Forms}  \\
% 		\hline
% 		Commercial & commercial Use, sell, offer of sale,resale, use\\ 
% 		Use & for any purpose \\
% 		\hline
% 		Trademark & use trademark, use service mark, use product\\
% 		Use & name, use authors' names for advertising \\
% 		\hline
% 		%\bottomrule
% 	\end{tabular}}
% % 	\begin{center}
% % 	    \scriptsize
% % 	    \textit{The full list can be found on our website [XX].}
% % 	\end{center}
% \end{table}

% \begin{table}[t]
% 	\centering
% 	\small
% 	\caption{\ling{XXX} Examples of Each Term for \textit{Semantic Matching}(Full list on our website~\citep{lidetector})}
% 	\label{table-sematicMatching-example}
% 	\scalebox{0.92}{\begin{tabular}{c|l}
% 		%\toprule
% 		\hline
% 		\textbf{Term}    & \textbf{Term Standard Samples}  \\
% 		\hline
% 		 & (1)	This license does not grant you rights to use\\
% 		Use & any contributors' name, logo, or trademarks.\\
% 		Trademark & (2) This License does not grant permission to use\\
% 		 & the trade names, or product names of the Licensor. \\
% 		 \hline
% 		%\bottomrule
% 	\end{tabular}}
% \end{table}

%Since our proposed term entity extraction model employs semi-supervised training, which can reduce the cost of labeling and improve the accuracy of the model to a certain extent, we prepared its labeled data and unlabeled data as follows. 

We conduct the comparative studies on the ground-truth dataset described in Section~\ref{sec:dataset} (i.e., 400 labelled licenses {with {21,844} labeled sentences}). We randomly split these samples into the training and testing datasets by 4:1, and further split the training dataset for training and validation by 4:1. {All the techniques and tools are evaluated on the same testing dataset.}

\smallskip
\subsubsection{\textbf{Results}}

\begin{figure}[t]
    \centering
    \includegraphics[width=0.9\textwidth]{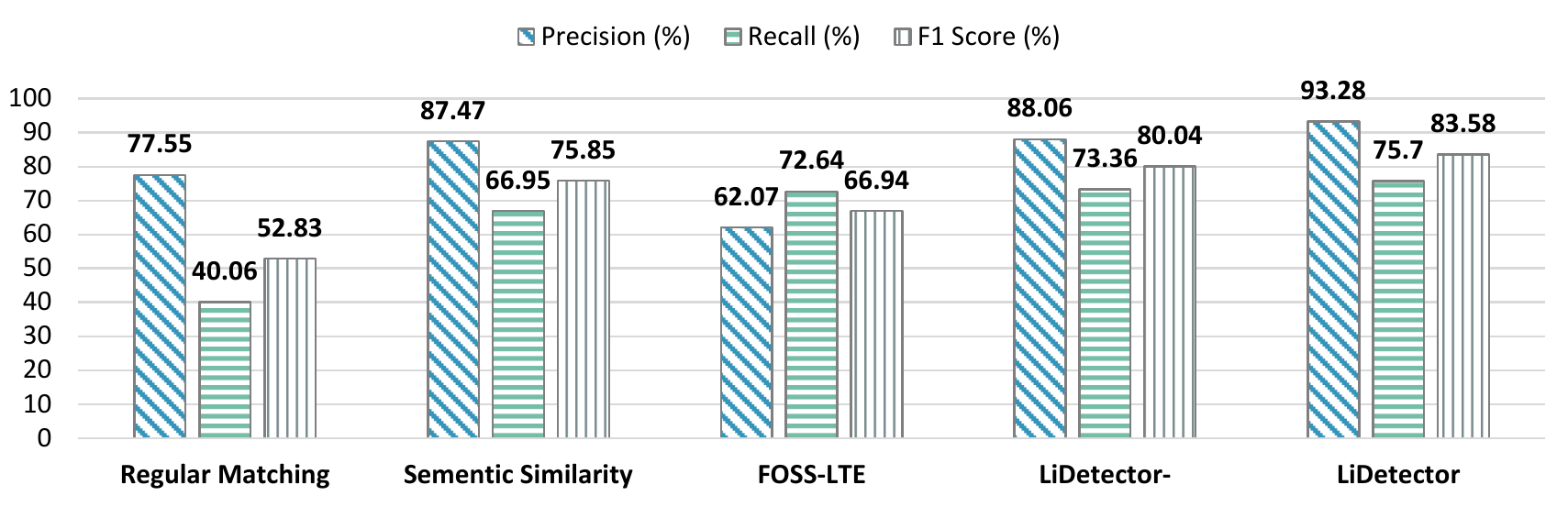}
    \caption{Comparison on Term Identification}
    \label{fig-phase1-eval}
\end{figure}

{Fig.~\ref{fig-phase1-eval} reports the results of each method. It can be seen that \tool outperforms the other methods, achieving 83.58\% F1-score, followed by \textsc{LiDetector$^-$} (80.04\%) and semantic similarity (75.85\%). Among five methods, regular matching has the worst performance with 52.83\% F1 score. The reason could be that} predefined patterns limit the flexibility of expressions, leading to a low recall. % not as flexible as natural language, especially for custom licwhich are limited and usually cannot cover all possible forms. \ling{the patterns are summarized by yourself? right? wired.}\textcolor{brown}{yes}
%The keyword matching method relies on the knowledge patterns obtained by manual study. 
Compared with recall, the precision of regular matching is relatively high due to the strict matching strategy. 
%Besides, it has a low overall evaluation results since the manual enumerated patterns are not enough to take in all possible forms in the data.  \ling{the patterns are summarized by yourself? right?}
It can also be seen that semantic similarity outperforms regular matching and FOSS-LTE. The results indicate that semantics of license sentences can be learned by the doc2vec model, so that sentences with similar representations contain similar license terms. %This is because the large data is learned and represented by the pretrained model first, and sentences with similar meanings can be captured. 
Moreover, we can also observe that the precision of FOSS-LTE is the lowest among all five methods. A possible reason could be that as a topic model, FOSS-LTE may be affected by noisy data during unsupervised learning.%Topic modeling is also involved in }
%The FOSS-LTE method also employs topic modeling method, while the accuracy appears slightly lower due to fewer types of terms identified by its model. 

Compared with FOSS-LTE, regular matching, and semantic similarity, both \tool and \textsc{LiDetector$^-$} have higher precision, i.e., 93.28\% and 88.06\%, respectively. The reasons behind are two folds: (1) the proposed method filters out license sentences which are irrelevant to license terms by preprocessing and clustering; (2) we employ a semi-supervised learning method that trains a sequential model on labelled sequences, which may lead to higher precision compared with unsupervised learning methods such as topic model in FOSS-LTE. %Finally, we conduct an ablation study to investigate the effectiveness of the 754 unlabelled samples. 
From the ablation study, it can be seen from {Fig.}~\ref{fig-phase1-eval} that \tool (semi-supervised learning with unlabelled samples) outperforms \textsc{LiDetector$^-$} (supervised learning without unlabelled samples) for both precision and recall. The results indicate that pseudo labelling of unlabelled samples are predictive, and it is worthy of incorporating these unlabelled samples for training.    

\revise{However, there are also corner cases that \tool fails to identify license terms. For example, {in the project Statsite~\citep{url-statsite}}, the license sentence \textit{``You may add your own copyright statement to your modifications and may provide additional or different license terms and conditions for use ....''} actually grant rights to Relicense (No.3 in Table~\ref{table-23terms}). However, \tool failed to identify this term due to the rareness of such expressions in the training dataset.}

\subsection{\textbf{Evaluation on Right and Obligation Inference}}

\subsubsection{\textbf{Setup}}
After identifying license terms, \tool aims to infer the attitudes towards these terms from license texts. To conduct a comprehensive study, in addition to FOSS-LTE~\citep{APSEC`17-Kapitsaki-termsIdentifying}, we also compare \tool with two NLP techniques, i.e.,
%keyword recognition~\citep{IAT'14-Solakidis-keyword} and \sihan{revise it: sentiment analysis}~\citep{etal`13-Socher-SST}. 
regular matching~\citep{IAT'14-Solakidis-keyword} and SST-based sentiment analysis~\citep{etal`13-Socher-SST}.
Again, these NLP techniques were implemented and adapted to the context of licenses, so as to infer rights and obligations. 
%Same as Section \ref{subsec:phase1}, concerning the research related to right and obligation inference in OSS licenses is not adequate, we manually construct the first two methods inspired by some text classification techniques involved in papers from other research tasks. 
%For the same reason mentioned in Section \ref{subsec:phase1}, we need to manually construct the first two methods.
%The descriptions of two baselines in addition to FOSS-LTE are as follows:
\begin{itemize}
    \item \textbf{Regular matching}~\citep{IAT'14-Solakidis-keyword}, which predefines a set of keywords to infer the attitudes implied by license sentences. To conduct a fair comparison, we use the same set of keywords (as listed in \Cref{table-polarity keywords}) for the baseline and \tool. Specifically, the baseline searches for the keywords that represent attitudes in the order of CANNOT, MUST, and CAN, which denote the prohibition, obligation, and right of a software product, respectively. %If none of keywords listed in \Cref{table-polarity keywords} were found, then the attitude the same with that express the license term attitude to perform regular matching, and then employ a progressive strategy to identify the attitude. That is, for each term, this method first checks the existence of restrictive attitude (i.e., check for the type \textit{Cannot} and \textit{Must} in turn), if without, it further checks the existence of permissive attitude (i.e., the type \textit{Can}). In any of the phase that the attitude type matches, it stops matching.
    %The list of the attitude keywords~\cref{table-polarity keywords} is the same as aforementioned. 
    %When a sentence input, which has been considered to contain a certain type of term after the locating term phase, this paper first judges whether it is type Cannot. If not, then we judge whether it is type Must, otherwise we judge whether it is type Can, that is, if the former authorization type has been matched successfully, the latter ones will no longer be matched. 
    %Finally, the attitude of terms is inferred as a license right, obligation or prohibition. 
    
    \item \textbf{SST-based sentiment analysis}~\citep{etal`13-Socher-SST}, which learns a classifier that predicts the positive or negative attitudes from the Stanford Sentiment Treebank (SST). Since licenses are written in natural language, we train a LSTM model based on the SST sentiment dataset, so as to predict the attitudes of authors behind licenses. %both aiming to extract the tendency of a sentence or text.     \textcolor{brown}{There are considerable possibility to transfer it on our task to infer the attitude among \textit{Can, Cannot, Must}, despite its more focus on personal emotion attitudes.    Particularly, \textit{SST} sentiment dataset~\citep{etal`13-Socher-SST} and LSTM network~\citep{CSAI`19-Xia-BiLSTM} are applied by us to train a sentiment analysis model to predict the attitudes of license terms. }
    
\end{itemize}

To investigate the effectiveness of \tool, we conduct a comparative study on the ground-truth dataset (i.e., 400 labelled licenses with 21,844 sentences), and randomly split the dataset into training, validation, and testing datasets as described in Section~\ref{subsec:phase1}. {All the methods are evaluated on the same testing dataset.} 
{Moreover, to conduct a fair comparison, we evaluate the effectiveness of these methods on the same set of license terms, which has been correctly identified in the previous phase. For this reason, here we only compare the \textit{accuracy} of these methods, since there are no false negatives in the evaluation and the \textit{recall} cannot be calculated.}

%Our evaluation is based on the ground-truth data (i.e., 400 licenses), and the test data set in this experiment is the same as that in the term identification phase in Section \ref{subsec:phase1}, i.e. 80 licenses. 
%We also use Precision, Recall and F1 score to measure the performance of each method.
%Note that, since this experiment aims to evaluate the identification ability for authorization polarity, the ground-truth of this phase is only the \textit{CAN}, \textit{CANNOT}, \textit{MUST} tags for each term, of these testing license texts. 

\subsubsection{\textbf{Results}}

%The evaluation results of the right and obligation inference phase is shown in Table~\ref{table-phase3-eval}.

\begin{table}[t]
	\centering
	\small
	\caption{Comparison on Right and Obligation Inference}
	\scalebox{0.9}{\begin{tabular}{cc}
		\toprule
		%\hline
		\textbf{Method} & \textbf{Accuracy (\%)} \\
		\midrule
		Regular Matching                      & 81.27 \\
		SST-based Sentiment Analysis                                & 82.88 \\
		FOSS-LTE                                 & 82.71 \\
		\textbf{\tool} & \textbf{91.09} \\
		%\hline
		\bottomrule
		\label{table-phase3-eval}
	\end{tabular}}
\end{table}

Table~\ref{table-phase3-eval} shows the results on right and obligation inference. 
%\sihan{Since that for each identified term, the inferred attitudes are either correct or incorrect, we only report the accuracy of inferences in this phase.} 
%From Table~\ref{table-phase3-eval}, 
It can be observed that 
%in the comparative experiment of the right and obligation inference phase, the syntax analysis method proposed in this paper 
\tool outperforms the other methods in attitude inference, achieving 91.09\% accuracy. SST-based sentiment analysis achieves a comparable performance with FOSS-LTE and regular matching, with the accuracy around 82\%. Note that we utilize a model that has been well trained over the SST sentiment dataset. However, the results reported in Table~\ref{table-phase3-eval} indicate that although license texts are written in natural language, methods of sentiment analysis cannot be directly applied to infer the stated conditions of software use. 

By analyzing the results of FOSS-LTE, we found that FOSS-LTE does not distinguish between license terms and their attitudes. Instead, it predefines a set of phrases that represent regulations (e.g., \textit{MustOfferSourceCode}), and utilizes a topic model to extract regulations from license texts. Compared with FOSS-LTE, \tool learns and extracts information from license texts with finer granularity. Specifically, it learns to identify license terms from texts, based on which it infers the implied attitudes towards the identified terms. By this means, \tool is capable of inferring regulations with more flexibility compared with predefined regulations. 

For regular matching and \tool, we define the same set of attitude keywords, as well as the same order to search for the attitude keywords. However, we observe that simple regular matching is less effective than \tool in predicting attitudes. The characteristics of license sentences may contribute to the gap between performances of regular matching and \tool. Specifically, license sentences are typically long and complicated, which may contain massive tokens that are irrelevant to the attitudes towards a target term. Regular matching equally compares each token in the sentence with predefined keywords, which may introduce noise. In contrast, \tool filters out irrelevant tokens by grammar parsing, narrows down the scope of searching for permissive and restrictive expressions, and only analyzes powerful tokens that may have an influence on the attitude toward a target entity.% Comparing the effectiveness keyword recognition method, since it relies on the vocabulary of predefined attitude keyword, the accuracy is limited by the keyword set. Besides,  

\noindent \textbf{The overall performance for license comprehension.}
To further investigate the performance of \tool combining the first two phases, we also report the results over the entire process of license term identification and attitude inference. Specifically, given a license, we first identify license terms listed in Table~\ref{table-23terms}. Then, the identified term entities, as well as the original license sentences, were fed into the second phase, so as to infer the attitudes towards these terms. We conduct the comparative study on the ground truth described in Section~\ref{sec:dataset}, and use precision, recall, and F1 score to evaluate the performances of the compared methods.
%The test data set here is the same as mentioned earlier Section \ref{subsec:phase1}. The ground-truth here comes from all the term tags, both terms extracted and their polarity tags, of each testing license text.
%
%We use precision, recall, and F1 score to evaluate the performance of each model. 
% Here, precision refers to the proportion of texts that are predicted correctly on the term type and the corresponding authority polarity type, to all the texts that are predicted containing related terms; recall refers to the proportion of texts that are predicted correctly on the term type and the corresponding authority polarity type, to all the texts that are labeled containing related terms; the F1 score is the harmonic average of precision and recall.
%
%Finally, the evaluation results of the overall model of term identification are shown in 
Table~\ref{table-phase2and3-eval} summarizes the results of each method. It can be seen that the combination of the methods used in \tool (i.e., sequence labelling and sentiment analysis) achieves the best performance among all combination methods, with 85.88\% precision, 69.70\% recall, and 76.95\% F1 score. 

\begin{table}[t]
	\centering
	\footnotesize
	\caption{Evaluation Results for License Comprehension}
	\label{table-phase2and3-eval}
	\begin{tabular}{ccccc}
		\toprule
		%\hline
		\textbf{Term Extraction} & \textbf{R. and O. Inference} & \textbf{P (\%)} & \textbf{R (\%)} & \textbf{F1 (\%)} \\
		\hline
		Regular Matching  & Regular Matching                     & 61.44 & 31.74 & 41.86 \\
		Semantic Similarity & Regular Matching                     & 70.94 & 54.30 & 61.51 \\
		%Topic Modeling  & Keyword Recognition                       & 53.02 & 66.80 & 59.12 \\
		Sequence Labelling & Regular Matching & 78.04 & 63.33 & 69.92\\
		Regular Matching  & Sentiment Analysis         & 68.81 & 37.12 & 48.22 \\
		FOSS-LTE   & FOSS-LTE                                       & 51.34 & 60.08 & 55.37 \\
		\textsc{LiDetector}$^-$ & \tool & {80.23} & {66.75} & {72.87}\\
		\rowcolor{gray!20} 
		%(Term Entity Extraction)
		\begin{tabular}[c]{@{}c@{}}\textbf{\tool} \\(\textbf{Sequence Labelling})\end{tabular}
		& %\textbf{\tool}
		%(Syntax Analysis)
		\begin{tabular}[c]{@{}c@{}}\textbf{\tool} \\(\textbf{Sentiment Analysis})\end{tabular}
		& \textbf{85.88} & \textbf{69.70} & \textbf{76.95}\\
		%\hline
		\bottomrule
	\end{tabular}
	\begin{center}
	    \footnotesize
	    \textit{R.: Right; O.: Obligation; P: Precision; R: Recall; F1: F1 score.}
	\end{center}
\end{table}

\iffalse
\begin{table}[t]
	\centering
	\scriptsize
	\caption{Evaluation Results for the Overall Model\sihan{overall model?}}
	\label{table-phase2and3-eval}
	\begin{tabular}{c|ccccc}
		\toprule
		%\hline
		\textbf{Method} & \textbf{Phase 1} & \textbf{Phase 2} & \textbf{P (\%)} & \textbf{R (\%)} & \textbf{F1 (\%)} \\
		\hline
		C1 & Regular M.  & Regular M.                     & 61.44 & 31.74 & 41.86 \\
		C2 & Semantic S. & Regular M.                     & 70.94 & 54.30 & 61.51 \\
		%Topic Modeling  & Keyword Recognition                       & 53.02 & 66.80 & 59.12 \\
		C3 & Sequence L. & Regular M.                & 78.04 & 63.33 & 69.92\\
		C4 & Regular M. & Syntax A.         & 68.81 & 37.12 & 48.22 \\
		FOSS-LTE & FOSS-LTE   & FOSS-LTE                                       & 51.34 & 60.08 & 55.37 \\
		\textsc{LiDetector}$^-$ & Sequence L.$^-$ & Sentiment A. & {80.23} & {66.75} & {72.87}\\
		\textbf{\tool} & \textbf{Sequence L.} & \textbf{Sentiment A.} & \textbf{85.88} & \textbf{69.70} & \textbf{76.95}\\
		\bottomrule
	\end{tabular}
	\begin{center}
	    \scriptsize
	    \textit{R.: Right; O.: Obligation; P: Precision; R: Recall; F1: F1 score.}
	\end{center}
\end{table}
\fi

%%%%%%%%%%%%%%%%%%%%%%%%%%%%%%%%%%%%%%%%%%%%%%%%%%%%%%%%%%%%%%%%%%%%%%%%%%%%%%%%%%%%%%%%%%%%%%%%%%%%%%%%%%%%%%%%%%%%%%%%%%%%%%%%%%%%%%%%%%%%%%%%%%%%%%%%%%%%%%%%%%%%%%%%%%%%%%%%%%%%%%%%%%%%%%%%%%%%%%%%%%%%%%%%%%%%%%%%%%%%%%%%%%%%%%%%%%
\subsection{\textbf{Evaluation on Incompatibility Detection}}
\subsubsection{\textbf{Setup}}
To investigate the overall effectiveness of \tool in incompatibility detection, we also compare \tool with {three} state-of-the-art tools, i.e., the SPDX Violation Tools (\textbf{SPDX-VT})~\citep{jss`17-Kapitsaki-SPDX,tse`18-Kapitsaki-findOSSLicense}, \textbf{Ninka}~\citep{ase`10-German-Ninka} equipped with \textit{tldrlegal}~\citep{url-tldrlegal}, and {\textbf{Librariesio}~\citep{url-librariesio}, a license compatibility checking tool for SPDX licenses}. 
\revise{Since the other three tools are all based on the licenses on SPDX due to their inability on custom licenses, to make a fair comparison, we conduct this experiment and show the result in two ways: (1) Only the official licenses listed in the Software Package Data Exchange (SPDX)~\citep{url-spdx} are considered for incompatibility detection. (2) All the licenses extracted from the project are considered (including the custom ones) for incompatibility detection.}

Specifically,
we randomly selected 200 projects from the 1,846 GitHub projects described in Section \ref{sec:motivatingstudy}, and constructed a ground-truth dataset manually verified and cross-validated by three authors and a lawyer. 
\revise{In total, we extracted 1,298 unique licenses (including 191 project licenses and 1,107 component licenses), with 1,041 official licenses and 257 custom ones.} {The ground-truth dataset of incompatible projects has been made available online~\citep{lidetector}.} 
For each project, we first extract {licenses in three forms (i.e., declared, referenced, and inline)}, and then use the aforementioned tools to detect the incompatibility issues.
 %both the declared and referenced licenses 
%\gy{It is noticing that, this paper has constructed a ground-truth on incompatibility projects, not the incompatibility issues((i.e., conflicts) and the conflicts are predicted and calculated without manually checking.  Therefore, in the following results, the \textit{FN} and \textit{FP} evaluation standards are based on OSS projects. }
{Note that, since we focus on license incompatibility detection for software projects, the FP and FN metrics are computed at the project level.} {Moreover, we also calculated the number of \textit{conflicts} and \textit{overlaps} for the these tools. In this paper, a \textit{conflict} denotes a specific incompatibility issue, which means two incompatible attitudes towards the same license term. For instance, \textit{cannot disclose source} versus \textit{must disclose source}.} We use \textit{overlap} to denote the number of incompatible projects/conflicts that can also be detected by \tool.

\begin{table}[t]
	\centering
	%\small
	\caption{Results for Incompatibility Detection (``Conflict'': two incompatible attitudes towards the same term.)}
	\label{table-compatibility}
	\scalebox{0.83}{\begin{tabular}{c|clccc}
		\toprule
		%\hline
		\textbf{Method} &\textbf{\#Pro.} & \textbf{\#Identified pro.} & \textbf{\#Conflicts} & \textbf{FP} & \textbf{FN}\\ \hline
		Ninka~\citep{ase`10-German-Ninka}  & 200 & {144 (117 overlap)}  & {13,978 (12,509 overlap)} & {23 (15.97\%)} & {35 (22.44\%)} \\
		SPDX-VT~\citep{jss`17-Kapitsaki-SPDX}  & 200 & {71 (71 overlap)}  & -- & {6 (8.45\%)} & {91 (58.33\%)} \\
	    Librariesio~\citep{url-librariesio}  & 200 & {169 (137 overlap)}  & -- & {40 (23.67\%)} & {27 (17.31\%)} \\
	    \textbf{\tool (SPDX)} &  \textbf{200} &  \revise{\textbf{157}}  & \revise{\textbf{14,586}} & \revise{\textbf{--}}  & \revise{\textbf{--}}\\
        \textbf{\tool (All)} &  \textbf{200} &  {\textbf{169}}  & {\textbf{22,941}} & {\textbf{17 (10.06\%)}}  & {\textbf{4 (2.56\%)}}\\
		%\hline
		\bottomrule
	\end{tabular}
	}
	%\vspace{-1.5mm}
	\begin{center}
	    \footnotesize
	    \revise{\textit{\textbf{\tool (SPDX)}: the result for licenses only on SPDX;} 
	    \textit{\textbf{\tool (All)}: the result for all the extracted licenses.}}
	\end{center}
% 	\begin{itemize}
% 	    \centering
% 	    \footnotesize
% 	    %\item \textit{``Conflict'' refers to two incompatible attitudes towards the same term.}
% 	    \item \revise{\textit{\textbf{\tool (SPDX)}: the result for licenses only on SPDX.}
% 	    \item \textit{\textbf{\tool (All)}: the result for all the extracted licenses.}}
% 	\end{itemize}
\end{table}

\subsubsection{\textbf{Results}}
\revise{Table~\ref{table-compatibility} shows the incompatibility detection results on SPDX licenses only and all the extracted licenses.
As for the results on SPDX licenses only, it can be seen that \tool detects 14,586 conflicts in 157 projects, while SPDX-VT only finds license incompatibility in {71} out of 200 projects, {all of which can be detected by \tool. }Note that \tool filters all official licenses whose regulations have already been known. For this reason, when only considering the official licenses in SPDX~\citep{url-spdx-licenses}, there are no false positives and false negatives.} 
{Since SPDX-VT only predefined a graph to denote the compliance relationships between a set of licenses, the output of SPDX-VT is relatively coarse-grained (i.e., without detailed explanation of how licenses conflict with each other). Therefore, the number of conflicts detected by SPDX-VT are not presented in~\Cref{table-compatibility}.} It can be seen that the number of false positives reported by SPDX-VT is 6. However, the FN rate of SPDX-VT is 58.33\%, \revise{nearly 30} times higher than that of \tool for all extracted licenses. By analyzing the process of SPDX-VT, we found that it designs a strict rule that only a set of license texts are identified and analyzed to detect incompatibility. However, the graph defined by SPDX-VT only includes 20 popular licenses, other licenses can not be detected by SPDX-VT. 

\begin{figure*}[t]
    \centering
    \includegraphics[width=0.78\textwidth]{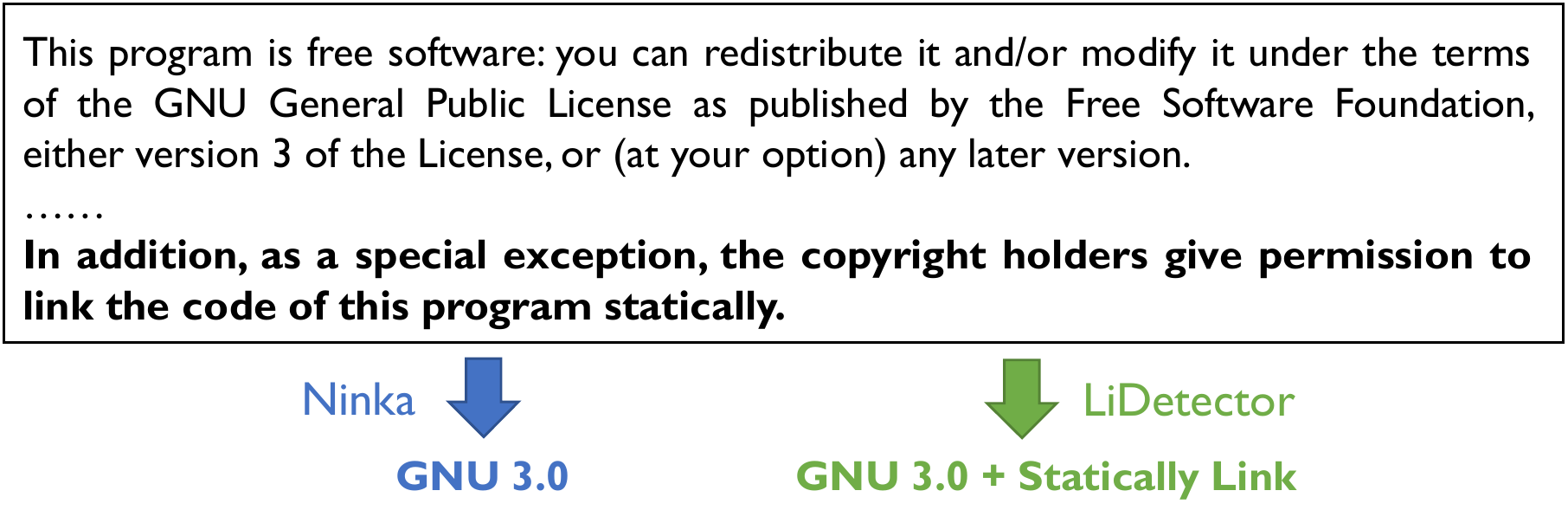}
    \vspace{-2mm}
    \caption{A License Exception in BDF~\citep{url-ninkacannot} Misidentified by Ninka}
    \label{fig_ninka}
\end{figure*}

For the results on all extracted licenses, among 200 projects, {\tool identified {22,941} conflicts in %168 
{169} projects, {with 10.06\% FP rate and 2.56\% FN rate}. Ninka equipped with tldrlegal identified 13,978 conflicts in 144 projects, with {15.97\%} FP rate and {22.44\%} FN rate.} It can be concluded that \tool has the superiority over Ninka (equipped with \textit{tldrlegal}) from the respects of both precision and recall. 
\revise{By analyzing the reported results of Ninka, we found that some licenses are customized by authors of software products, which cannot be identified by Ninka. 
For instance, the component license in Fig.~\ref{fig:example1} is a custom license that cannot be identified by Ninka. Therefore, Ninka equipped with \textit{tldrlegal} cannot detect the incompatibility issue in Fig.~\ref{fig:example1}. In addition, some licenses are \textit{exceptions} of common open source licenses (i.e., license variants generated by developers based on popular licenses, thus resulting in the modification of its term attitudes), which regulates different rights and obligations but misidentified by Ninka. Fig.~\ref{fig_ninka} shows a custom license which is an exception of GNU 3.0. In this case, \tool first identified that the license referenced GNU 3.0; then, it fed the rest text into the probabilistic model and inferred {``CAN Statistically Link''} from the text. However, Ninka only identified the license as GNU 3.0 ignoring the custom exceptions. As a result, the accuracy of Ninka limits its effectiveness in detecting license incompatibility.}

%Although SPDX-VT has more false negatives than \tool, 
%However, 

{Librariesio~\citep{url-librariesio} identified 169 incompatible projects, with {23.67\%} FP rate
and {17.31\%} FN rate. Similar to SPDX-VT, the number of conflicts detected by Librariesio are also not presented in Table~\ref{table-compatibility}, since Librariesio provides no detailed explanation about how licenses conflict with each other. By analyzing the results of Librariesio and \tool, we found that the FP rate of Librariesio is more than twice as high than that of \tool and the FN rate of Librariesio is almost {8} times higher. The reason are two folds. First, Librariesio detects incompatibility between a set of SPDX licenses, while other licenses are ignored. Second, Librariesio detects incompatibility with a predefined set of heuristic rules, which might not be suited for large amounts of licenses which are flexible in expressions. %with heristics rules is that Librariesio divided 65 popular licenses into 5 types according to the degree of the strictness of their licenses, and ignored other popular and custom licenses. Furthermore, Librariesio listed limited situation between two license sets, obtaining relatively coarse-grained incompatibility analysis. 
}

{
%We also analyzed the overlap between the true positives detected by SPDX-VT and \tool. We found that there exists only one project, named \textit{OpenFermion}~\citep{url-OpenFermion}, that can be only detected by SPDX. The reason why \tool failed to detect the license incompatibility of this project is that \textit{CoreNLP}~\citep{url-corenlp} failed to parse the whole sentence as it is too long and complicated. 
It can also be seen that the FP rate of \tool is {10.06\%}. %(i.e., 17 out of 200 projects that were misclassified as incompatible projects)}. 
By analyzing the false positives, we found that some license terms were not correctly identified from license texts, especially for license terms such as \textit{Statically Link} and \textit{Relicense}.
The reason behind is that these license terms do not as frequently occur as other terms in the training dataset, which poses challenges in identifying these terms.}
%Since \tool regards the default attitude towards a license term as CANNOT if the term is not mentioned, license terms which are not identified may lead {to} false positives.} 
\revise{We note that false negatives in the first phrase (i.e., license term identification) could lead to both false positives and false negatives in incompatibility detection.
Another reason for false positives is the limitation of Stanford CoreNLP~\citep{url-corenlp}. For instance, in the project Blosc~\citep{url-blosc}, there is a license sentence:}

\noindent \revise{\textit{``... In any action to enforce the terms of this License or seeking damages relating thereto, its costs and expenses, including, without limitation, reasonable attorneys' fees and costs incurred in connection with such action, including any appeal of such action, shall be recovered for the prevailing party ...''}}

\noindent \revise{From this sentence, \tool identified the license term \textit{Compensate for Damages}. However, when inferring the attitudes towards this license term, Stanford CoreNLP~\citep{url-corenlp} failed to parse the whole sentence, since the sentence is too long. The part of sentence \textit{``shall be recovered for the prevailing party''} was ignored by the tool, leading to the inference results to be CAN instead of MUST. Due to the wrong parsing result, \tool failed to extract the obligation ``MUST Compensate for Damages'' from the project license, leading to a false positive.
}

%%%%%%%%%%%%%%%%%%%%%%%%%%%%%%%%%%%%%

\section{Empirical Study on Compatibility Analysis}
\label{sec:emp}
%On one hand, based on Section \ref{sec:motivatingstudy}, we applied the license term extraction model proposed in this paper to analyze their included licenses, for a more accurate analysis of license incompatibility in OSS projects. Finally, the total ratio of projects with license incompatibility increases from 29.37\% stated in the previous study to 47.51\% now, as Table~\ref{table-incompatibility-2-study} showed. 

By leveraging \tool, we further conduct an empirical study on the 1,846 projects collected from Github, so as to investigate incompatibility issues in real-world OSS licenses.
%
%Among 1,846 projects, \tool detected 5,112 conflicts in 877 projects, i.e., \textbf{47.51\%} projects suffer from license incompatibility issues according to \tool. Compared with the incompatible project rate (i.e., 29.37\%) in the motivating study (Section \ref{sec:motivatingstudy}), \tool is capable of detecting more incompatibility issues in OSS licenses owing to its ability for custom licenses.

% \begin{table}[t]
% 	\centering
% 	\small
% 	\caption{The prevalence of license incompatibility}
% 	\label{table-incompatibility-2-study}
% 	\begin{tabular}{c|c|c}
% 		\toprule
% 		%\hline
% 		\multicolumn{2}{c|}{Projects Number}              & 1,846  \\ 
% 		\hline
% 		Referenced Licenses & Imported Packages           & 96,606 \\
% 		                    & Different Licenses          & 37 \\
% 		\hline
% 		Declared Licenses  & License Texts               & 11,578 \\
% 		                    & Term Entities              & 121,264 \\ 
% 		\hline
% 	    \multicolumn{2}{c|}{Total Ratio}                & 47.51\% \\
% 	    %\hline
% 		\bottomrule
% 	\end{tabular}
% \end{table}

%\ling{add the top 10 conflict terms, and the affected number of projects. and analyze what kind of incompatibility.} \textcolor{brown}{which kind ?}

\subsection{Overall status of incompatibility issues} 
Among 1,846 projects, \tool detected {75,249} conflicts in {1,346} projects, i.e., {\textbf{72.91\%}} projects suffer from license incompatibility issues according to \tool. The evaluation results can be seen in Table~\ref{table-project_vs_component} \revise{and Table~\ref{table-official_vs_custom}}. Compared with the incompatible project rate (i.e., {48.86\%}) in the motivating study (Section \ref{sec:motivatingstudy}), \tool is capable of detecting more incompatibility issues in OSS licenses owing to its ability for custom licenses. %\sihan{checked? how to convince the reviewers that we find more ``real'' incompatibility issues? maybe we can demonstrate it by sampling.}
%\ling{For example, the project, XXX [?], contains referenced license (e.g., XX), declared license (e.g., XX), and custom licenses in XX file. In the motivating study, only the conflicts between XX and XX are identified, while \tool identifies XX more conflicts between the three kinds of licenses.}

To better understand the incompatibility status, \revise{we take an in-depth analysis on the incompatibility issues detected by \tool from two aspects: (1) incompatibility involving project licenses and component licenses (Table~\ref{table-project_vs_component}); and (2) incompatibility involving official licenses and custom licenses (Table~\ref{table-official_vs_custom}). For the first aspect, }
we divide the incompatibility issues into two categories: a) incompatibility between component licenses, and b) incompatibility between a project license and a component license. 
We can see that {1,087} out of 1,846 projects were found to have compliance issues between their project licenses and their component licenses, while only {259} projects contain incompatibility issues between component licenses. We also show the exact number of conflicts detected by \tool in Table~\ref{table-project_vs_component}. 
%We use PL and CL to denote \textit{project license} and \textit{component license}, respectively. 
The values in the column \textit{PL vs CL} are the numbers of incompatibility issues between a project license and a component license. It can be seen that among all investigated projects, conflicts often occur between a project license and a component license. In other words, given an open source project, it is often the case where users who conform to the license of the whole project may violate licenses of some components. The difficulty to combine licenses from different components into the whole project license may account for this phenomenon. The investigation results encourage developers to pay more attention to creating the license for the whole project especially when they integrate third-party software components with licenses in their projects.

\revise{Table~\ref{table-official_vs_custom} shows the detailed result about the number of official licenses and custom licenses involved in license incompatibility. It can be seen that 1,324 projects have incompatibility issues between a pair of official licenses; 384 projects have incompatibility issues between an official license and a custom license; only 13 projects contains conflicts between a pair of custom licenses. The results are consistent with the observation that a majority of licenses are based on official licenses. We can also observe that custom licenses are involved in 31,469 conflicts and 397 projects. In other words, even all official licenses are compatible with each other, there are still near 21.5\% of 1,846 projects that have incompatibility issues.}
%\gy{The reason why incompatibility between the project and component licenses contributes to more than 80\% is twofold: first, as we investigated, 80.39\% projects have project licenses, the incompatibility of which is detected from the comparison between their project license and component licenses to explore the ability to combine multiple licenses into the same software product. Second, most projects with project licenses have more component licenses compared with ones without project licenses, thus more likely to involve license incompatibility. The result indicates that the project licenses have quite difficulty in successfully combining all of their component licenses, and software developers need more effort to consider the rights, obligations and prohibitions of their final project licenses carefully. }

%\cmt{Add more description and findings about the result, like other subsections.}

\begin{table}[t]
	\centering
	%\small
	\caption{{Overall Status of License Incompatibility}}
	\label{table-project_vs_component}
	\scalebox{0.83}
	{
	\begin{tabular}{c|cc|cc}
		\toprule
		%\hline
		\textbf{\# Total projects} & \multicolumn{2}{c|}{\textbf{\# Incompatible projects}} & \multicolumn{2}{c}{\textbf{\# Conflicts}} \\ 
		\hline
		\multirow{3}*{1,846} & \multicolumn{2}{c|}{{1,346}} & \multicolumn{2}{c}{{75,249}} \\
		\cline{2-5}
		~ & \textbf{PL vs CL} & \textbf{CL vs CL} & \textbf{PL vs CL} & \textbf{CL vs CL} \\
		\cline{2-5}
		~ & {1,087 (80.76\%)} & {259 (19.24\%)} & {66,447 (88.30\%)} & {8,802 (11.70\%)} \\
		%\hline
		\bottomrule
	\end{tabular}
	}
	\begin{center}
	\footnotesize
     \textit{PL: Project License; CL: Component License}
	\end{center}
\end{table}

\begin{table}[t]
	\centering
	%\small
	%\color{blue}
	\caption{{The Number of Official and Custom Licenses Involved in Conflicts}}
	\label{table-official_vs_custom}
	\scalebox{0.83}
	{
	\begin{tabular}{c|c|c|c}
		\toprule
		%\hline
		 & \textbf{Official vs. Official} & \textbf{Custom vs. Official} & \textbf{Custom vs. Custom} \\ 
		\hline
		\textbf{\# Conflicts} & \revise{43,780} & \revise{29,835} & \revise{1,634}\\\hline
		\textbf{\# Projects} & \revise{1,324} & \revise{384} & \revise{13}\\
		%\hline
		\bottomrule
	\end{tabular}
	} 
\end{table}

\noindent \textbf{Impact of the default attitude towards an absence term.}
%\gy{need to delete?} \ling{A: no. remain it.}
As aforementioned in Section~\ref{subsec:incom}, following the rule declared by \textit{choosealicense}~\citep{url-choosealicense}, if a license term does not appear in the license according to \tool, the default attitude towards this term is set to CANNOT, which implies that nobody can copy, distribute, or modify the work. To further investigate the impact of such a default setting, we also conduct an ablation study where the default attitude of absent terms is set to {CAN}. 
{The results show that when the default attitude of absent terms is CAN, \tool detected 1,104 incompatible projects with 10,507 license incompatibility issues. When the default setting is CANNOT, \tool detected {1,346} incompatible projects with {75,249} incompatibility issues. It can be seen that when the default attitude is set to CANNOT, \tool detected more incompatible projects with seven times of the number of incompatibility issues. A possible reason is that there exist more restrictive statements than permissive statements when the default setting of absent terms is set to CANNOT. Nevertheless, it has more influence on the number of detected incompatibility issues than incompatible projects. A large number of absent license terms in component licenses might account for such difference.%it is more  The variety in 220 projects on license incompatibility may result from some terms of their component licenses appear in absence attitude, which are not compatible with the CAN attitude in their project licenses. In addition, the great change of the amount of such conflicts may be influenced by numerous huge OSS projects with a lot components. It can be concluded that the default setting as CANNOT does have a great impact on the license conflicts, but have a relatively weak influence on the final result of incompatible projects. }
}
%\cmt{add numbers and discussion..and possible conclusion.}
%\noindent \textbf{Top licenses with incompatibility issues.}
%

% \begin{table}[t]
% 	\centering
% 	\color{blue}
% 	%\small
% 	\caption{Different dimensions towards incompatibility issues.}
% 	\label{table-another_dimension_size}
% 	\scalebox{0.83}{\begin{tabular}{c|ccc}
% 		\toprule
% 		%\hline
% 		\textbf{Project Size} &\textbf{\#Pro.} & \textbf{\#Incompatible pro.} & \textbf{\#Conflicts} \\ \hline
% 		<= 999 & 840 & 550 (65.48\%)  & 12,899 \\
% 		1,000$\sim$4,999  & 551 & 426 (77.31\%)  & 20,757  \\
% 	    5,000$\sim$9,999  & 281 & 219 (77.94\%)  & 12,081  \\
% 	    >= 10,000 &  174 &  151 (86.78\%)  & 29,512 \\
% 	    \hline
%         \textbf{Total} &  1,846 &  1,346  & 75,249 \\
% 		%\hline
% 		\bottomrule
% 	\end{tabular}
% 	}
% \end{table}

% \begin{table}[t]
% 	\centering
% 	\color{blue}
% 	%\small
% 	\caption{Different dimensions towards incompatibility issues.}
% 	\label{table-another_dimension_CL}
% 	\scalebox{0.83}{\begin{tabular}{c|ccc}
% 		\toprule
% 		%\hline
% 		\textbf{\# Component licenses} &\textbf{\#Pro.} & \textbf{\#Incompatible pro.} & \textbf{\#Conflicts} \\ \hline
% 		<= 24 & 887 & 479 (54.00 \%)  & 7,283 \\
% 		25$\sim$49  & 495 & 428 (86.46\%)  & 11,272  \\
% 	    50$\sim$99  & 316 & 294 (93.04\%)  & 12,503  \\
% 	    >= 100 &  148 &  145 (97.97\%)  & 44,191 \\
% 	    \hline
%         \textbf{Total} &  1,846 &  1,346  & 75,249 \\
% 		%\hline
% 		\bottomrule
% 	\end{tabular}
% 	}
% \end{table}

\begin{table}[]
%\color{blue}
\begin{minipage}[t]{0.49\textwidth}
\centering
\footnotesize
\caption{Incompatible Projects with Different Sizes.}
	\label{table-another_dimension_size}
	\begin{tabular}{c|ccc}
		\toprule
		%\hline
		\textbf{Pro. Size (KB)} &\textbf{\#Pro.} & \textbf{\#Incomp. pro.} & \textbf{\#Conflicts} \\ \hline
		<= 999 & 840 & 550 (65.48\%)  & 12,899 \\
		1,000$\sim$4,999  & 551 & 426 (77.31\%)  & 20,757  \\
	    5,000$\sim$9,999  & 281 & 219 (77.94\%)  & 12,081  \\
	    >= 10,000 &  174 &  151 (86.78\%)  & 29,512 \\
	    \hline
        \textbf{Total} &  1,846 &  1,346  & 75,249 \\
		%\hline
		\bottomrule
	\end{tabular}
\end{minipage}
\hfill
\begin{minipage}[t]{0.49\textwidth}
\centering
\footnotesize
\caption{Projects with Different Numbers of CLs.}
	\label{table-another_dimension_CL}
	\begin{tabular}{c|ccc}
		\toprule
		%\hline
		\textbf{\#CL} &\textbf{\#Pro.} & \textbf{\#Incomp. pro.} & \textbf{\#Conflicts} \\ \hline
		<= 24 & 887 & 479 (54.00 \%)  & 7,283 \\
		25$\sim$49  & 495 & 428 (86.46\%)  & 11,272  \\
	    50$\sim$99  & 316 & 294 (93.04\%)  & 12,503  \\
	    >= 100 &  148 &  145 (97.97\%)  & 44,191 \\
	    \hline
        \textbf{Total} &  1,846 &  1,346  & 75,249 \\
		%\hline
		\bottomrule
	\end{tabular}
\end{minipage}
\end{table}

\noindent \revise{\textbf{The impacts of project size and the number of component licenses.}
To analyze the distribution of incompatibility issues from other dimensions, e.g., project size and the number of component licenses, we divide projects into different groups and show the results in Table~\ref{table-another_dimension_size} and Table~\ref{table-another_dimension_CL}. 
From Table~\ref{table-another_dimension_size}, we can see that projects that have larger size are more likely to trigger license incompatibility, which is because larger projects possibly own more imported packages and more complex dependency construction \citep{liu2022demystifying}. 
From Table~\ref{table-another_dimension_CL}, we can see that projects that have more component licenses are more likely to trigger license incompatibility. Especially for projects containing more than 100 component licenses, almost all of these projects (i.e., 97.97\%) have license incompatibility issues. The results show that projects with many components need to be noticed for the potential risk towards license incompatibility.}

\subsection{Top licenses with incompatibility issues}
To analyze the relationships between incompatibility and license types, we counted the number of licenses involved in the incompatibility issues. Specifically, we found that common open source licenses (\textit{official licences}) are incompatible with other licenses in {1,346} projects with {117,395} conflicts, and custom licenses are incompatible with other licenses in {384} projects with {33,103} conflicts. %\sihan{again, how to decide whether it is a well-known license or a custom exception?} 
{Note that we used Ninka~\citep{ase`10-German-Ninka} to identify well-known licenses.}
{Each conflict occurs between two licenses, and the total number of conflicts detected by \tool is still {75,249}} %\cmt{may be we should only count it once.}
The reason why well-known licenses contribute to more than half of incompatibility issues is twofold: first, well-known licenses are frequently used in OSS; second, third-party packages and libraries incorporated by developers are often accompanied with common licenses, which may induces incompatibility with the project license. Nevertheless, custom licenses and exceptions also account for a considerable number of incompatibility issues, which indicates that relying on well-known licenses to detect incompatibility is not effective enough. A tool that is capable of analyzing compatibility for both well-known and custom licenses are needed for practical use. %incompatibility between licenses Based on the empirical study, we investigate the possible difference between well-known licenses and custom licenses on license incompatibility issues. The results showed that the well-known licenses involve 873 projects with 6032 conflict cases, while the custom licenses involve 493 projects with 4192 conflicts. The prevalence of incompatibility well-known licenses related to is mainly because of their broad popularity in \textit{component licenses} in referenced forms, which may be largely incompatible with more permissive \textit{project licenses}. 
Finally, we counted the number of each license to be involved in the incompatibility issues, and sorted them from high to low. We found that the top 5 common open source licenses are \textit{MIT License}~\citep{url-MIT}, \textit{Zope Public License 2.1}~\citep{url-ZPL-2.1}, \textit{Apache License 2.0}~\citep{url-Apache-2.0}, \textit{GNU Lesser General Public License v3}~\citep{url-LGPL-3.0-only}, and \textit{BSD 3-Clause License}~\citep{url-BSD-3-Clause}.  
\revise{For example, in the project HaboMalHunter~\citep{url-mitcannot}
, the project license is the MIT License that states {``CAN Sublicense''}, while one of the component licenses is the Wizardry License that regulates {``CANNOT Sublicense''}. In this case, developers who conform to the project license may still violate the component license, which leads to incompatibility issues.}

 \begin{table}[t]
 	\centering
 	\footnotesize
 	\caption{Top 10 Terms with Incompatibility Issues}
 	\label{table-top}
 	\begin{tabular}{ccccc}
 		\toprule
 		%\hline
 		\textbf{ID}&\textbf{Term} & \textbf{\#Conflicts} & \textbf{\#Pro.} & \textbf{Incompatibility Type} \\ %\hline 
 		\midrule
 	1	& Sublicense & {7,801} & {604} & Can$\leftrightarrow$Cannot \\\hline
 	2	& Use Trademark & {7761} & {933} & \begin{tabular}[c]{@{}c@{}}Can$\leftrightarrow$Cannot, Can$\leftrightarrow$Must, \\Cannot$\leftrightarrow$Must\end{tabular} \\\hline
 	3	& Commercial Use & {7,418} & {720} & Can$\leftrightarrow$Cannot \\\hline
 	4	& Distribute & {6,716} & {824} & Can$\leftrightarrow$Cannot \\\hline
 	5	& Modify & {6,564} & {749} & Can$\leftrightarrow$Cannot \\\hline
 	6	& Place Warranty & {5,544} & {489} & Can$\leftrightarrow$Cannot \\\hline
 	7	& Include Copyright & {4,741} & {1,303} &  Cannot$\leftrightarrow$Must \\\hline
 	8	& Include License & {4,312} & {1,306} & Cannot$\leftrightarrow$Must	\\\hline
 	9	& State Changes & {2,714} & {969} & Can$\leftrightarrow$Must, Cannot$\leftrightarrow$Must  \\\hline
 	10	& Disclose Source & {2,670} & {1,085} & \begin{tabular}[c]{@{}c@{}}Can$\leftrightarrow$Cannot, Can$\leftrightarrow$Must, \\Cannot$\leftrightarrow$Must\end{tabular} \\
 		%\hline
 		\bottomrule
 	\end{tabular}
 \end{table}

\subsection{Top 10 license terms with incompatibility issues}%\gy{need to revise}
%\cmt{differentiate official and custom?}
To further analyze the reasons of the incompatibility issues, we list the top 10 license terms involved in license incompatibility in Table~\ref{table-top}. For each license term, we show the number of incompatible projects ({\#Pro.}), the number of conflicts ({\#Conflicts}), and the %attitude pairs that induce incompatibility
type of incompatibility ({Incompatibility Type}).
It can be seen that \textit{Sublicense}, \textit{Use Trademark}, and \textit{Commercial Use} are the top 3 license terms that lead to incompatibility issues, affecting {7,801, 7,761, and 7,418} conflicts, respectively. 
For \textit{Sublicense}, we find that the incompatibility typically occurs when a {project license} indicates that users CAN incorporate the work into works with more restrictive licenses, while its {component license} declares CANNOT. The same can be seen in \textit{Commercial Use}, where a {project license} allows using the project for commercial purposes, while its {component licenses} declares CANNOT. 
Developers need to pay attention to these license terms especially when they incorporate third-party software packages accompanied with licenses, since unauthorized use of such packages may lead to the legal and financial risks.

We can also observe that \textit{Include License} is involved in projects most frequently (i.e., in 1,306 projects with license incompatibility), which is very common license term among most open source software licenses. For \textit{Include License}, the incompatibility usually occurs between the two component licenses; one license component declares that users MUST include the full text of license in modified software, while another component license declares CANNOT.

\revise{We also investigated the top 5 license terms with incompatibility issues for official licenses and custom licenses, respectively. As shown in Table~\ref{table-top5}, there are some differences between the top license terms with conflicts of official licenses and custom licenses. Specifically, the top 3 terms of custom licenses are \textit{Distribute}, \textit{Commercial Use}, and \textit{Modify}, while the top 3 terms of official licenses are \textit{Use Trademark}, \textit{Sublicense}, and \textit{Commercial Use}. We can also observe that in both official and custom licenses, top 5 conflict terms are all right-related terms. The results encourage developers to pay more attention on the rights conveyed by licenses. Finally, some license terms are not frequently involved in conflicts. It does not mean that these license terms are more reliable than others, since the reason behind might be that they are rarely used in licenses.} %\textit{Use Trademark}The reason behind that is numerous custom licenses are more likely to be created for clarifying the developer's licensing purposes on those significant software behaviours, e.g., modify the code of open source software, distribute the derivative works and use the open source software for commercial purposes, which are more often to appear in custom licenses. For official licenses, their top terms list seems more similar to that of the whole licenses, which is mainly because official licenses take up the majority of available licenses and also have main influence on license incompatibility issues. 
%Moreover, the terms appearing in above lists have the similar compositions, which not equals to the other license terms are completely safe and reliable. After analyzing the incompatibility issues and related license content, we found that these license terms are relatively rare in license dataset, which finally cause less term conflicts. }

\begin{table}[ht]
\centering
%\color{blue}
	\footnotesize
	 	\caption{Top 5 Terms with Incompatibility Issues in Official Licenses and Custom Licenses}
\begin{tabular}{c|ccccc}
\toprule
No. & 1              & 2                 & 3               & 4             & 5               \\\hline
Official & Use Trademark  & Sublicense        & Commercial Use  & Modify        & Distribute      \\\hline
Custom & Distribute  & Commercial Use       & Modify  & Sublicense        & Use Trademark      \\\bottomrule
\end{tabular}
\label{table-top5}
\end{table}

\section{Discussion}

In this section, we first discuss the lessons learned from perspectives of different stakeholders (i.e., developers, commercial product managers), and then discuss the limitations and threats to validity.

\subsection{Lessons Learned}
\noindent \textbf{From the perspective of developers}, although open source software facilitates software development, developers need to pay more attention to the compatibility between multiple licenses, especially when incorporating third-party software packages. \revise{\textbf{First}, as revealed in the empirical study (Section~\ref{sec:emp}), {over 70 percent} of popular projects in GitHub suffer from the problem of license incompatibility, each of which contains 56 pairs of conflict licenses in average. \textbf{Second}, developers need to carefully select the project license, since more than 80\% conflicts occur between the project license and other licenses. Especially for projects incorporating third-party software packages, developers are expected to check all component licenses and select the license that are more restrictive than component licenses as the project license. For instance, although widely-used and very permissive, licenses such as the MIT License are often involved in incompatibility issues when they are used as a project license. \textbf{Third}, licenses may have different versions and exceptions that regulate different rights and obligations. Developers are also allowed to create their own licenses based on some official licenses (i.e., exceptions). Relying on license identification tools such as Ninka~\citep{ase`10-German-Ninka} may lead to misunderstandings of license texts, resulting in license incompatibility issues.} %popular  Well-known licenses such as the MIT license, the Zope Public License 2.1, and Apache License 2.0 License contribute to more than half of license conflicts. License texts are typically long and complicated, but developers need to figure out the relationships between licenses. 
\tool can help developers understand the semantics of license texts automatically and then detect license incompatibility within a project. With the assistant of \tool, developers can avoid infringements of laws and rules when reusing existing knowledge.  %also recommend one compatible license for the project or even generate one by combining the sentences with compatible terms and attitudes.
%\textcolor{brown}{\tool can assist them in understanding OSS licenses easier and making choices more efficiently. Faced with quantities of long and complicated licenses, software developers encounter a difficulty in understanding them. In this paper, through extracting crucial license terms and term attitudes, \tool can help OSS developers grasp the key information quickly, based on which, they can better handle the software reuse situation and their selection among numerous licenses will thereby come easier. }

\noindent \textbf{From the perspective of commercial product manager}, although common open source licenses are popular in the OSS community, authors can also create their own licenses, which are difficult to be identified with existing tools. When reusing knowledge from OSS, companies should be aware of the risk of incompatible licenses, especially the attitudes of authors towards \textit{Sublicense}, \textit{Use Trademark}, and \textit{Commercial Use} as revealed in Section~\ref{sec:emp}. With the ability to detect license incompatibility, \tool can help commercial companies examine the compatibility between multiple licenses accurately, so as to avoid financial and legal risks. In addition, with \tool, product managers can acquire license terms and attitudes implied by each license, which provide useful information for product managers to select an appropriate license or create a new license for their software products. %which is compatible with existing component licenses for the whole project.% that their software products do not violate any licenses, and  and Under current open source software environment, license compliance and compatibility of software are receiving increasing attention. And \tool is born for incompatibility detection with a relatively higher accuracy, and can benefit software product managers a lot for avoiding involving license incompatibility issues. legality
%\textcolor{brown}{\tool can help them check license compatibility issues more accurately thereby avoiding potential commercial or legal risks. License compliance and compatibility of software are drawing more and more attention in current open source circumstances. We propose \tool, a license incompatibility detector, with a relatively higher accuracy and more thorough detecting process, can benefit software product manager a lot for preventing from license incompatibility adventures. }
%In addition, \tool can also help software companies to ensure the legality of their software products, so that any licenses are not violated by their products. 

\subsection{Limitations and Threats to Validity}
\noindent \textbf{Limitations}.
\tool utilizes the \textit{Stanford CoreNLP} tool~\citep{url-corenlp} to preprocess and parse license sentences. For this reason, the effectiveness of \textit{Stanford CoreNLP} affects the accuracy of \tool. For instance, by analyzing the evaluation results, we found that when the target sentence is too long and complicated, \textit{Stanford CoreNLP} only parsed parts of the sentence, which affects the effectiveness of \tool. In addition, we construct a probabilistic model to identify license terms, therefore it cannot be ensured that all license terms can be identified by the proposed model. Nevertheless, from our experimental results, we can see \tool failed to detect incompatibility issues {in only {17} cases, with {10.06\%} FP rate and {2.56\%} FN rate.} Therefore, \tool is still effective to detect license incompatibility for most projects. 
%Finally, the absence of a license or a license term denotes that users are not permitted to reuse, modify or distribute the works. Nevertheless, such assumption may not hold in all cases, which limits the effectiveness of \tool.
%utilize many NLP techniques. In the background of the still active NLP research field, the performance of our approches is limited by the development of NLP. 
Another limitation is that \tool cannot filter out irrelevant licenses. For example, for files that are only for testing purposes and are not really component files, if they contain licenses and should not be considered for incompatibility detection, \tool cannot filter them out. {\color{black}{Moreover, for licenses that are neither designated in the target projects nor obtained by \tool from external resources, \tool fails to collect these licenses and detect incompatibility. For instance, although \tool collects licenses accompanied with imported packages, for packages whose sources cannot be found, \tool can not collect them and detect incompatibility. Another example is that if derivative works only reuse OSS code without any licenses or references, \tool cannot obtain licenses and thus cannot detect license incompatibility in this case. Some related research directions such as clone detection~\citep{li2017cclearner} and license compliance detection~\citep{german2012method} address such issues, while this paper only focuses on incompatibility between multiple licenses within the same project. Finally, for referenced licenses, we collect licenses accompanied with imported packages, while ignoring packages which are further imported by these packages.}}

\noindent \textbf{Threats to Validity.}
(1) \tool identifies 23 types of license terms from each license to detect incompatibility. 
%As previous studies~\citep{url-tldrlegal}, we expect these license terms can cover all rights, obligations and stated conditions. 
Nevertheless, there could be a few cases where authors of software products have special requirements outside the scope of \tool. %{For example, some developers may have special conditions of software use, which can not be simply summarized by CAN, CANNOT, or MUST. However, we investigated XXX projects and XXX licenses, and found that only XXX licenses state such conditions of software use.} 
%In addition, the empirical study in this paper may suffer from the problem of data imbalance. To mitigate it, we collected a large-scale dataset (i.e., {507,688} license texts from 1,846 projects) to investigate the incompatibility issues in real-world OSS. 
(2) \revise{In addition, \tool may detect some incompatibility issues based on the inaccurate results from the previous two phases (i.e., term identification, and right and obligation inferring), leading to false positives. However, according to our experiments, the accuracy of the two phases achieves over 90\%, and the incompatibility detection accuracy also reaches over 90\%, which is much higher than state-of-the-art tools. 
%Even if some false positives occur in term identification or obligation inferring phases, there is still a chance that \tool can correctly detect the incompatibility issues since incompatibility usually not only involves one conflict.
}
(3) Another threat may be that due to the substantial manual effort of labelling, we were not able to collect a large-scale dataset for license entity tagging. However, we employ a semi-supervised learning method that incorporates unlabelled samples for training. The experimental results show that \tool achieves 93.28\% precision and 83.58\% F1 score %\sihan{(confirm)} 
when identifying license terms, which outperforms the baselines in this phase. (4) The performance of \tool was evaluated on a ground-truth dataset comprising 200 popular projects, and the quality of the test set may threaten the results. To mitigate this problem, three experts manually verified and cross-validated the dataset, which has been made available online for further study~\citep{lidetector}.

\section{Related Work}

\subsection{Semantic Extraction% from License Texts
}
\balance
%Most %official 
License texts are typically long and complicated, which are not straightforward for developers to understand.
%and the words are relatively academic, 
%which creates a big obstacle for developers to read and understand. 
To facilitate the process of understanding and choosing licenses, much research has been done to extract license semantics. The previous studies can be categorized into two groups.
%Studies in this subsection can be classified into two categories: ontology study and term extraction. 

\noindent \textbf{The ontology study}. Alspaugh et al.~\citep{IREC`09-Alspaugh-Intellectual}~\citep{AIS`10-Alspaugh-Challenge} extracted tuples (e.g., actor, action, and object) from license texts to model 10 licenses. 
Gordon et al.~\citep{Qualipso`10-gordon-prototype} used the Web Ontology Language (OWL) tool to build the ontology for OSS licenses and projects. With manual analysis, the constructed ontology contains knowledge from 8 popular licenses. Based on this ontology, the authors further analyzed license compatibility for open source software~\citep{ICAIL`11-Gordon-Analyzing}, and developed the MARKOS license analyzer~\citep{ICAIL`13-Gordon-Introducing}~\citep{COMMA`14-Gordon-Demonstration}. These studies typically extracted information from license texts by manually analyzing several licenses, which limits its application scope. 

\noindent \textbf{Term extraction}. 
%the FindOSSLicense tool developed by Kapitsaki et al.~\citep{tse`18-Kapitsaki-findOSSLicense} also involves the modeling of license texts. 
FindOSSLicense~\citep{tse`18-Kapitsaki-findOSSLicense} classified and summarized license sentences in 24 license texts through manual analysis, and obtained terms to model licenses. 
Based on a topic model, FOSS-LTE ~\citep{APSEC`17-Kapitsaki-termsIdentifying} identified license terms by mapping licenses terms with sentences, and mapping sentences with topics by Latent Dirichlet Allocation (LDA). By this means, it built the relationships between terms and topics. 
%In general, the research of Alspaugh et al.~\citep{IREC`09-Alspaugh-Intellectual}~\citep{AIS`10-Alspaugh-Challenge}, Gordon et al.~\citep{Qualipso`10-gordon-prototype}, and findOSSlicense~\citep{tse`18-Kapitsaki-findOSSLicense} understand licenses through manual inspection on a small number of licenses, and are not easy to migrate.
%and summarization within a small license scope, and are not easy to migrate. 
FOSS-LTE is most related to this work, focusing on the automated extraction of license terms to help developers better understand of licenses. 
%and our task are more similar, both focus on the automatic extraction of license terms, 
%aiming to achieve an automated method of terms extraction with stronger migratory to better deal with a variety of license types and better help developers understand the key information of licenses. 
However, the simple topic model employed by FOSS-LTE may introduce much noise, which could be a possible explanation for the low accuracy of FOSS-LTE. 
%resulting partly from 
%due to its thoroughly unsupervised learning, 
%and the effectiveness of its model needs further verification with only 15 licenses contained in the test dataset. 
In contrast, \tool prepossesses licenses with a clustering technique, followed by a two-phase learning method (i.e., term entity extraction, right and obligation inference), which is capable of extracting licenses flexible in expressions. 
%license modelling process into two steps, the term entity extraction, and the authorization type identification, which is easier to understand and covers more possibilities (but the \textit{term} in FOSS-LTE actually includes both terms entities and authorization types). 
In addition, \tool employs semi-supervised learning to train the term entity extraction model so as to save the labelling effort and enhance performance.
%on more than 600 licenses and finally uses 80 license texts as a test set to verify the effectiveness of our approach. 

\subsection{License Identification% for Name and Version
}
In order to facilitate the understanding of licenses, some studies attempted to identify common licenses automatically. 
%with a fast recognition speed but low accuracy, and they are highly dependent on prior knowledge. 
Gobeille et al.~\citep{msr`08-Gobeille-fossology} implemented FOSSology that uses a binary Symbol Alignment Matrix algorithm (bSAM) to identify licenses.
%which is more accurate than the former two but slower. 
\revise{Tuunanen et al.~\citep{ase`09-Timo-ASLA} developed a tool called ASLA to identify software licenses based on regular expressions, and analyze the interactions between a set of licenses.} Similarly, Xu et al.~\citep{cise`10-Xu-OSLC} proposed regular matching to identify licenses including their names and versions. 
%is also a GUI environment providing analysis of interactions between different licenses in FOSS components of heterogeneous systems. 
German et al.~\citep{ase`10-German-Ninka} developed Ninka for automated license identification based on sentence matching.
%involving the filtering of license keywords, synonyms, rules and other information. 
%It is currently one of the tools with high recognition accuracy and recall rates. Ninka can identify 112 popular OSS licenses, but its recognition range is relatively limited due to the increasing new or custom licenses.
%that continue to appear today. 
Higashi et al.~\citep{IWESEP`16-Higashi-clusteringForNinka} exploited the cluster learning method to further identify licenses marked as \textit{unknown} by Ninka, as a supplementary study. 

Despite the progress, previous studies on license identification requires much prior knowledge from experts, and can only be applied on a predefined set of licenses. In contrast, \tool learns to extract the semantics of licenses from a large corpus of license sentences, which equips it with the capability to analyze arbitrary licenses which are flexible in expressions. 
%As a large number of newly generated licenses appears, the license identification tool above can only cover a few types in it, which can hardly satisfy the more flexible needs of OSS developers. Our approach, however, focuses on the license content itself 
%and deconstructs it under a widely accepted terms frame and through a unified extraction method, so as to obtain larger automation capability and adaptability. ???
%Moreover, our research explains the licenses by extracting terms and their attitudes from license texts to better understand
%licenses, whereas the aforementioned license identification tools focus mainly on identifying license names and versions from source code, which are not detailed enough for OSS developers. 

\subsection{License Incompatibility Detection}
To avoid legal and financial risks, recently, much research has been done to detect license incompatibility. A major of studies on license incompatibility detection are graph-based approaches~\citep{jss`17-Kapitsaki-SPDX,SRDSCB`14-Kapitsaki-SPDX,Wheeler`07-Wheeler,paschalides2016validate}. 
%such as Kapitsaki et al.~\citep{jss`17-Kapitsaki-SPDX}~\citep{SEKE`12-Kapitsaki-SPDX}~\citep{SRDSCB`14-Kapitsaki-SPDX} and Wheeler’s~\citep{Wheeler`07-Wheeler}.
Typically, these studies manually constructed a graph to describe the compatibility relationships between a set of licenses, and detected incompatibility between these licenses by traversing the nodes and edges of the graph. \revise{For instance, Paschalides and Kapitsaki~\citep{paschalides2016validate} proposed a tool named SLVT, which is based on the
directed acyclic license graph, to examine license violations that may
exist in a single or multiple SPDX files.} Although strict and efficient, there exist a large number of licenses in real-word OSS, including various versions, exceptions, and custom licenses. Therefore, it is difficult to manually analyze the relationships between all licenses. Unlike graph-based approaches, \tool is a flexible and extensible tool that can be applied on an arbitrary license without prior knowledge, which enlarges its application scope.

\section{Conclusion}

In this paper, we propose \tool, an automated and effective tool to detect license incompatibility for open source software. It first identifies license terms and then infers the attitudes of authors towards identified terms. The experimental results demonstrate the effectiveness of \tool in incompatibility detection for OSS.
%that \tool is capable of extracting terms and inferring rights and obligations with high performance, i.e., 76.95\% F1 score, and can effectively identify license incompatibility in OSS with low FP and FN.
%only 2\% FP and 1.5\% FN. 
We also study the license incompatibility status on 1,846 real-world open source projects by leveraging \tool, and analyze the characteristics of incompatibility issues and licenses. Our large-scale empirical study on 1,846 projects reveals that {72.91\%} of the projects are suffering from license incompatibility, including popular ones such as the {MIT} License and the {Apache License}. We highlighted lessons learned from perspectives of different stakeholders. {\color{black}{All the datasets~\citep{lidetector} and the replication package~\citep{lidetector-github} are released for follow-up research}}.
We believe \tool can benefit different stakeholders (e.g., developers) in terms of license compatibility. 

\section*{Acknowledgement}
This work was supported by National Natural Science Foundation of China (No. 62102197) and National Key Research Project of China (No. 2021YFF0307202 and No. 2020YFB1005700).

%\clearpage
\bibliographystyle{ACM-Reference-Format}
\bibliography{main.bib}

\begin{thebibliography}{70}

%%% the Web address.

\ifx \showCODEN    \undefined \def \showCODEN     #1{\unskip}     \fi
\ifx \showDOI      \undefined \def \showDOI       #1{#1}\fi
\ifx \showISBNx    \undefined \def \showISBNx     #1{\unskip}     \fi
\ifx \showISBNxiii \undefined \def \showISBNxiii  #1{\unskip}     \fi
\ifx \showISSN     \undefined \def \showISSN      #1{\unskip}     \fi
\ifx \showLCCN     \undefined \def \showLCCN      #1{\unskip}     \fi
\ifx \shownote     \undefined \def \shownote      #1{#1}          \fi
\ifx \showarticletitle \undefined \def \showarticletitle #1{#1}   \fi
\ifx \showURL      \undefined \def \showURL       {\relax}        \fi
% invisible to TeX
\providecommand\bibfield[2]{#2}
\providecommand\bibinfo[2]{#2}
\providecommand\natexlab[1]{#1}
\providecommand\showeprint[2][]{arXiv:#2}

\bibitem[\protect\citeauthoryear{Alspaugh, Asuncion, and Scacchi}{Alspaugh
  et~al\mbox{.}}{2009}]%
        {IREC`09-Alspaugh-Intellectual}
\bibfield{author}{\bibinfo{person}{Thomas~A Alspaugh},
  \bibinfo{person}{Hazeline~U Asuncion}, {and} \bibinfo{person}{Walt Scacchi}.}
  \bibinfo{year}{2009}\natexlab{}.
\newblock \showarticletitle{Intellectual property rights requirements for
  heterogeneously-licensed systems}. In \bibinfo{booktitle}{\emph{Proceedings
  of the 17th IEEE International Requirements Engineering Conference}}.
  \bibinfo{pages}{24--33}.
\newblock


\bibitem[\protect\citeauthoryear{Alspaugh, Scacchi, and Asuncion}{Alspaugh
  et~al\mbox{.}}{2010}]%
        {AIS`10-Alspaugh-Challenge}
\bibfield{author}{\bibinfo{person}{Thomas~A Alspaugh}, \bibinfo{person}{Walt
  Scacchi}, {and} \bibinfo{person}{Hazeline~U Asuncion}.}
  \bibinfo{year}{2010}\natexlab{}.
\newblock \showarticletitle{Software licenses in context: The challenge of
  heterogeneously-licensed systems}.
\newblock \bibinfo{journal}{\emph{Journal of the Association for Information
  Systems}} \bibinfo{volume}{11}, \bibinfo{number}{11} (\bibinfo{year}{2010}),
  \bibinfo{pages}{2}.
\newblock


\bibitem[\protect\citeauthoryear{Andow, Mahmud, Wang, Whitaker, Enck, Reaves,
  Singh, and Xie}{Andow et~al\mbox{.}}{2019}]%
        {USENIX`19-Andow-Policylint}
\bibfield{author}{\bibinfo{person}{Benjamin Andow},
  \bibinfo{person}{Samin~Yaseer Mahmud}, \bibinfo{person}{Wenyu Wang},
  \bibinfo{person}{Justin Whitaker}, \bibinfo{person}{William Enck},
  \bibinfo{person}{Bradley Reaves}, \bibinfo{person}{Kapil Singh}, {and}
  \bibinfo{person}{Tao Xie}.} \bibinfo{year}{2019}\natexlab{}.
\newblock \showarticletitle{Policylint: investigating internal privacy policy
  contradictions on Google Play}. In \bibinfo{booktitle}{\emph{Proceedings of
  the 28th USENIX Conference on Security Symposium}}.
  \bibinfo{pages}{585–602}.
\newblock


\bibitem[\protect\citeauthoryear{BDF}{BDF}{2021}]%
        {url-ninkacannot}
\bibfield{author}{\bibinfo{person}{BDF}.} \bibinfo{year}{2021}\natexlab{}.
\newblock \bibinfo{title}{The Backdoor Factory}.
\newblock
  \bibinfo{howpublished}{https://github.com/secretsquirrel/the-backdoor-factory}.
\newblock


\bibitem[\protect\citeauthoryear{Blosc}{Blosc}{2021}]%
        {url-blosc}
\bibfield{author}{\bibinfo{person}{Blosc}.} \bibinfo{year}{2021}\natexlab{}.
\newblock \bibinfo{title}{A blocking, shuffling and lossless compression
  library}.
\newblock \bibinfo{howpublished}{https://github.com/Blosc/c-blosc}.
\newblock


\bibitem[\protect\citeauthoryear{Chen, Fan, Meng, Su, Xue, Xue, Liu, and
  Xu}{Chen et~al\mbox{.}}{2020}]%
        {chen2020empirical}
\bibfield{author}{\bibinfo{person}{Sen Chen}, \bibinfo{person}{Lingling Fan},
  \bibinfo{person}{Guozhu Meng}, \bibinfo{person}{Ting Su},
  \bibinfo{person}{Minhui Xue}, \bibinfo{person}{Yinxing Xue},
  \bibinfo{person}{Yang Liu}, {and} \bibinfo{person}{Lihua Xu}.}
  \bibinfo{year}{2020}\natexlab{}.
\newblock \showarticletitle{An empirical assessment of security risks of global
  android banking apps}. In \bibinfo{booktitle}{\emph{2020 IEEE/ACM 42nd
  International Conference on Software Engineering (ICSE)}}. IEEE,
  \bibinfo{pages}{1310--1322}.
\newblock


\bibitem[\protect\citeauthoryear{Chen, Su, Fan, Meng, Xue, Liu, and Xu}{Chen
  et~al\mbox{.}}{2018}]%
        {chen2018mobile}
\bibfield{author}{\bibinfo{person}{Sen Chen}, \bibinfo{person}{Ting Su},
  \bibinfo{person}{Lingling Fan}, \bibinfo{person}{Guozhu Meng},
  \bibinfo{person}{Minhui Xue}, \bibinfo{person}{Yang Liu}, {and}
  \bibinfo{person}{Lihua Xu}.} \bibinfo{year}{2018}\natexlab{}.
\newblock \showarticletitle{Are mobile banking apps secure? what can be
  improved?}. In \bibinfo{booktitle}{\emph{Proceedings of the 2018 26th ACM
  Joint Meeting on European Software Engineering Conference and Symposium on
  the Foundations of Software Engineering}}. \bibinfo{pages}{797--802}.
\newblock


\bibitem[\protect\citeauthoryear{choosealicense}{choosealicense}{2012}]%
        {url-choosealicense}
\bibfield{author}{\bibinfo{person}{choosealicense}.}
  \bibinfo{year}{2012}\natexlab{}.
\newblock \bibinfo{title}{Choose an open source license}.
\newblock \bibinfo{howpublished}{https://choosealicense.com/no-permission/}.
\newblock


\bibitem[\protect\citeauthoryear{F.}{F.}{2010}]%
        {Qualipso`10-gordon-prototype}
\bibfield{author}{\bibinfo{person}{Gordon~Thomas F.}}
  \bibinfo{year}{2010}\natexlab{}.
\newblock \showarticletitle{Report on prototype decision support system for oss
  license compatibility issues}.
\newblock \bibinfo{journal}{\emph{Qualipso}} (\bibinfo{year}{2010}),
  \bibinfo{pages}{80}.
\newblock


\bibitem[\protect\citeauthoryear{facebookarchive}{facebookarchive}{2021}]%
        {url-example1}
\bibfield{author}{\bibinfo{person}{facebookarchive}.}
  \bibinfo{year}{2021}\natexlab{}.
\newblock \bibinfo{title}{Augmented Traffic Control}.
\newblock
  \bibinfo{howpublished}{https://github.com/facebookarchive/augmented-traffic-control}.
\newblock


\bibitem[\protect\citeauthoryear{Fan, Wang, Yan, Song, Zhu, and Chen}{Fan
  et~al\mbox{.}}{2020}]%
        {ISPRS`20-Runyu-NER}
\bibfield{author}{\bibinfo{person}{Runyu Fan}, \bibinfo{person}{Lizhe Wang},
  \bibinfo{person}{Jining Yan}, \bibinfo{person}{Weijing Song},
  \bibinfo{person}{Yingqian Zhu}, {and} \bibinfo{person}{Xiaodao Chen}.}
  \bibinfo{year}{2020}\natexlab{}.
\newblock \showarticletitle{Deep learning-based named entity recognition and
  knowledge graph construction for geological hazards}.
\newblock \bibinfo{journal}{\emph{ISPRS International Journal of
  Geo-Information}} (\bibinfo{year}{2020}).
\newblock


\bibitem[\protect\citeauthoryear{Foundation}{Foundation}{2018}]%
        {url-spdx}
\bibfield{author}{\bibinfo{person}{Linux Foundation}.}
  \bibinfo{year}{2018}\natexlab{}.
\newblock \bibinfo{title}{The Software Package Data Exchange}.
\newblock \bibinfo{howpublished}{https://spdx.dev/}.
\newblock


\bibitem[\protect\citeauthoryear{Gangadharan, D’Andrea, De~Paoli, and
  Weiss}{Gangadharan et~al\mbox{.}}{2012}]%
        {gangadharan2012managing}
\bibfield{author}{\bibinfo{person}{GR Gangadharan}, \bibinfo{person}{Vincenzo
  D’Andrea}, \bibinfo{person}{Stefano De~Paoli}, {and}
  \bibinfo{person}{Michael Weiss}.} \bibinfo{year}{2012}\natexlab{}.
\newblock \showarticletitle{Managing license compliance in free and open source
  software development}.
\newblock \bibinfo{journal}{\emph{Information Systems Frontiers}}
  (\bibinfo{year}{2012}), \bibinfo{pages}{143--154}.
\newblock


\bibitem[\protect\citeauthoryear{German and Di~Penta}{German and
  Di~Penta}{2012}]%
        {german2012method}
\bibfield{author}{\bibinfo{person}{Daniel German} {and}
  \bibinfo{person}{Massimiliano Di~Penta}.} \bibinfo{year}{2012}\natexlab{}.
\newblock \showarticletitle{A method for open source license compliance of java
  applications}.
\newblock \bibinfo{journal}{\emph{IEEE software}} \bibinfo{volume}{29},
  \bibinfo{number}{3} (\bibinfo{year}{2012}), \bibinfo{pages}{58--63}.
\newblock


\bibitem[\protect\citeauthoryear{German, Manabe, and Inoue}{German
  et~al\mbox{.}}{2010}]%
        {ase`10-German-Ninka}
\bibfield{author}{\bibinfo{person}{Daniel~M. German}, \bibinfo{person}{Yuki
  Manabe}, {and} \bibinfo{person}{Katsuro Inoue}.}
  \bibinfo{year}{2010}\natexlab{}.
\newblock \showarticletitle{A sentence-matching method for automatic license
  identification of source code files}. In
  \bibinfo{booktitle}{\emph{Proceedings of the IEEE/ACM International
  Conference on Automated Software Engineering}}. \bibinfo{pages}{437–446}.
\newblock


\bibitem[\protect\citeauthoryear{Gobeille}{Gobeille}{2008}]%
        {msr`08-Gobeille-fossology}
\bibfield{author}{\bibinfo{person}{Robert Gobeille}.}
  \bibinfo{year}{2008}\natexlab{}.
\newblock \showarticletitle{The fossology project}. In
  \bibinfo{booktitle}{\emph{Proceedings of the 2008 International Working
  Conference on Mining Software Repositories}}. \bibinfo{pages}{47--50}.
\newblock


\bibitem[\protect\citeauthoryear{Gordon}{Gordon}{2011}]%
        {ICAIL`11-Gordon-Analyzing}
\bibfield{author}{\bibinfo{person}{Thomas~F. Gordon}.}
  \bibinfo{year}{2011}\natexlab{}.
\newblock \showarticletitle{Analyzing open Source license compatibility issues
  with Carneades}. In \bibinfo{booktitle}{\emph{Proceedings of the 13th
  International Conference on Artificial Intelligence and Law}}.
  \bibinfo{pages}{51–55}.
\newblock


\bibitem[\protect\citeauthoryear{Gordon}{Gordon}{2013}]%
        {ICAIL`13-Gordon-Introducing}
\bibfield{author}{\bibinfo{person}{Thomas~F. Gordon}.}
  \bibinfo{year}{2013}\natexlab{}.
\newblock \showarticletitle{Introducing the Carneades web application}. In
  \bibinfo{booktitle}{\emph{Proceedings of the Fourteenth International
  Conference on Artificial Intelligence and Law}}. \bibinfo{pages}{243–244}.
\newblock


\bibitem[\protect\citeauthoryear{Gordon}{Gordon}{2014}]%
        {COMMA`14-Gordon-Demonstration}
\bibfield{author}{\bibinfo{person}{Thomas~F. Gordon}.}
  \bibinfo{year}{2014}\natexlab{}.
\newblock \showarticletitle{A demonstration of the MARKOS license analyser}. In
  \bibinfo{booktitle}{\emph{Proceedings of the 5th International Conference on
  Computational Models of Argument}}. \bibinfo{pages}{461--462}.
\newblock


\bibitem[\protect\citeauthoryear{Group}{Group}{2020}]%
        {url-corenlp}
\bibfield{author}{\bibinfo{person}{Stanford~NLP Group}.}
  \bibinfo{year}{2020}\natexlab{}.
\newblock \bibinfo{title}{corenlp}.
\newblock \bibinfo{howpublished}{https://stanfordnlp.github.io/CoreNLP/}.
\newblock


\bibitem[\protect\citeauthoryear{Guo, Chen, Xing, Li, Bai, and Sun}{Guo
  et~al\mbox{.}}{2021a}]%
        {guo2021detecting}
\bibfield{author}{\bibinfo{person}{Hao Guo}, \bibinfo{person}{Sen Chen},
  \bibinfo{person}{Zhenchang Xing}, \bibinfo{person}{Xiaohong Li},
  \bibinfo{person}{Yude Bai}, {and} \bibinfo{person}{Jiamou Sun}.}
  \bibinfo{year}{2021}\natexlab{a}.
\newblock \showarticletitle{Detecting and Augmenting Missing Key Aspects in
  Vulnerability Descriptions}.
\newblock \bibinfo{journal}{\emph{ACM Transactions on Software Engineering and
  Methodology (TOSEM)}} (\bibinfo{year}{2021}).
\newblock


\bibitem[\protect\citeauthoryear{Guo, Xing, Chen, Li, Bai, and Zhang}{Guo
  et~al\mbox{.}}{2021b}]%
        {guo2021key}
\bibfield{author}{\bibinfo{person}{Hao Guo}, \bibinfo{person}{Zhenchang Xing},
  \bibinfo{person}{Sen Chen}, \bibinfo{person}{Xiaohong Li},
  \bibinfo{person}{Yude Bai}, {and} \bibinfo{person}{Hu Zhang}.}
  \bibinfo{year}{2021}\natexlab{b}.
\newblock \showarticletitle{Key aspects augmentation of vulnerability
  description based on multiple security databases}. In
  \bibinfo{booktitle}{\emph{2021 IEEE 45th Annual Computers, Software, and
  Applications Conference (COMPSAC)}}. IEEE, \bibinfo{pages}{1020--1025}.
\newblock


\bibitem[\protect\citeauthoryear{HaboMalHunter}{HaboMalHunter}{2021}]%
        {url-mitcannot}
\bibfield{author}{\bibinfo{person}{HaboMalHunter}.}
  \bibinfo{year}{2021}\natexlab{}.
\newblock \bibinfo{title}{Habo Linux Malware Analysis System}.
\newblock \bibinfo{howpublished}{https://github.com/Tencent/HaboMalHunter}.
\newblock


\bibitem[\protect\citeauthoryear{Higashi, Manabe, and Ohira}{Higashi
  et~al\mbox{.}}{2016}]%
        {IWESEP`16-Higashi-clusteringForNinka}
\bibfield{author}{\bibinfo{person}{Yunosuke Higashi}, \bibinfo{person}{Yuki
  Manabe}, {and} \bibinfo{person}{Masao Ohira}.}
  \bibinfo{year}{2016}\natexlab{}.
\newblock \showarticletitle{Clustering OSS license statements toward automatic
  generation of license rules}. In \bibinfo{booktitle}{\emph{Proceddings of the
  7th International Workshop on Empirical Software Engineering in Practice}}.
  \bibinfo{pages}{30--35}.
\newblock


\bibitem[\protect\citeauthoryear{Kapitsaki and Charalambous}{Kapitsaki and
  Charalambous}{2019}]%
        {tse`18-Kapitsaki-findOSSLicense}
\bibfield{author}{\bibinfo{person}{Georgia Kapitsaki} {and}
  \bibinfo{person}{Georgia Charalambous}.} \bibinfo{year}{2019}\natexlab{}.
\newblock \showarticletitle{Modeling and recommending open source licenses with
  findOSSLicense}.
\newblock \bibinfo{journal}{\emph{IEEE Transactions on Software Engineering}}
  (\bibinfo{year}{2019}).
\newblock


\bibitem[\protect\citeauthoryear{Kapitsaki and Kramer}{Kapitsaki and
  Kramer}{2014}]%
        {SRDSCB`14-Kapitsaki-SPDX}
\bibfield{author}{\bibinfo{person}{Georgia~M. Kapitsaki} {and}
  \bibinfo{person}{Frederik Kramer}.} \bibinfo{year}{2014}\natexlab{}.
\newblock \showarticletitle{Open source license violation check for SPDX
  files}. In \bibinfo{booktitle}{\emph{Software Reuse for Dynamic Systems in
  the Cloud and Beyond}}. \bibinfo{pages}{90--105}.
\newblock


\bibitem[\protect\citeauthoryear{Kapitsaki, Kramer, and Tselikas}{Kapitsaki
  et~al\mbox{.}}{2017}]%
        {jss`17-Kapitsaki-SPDX}
\bibfield{author}{\bibinfo{person}{Georgia~M. Kapitsaki},
  \bibinfo{person}{Frederik Kramer}, {and} \bibinfo{person}{Nikolaos~D.
  Tselikas}.} \bibinfo{year}{2017}\natexlab{}.
\newblock \showarticletitle{Automating the license compatibility process in
  open source software with SPDX}.
\newblock \bibinfo{journal}{\emph{Journal of Systems and Software}}
  (\bibinfo{year}{2017}), \bibinfo{pages}{386 -- 401}.
\newblock


\bibitem[\protect\citeauthoryear{Kapitsaki and Paschalides}{Kapitsaki and
  Paschalides}{2017}]%
        {APSEC`17-Kapitsaki-termsIdentifying}
\bibfield{author}{\bibinfo{person}{Georgia~M. Kapitsaki} {and}
  \bibinfo{person}{Demetris Paschalides}.} \bibinfo{year}{2017}\natexlab{}.
\newblock \showarticletitle{Identifying terms in open source software license
  texts}. In \bibinfo{booktitle}{\emph{Proceedigns of the 24th Asia-Pacific
  Software Engineering Conference}}. \bibinfo{pages}{540--545}.
\newblock


\bibitem[\protect\citeauthoryear{Karvelis, Gavrilis, Georgoulas, and
  Stylios}{Karvelis et~al\mbox{.}}{2018}]%
        {IJCNN`18-Doc2Vec-method}
\bibfield{author}{\bibinfo{person}{Petros Karvelis}, \bibinfo{person}{Dimitris
  Gavrilis}, \bibinfo{person}{George Georgoulas}, {and}
  \bibinfo{person}{Chrysostomos Stylios}.} \bibinfo{year}{2018}\natexlab{}.
\newblock \showarticletitle{Topic recommendation using Doc2Vec}. In
  \bibinfo{booktitle}{\emph{2018 International Joint Conference on Neural
  Networks}}. \bibinfo{pages}{1--6}.
\newblock


\bibitem[\protect\citeauthoryear{kevin}{kevin}{2012}]%
        {url-tldrlegal}
\bibfield{author}{\bibinfo{person}{kevin}.} \bibinfo{year}{2012}\natexlab{}.
\newblock \bibinfo{title}{Software Licenses in Plain English}.
\newblock \bibinfo{howpublished}{https://tldrlegal.com/}.
\newblock


\bibitem[\protect\citeauthoryear{Le and Mikolov}{Le and Mikolov}{2014}]%
        {ICML'14-Quoc-doc2vec}
\bibfield{author}{\bibinfo{person}{Quoc Le} {and} \bibinfo{person}{Tomas
  Mikolov}.} \bibinfo{year}{2014}\natexlab{}.
\newblock \showarticletitle{Distributed representations of sentences and
  documents}. In \bibinfo{booktitle}{\emph{Proceedings of the 31st
  International Conference on Machine Learning}}. \bibinfo{pages}{1188--1196}.
\newblock


\bibitem[\protect\citeauthoryear{Li, Feng, Zhuang, Meng, and Ryder}{Li
  et~al\mbox{.}}{2017}]%
        {li2017cclearner}
\bibfield{author}{\bibinfo{person}{Liuqing Li}, \bibinfo{person}{He Feng},
  \bibinfo{person}{Wenjie Zhuang}, \bibinfo{person}{Na Meng}, {and}
  \bibinfo{person}{Barbara Ryder}.} \bibinfo{year}{2017}\natexlab{}.
\newblock \showarticletitle{Cclearner: A deep learning-based clone detection
  approach}. In \bibinfo{booktitle}{\emph{2017 IEEE International Conference on
  Software Maintenance and Evolution (ICSME)}}. IEEE,
  \bibinfo{pages}{249--260}.
\newblock


\bibitem[\protect\citeauthoryear{librariesio}{librariesio}{2015}]%
        {url-librariesio}
\bibfield{author}{\bibinfo{person}{librariesio}.}
  \bibinfo{year}{2015}\natexlab{}.
\newblock \bibinfo{title}{Check compatibility between different SPDX licenses
  for checking dependency license compatibility.}
\newblock
  \bibinfo{howpublished}{https://github.com/librariesio/license-compatibility}.
\newblock


\bibitem[\protect\citeauthoryear{Licensing}{Licensing}{2004}]%
        {licensing2004software}
\bibfield{author}{\bibinfo{person}{Open~Source Licensing}.}
  \bibinfo{year}{2004}\natexlab{}.
\newblock \bibinfo{booktitle}{\emph{Software freedom and intellectual property
  law}}.
\newblock


\bibitem[\protect\citeauthoryear{Ling}{Ling}{2003}]%
        {url-postagslist}
\bibfield{author}{\bibinfo{person}{Ling}.} \bibinfo{year}{2003}\natexlab{}.
\newblock \bibinfo{title}{Alphabetical list of part-of-speech tags used in the
  Penn Treebank Project}.
\newblock
  \bibinfo{howpublished}{https://www.ling.upenn.edu/courses/Fall\_2003/ling001/penn\_treebank\_pos.html}.
\newblock


\bibitem[\protect\citeauthoryear{Liu, Chen, Fan, Chen, Liu, and Peng}{Liu
  et~al\mbox{.}}{2022}]%
        {liu2022demystifying}
\bibfield{author}{\bibinfo{person}{Chengwei Liu}, \bibinfo{person}{Sen Chen},
  \bibinfo{person}{Lingling Fan}, \bibinfo{person}{Bihuan Chen},
  \bibinfo{person}{Yang Liu}, {and} \bibinfo{person}{Xin Peng}.}
  \bibinfo{year}{2022}\natexlab{}.
\newblock \showarticletitle{Demystifying the Vulnerability Propagation and Its
  Evolution via Dependency Trees in the NPM Ecosystem}. In
  \bibinfo{booktitle}{\emph{2022 IEEE/ACM 44nd International Conference on
  Software Engineering (ICSE)}}. IEEE.
\newblock


\bibitem[\protect\citeauthoryear{Mancinelli, Boender, Di~Cosmo, Vouillon,
  Durak, Leroy, and Treinen}{Mancinelli et~al\mbox{.}}{2006}]%
        {mancinelli2006managing}
\bibfield{author}{\bibinfo{person}{Fabio Mancinelli}, \bibinfo{person}{Jaap
  Boender}, \bibinfo{person}{Roberto Di~Cosmo}, \bibinfo{person}{Jerome
  Vouillon}, \bibinfo{person}{Berke Durak}, \bibinfo{person}{Xavier Leroy},
  {and} \bibinfo{person}{Ralf Treinen}.} \bibinfo{year}{2006}\natexlab{}.
\newblock \showarticletitle{Managing the complexity of large free and open
  source package-based software distributions}. In
  \bibinfo{booktitle}{\emph{21st IEEE/ACM International Conference on Automated
  Software Engineering}}. \bibinfo{pages}{199--208}.
\newblock


\bibitem[\protect\citeauthoryear{Mathur, Choudhary, Vashist, Thies, and
  Thilagam}{Mathur et~al\mbox{.}}{2012}]%
        {AISEW`12-Mathur-Empirical}
\bibfield{author}{\bibinfo{person}{Arunesh Mathur}, \bibinfo{person}{Harshal
  Choudhary}, \bibinfo{person}{Priyank Vashist}, \bibinfo{person}{William
  Thies}, {and} \bibinfo{person}{Santhi Thilagam}.}
  \bibinfo{year}{2012}\natexlab{}.
\newblock \showarticletitle{An empirical study of license violations in open
  source projects}. In \bibinfo{booktitle}{\emph{Proceedings of the 35th Annual
  IEEE Software Engineering Workshop}}. \bibinfo{pages}{168--176}.
\newblock


\bibitem[\protect\citeauthoryear{nltk}{nltk}{2021}]%
        {url-nltk}
\bibfield{author}{\bibinfo{person}{nltk}.} \bibinfo{year}{2021}\natexlab{}.
\newblock \bibinfo{title}{Natural Language Toolkit}.
\newblock \bibinfo{howpublished}{https://www.nltk.org/}.
\newblock


\bibitem[\protect\citeauthoryear{Opensource}{Opensource}{2021}]%
        {url-oss}
\bibfield{author}{\bibinfo{person}{Opensource}.}
  \bibinfo{year}{2021}\natexlab{}.
\newblock \bibinfo{title}{What is open source?}
\newblock
  \bibinfo{howpublished}{https://opensource.com/resources/what-open-source}.
\newblock


\bibitem[\protect\citeauthoryear{Paschalides and Kapitsaki}{Paschalides and
  Kapitsaki}{2016}]%
        {paschalides2016validate}
\bibfield{author}{\bibinfo{person}{Demetris Paschalides} {and}
  \bibinfo{person}{Georgia~M Kapitsaki}.} \bibinfo{year}{2016}\natexlab{}.
\newblock \showarticletitle{Validate your SPDX files for open source license
  violations}. In \bibinfo{booktitle}{\emph{Proceedings of the 2016 24th ACM
  SIGSOFT International Symposium on Foundations of Software Engineering}}.
  \bibinfo{pages}{1047--1051}.
\newblock


\bibitem[\protect\citeauthoryear{paul}{paul}{2021}]%
        {url-qdox}
\bibfield{author}{\bibinfo{person}{paul}.} \bibinfo{year}{2021}\natexlab{}.
\newblock \bibinfo{title}{Full extractor of class/interface/method
  definitions}.
\newblock \bibinfo{howpublished}{https://github.com/paul-hammant/qdox}.
\newblock


\bibitem[\protect\citeauthoryear{pivotal}{pivotal}{2021}]%
        {url-LicenseFinder}
\bibfield{author}{\bibinfo{person}{pivotal}.} \bibinfo{year}{2021}\natexlab{}.
\newblock \bibinfo{title}{Find licenses for your project's dependencies.}
\newblock \bibinfo{howpublished}{https://github.com/pivotal/LicenseFinder}.
\newblock


\bibitem[\protect\citeauthoryear{ProgrammerSought}{ProgrammerSought}{2021}]%
        {chinacase}
\bibfield{author}{\bibinfo{person}{ProgrammerSought}.}
  \bibinfo{year}{2021}\natexlab{}.
\newblock \bibinfo{booktitle}{\emph{The first case of GPL agreement in China is
  settled. How should the relevant open source software be controlled?}}
\newblock


\bibitem[\protect\citeauthoryear{PyPi}{PyPi}{2021}]%
        {url-pypi}
\bibfield{author}{\bibinfo{person}{PyPi}.} \bibinfo{year}{2021}\natexlab{}.
\newblock \bibinfo{title}{Find, install and publish Python packages with the
  Python Package Index}.
\newblock \bibinfo{howpublished}{https://pypi.org/}.
\newblock


\bibitem[\protect\citeauthoryear{Reddy}{Reddy}{2015}]%
        {infringement-case}
\bibfield{author}{\bibinfo{person}{Jaideep Reddy}.}
  \bibinfo{year}{2015}\natexlab{}.
\newblock \bibinfo{title}{The Consequences of Violating Open Source Licenses}.
\newblock
  \bibinfo{howpublished}{https://btlj.org/2015/11/consequences-violating-open-source-licenses/}.
\newblock


\bibitem[\protect\citeauthoryear{{Reimers} and {Gurevych}}{{Reimers} and
  {Gurevych}}{2017}]%
        {CSCL`17-Reimers-BIO}
\bibfield{author}{\bibinfo{person}{Nils {Reimers}} {and} \bibinfo{person}{Iryna
  {Gurevych}}.} \bibinfo{year}{2017}\natexlab{}.
\newblock \showarticletitle{Optimal hyperparameters for deep LSTM-networks for
  sequence labeling tasks}.
\newblock \bibinfo{journal}{\emph{arXiv e-prints}} (\bibinfo{year}{2017}).
\newblock
\showeprint[arxiv]{1707.06799}


\bibitem[\protect\citeauthoryear{Richard~Socher}{Richard~Socher}{2014}]%
        {url-glove}
\bibfield{author}{\bibinfo{person}{Christopher D.~Manning Richard~Socher}.}
  \bibinfo{year}{2014}\natexlab{}.
\newblock \bibinfo{title}{GloVe: Global Vectors for Word Representation}.
\newblock \bibinfo{howpublished}{https://nlp.stanford.edu/projects/glove/}.
\newblock


\bibitem[\protect\citeauthoryear{robinhood}{robinhood}{2021}]%
        {url-example2}
\bibfield{author}{\bibinfo{person}{robinhood}.}
  \bibinfo{year}{2021}\natexlab{}.
\newblock \bibinfo{title}{Faust}.
\newblock \bibinfo{howpublished}{https://github.com/robinhood/faust}.
\newblock


\bibitem[\protect\citeauthoryear{Scicluna}{Scicluna}{2016}]%
        {FI`16-Scicluna-PCFG}
\bibfield{author}{\bibinfo{person}{Colin Scicluna, James de la~Higuera}.}
  \bibinfo{year}{2016}\natexlab{}.
\newblock \showarticletitle{Grammatical inference of PCFGs applied to language
  modelling and unsupervised parsing}.
\newblock \bibinfo{journal}{\emph{Fundamenta Informaticae}}
  (\bibinfo{year}{2016}), \bibinfo{pages}{379--402}.
\newblock


\bibitem[\protect\citeauthoryear{Socher, Perelygin, Wu, Chuang, Manning, Ng,
  and Potts}{Socher et~al\mbox{.}}{2013}]%
        {etal`13-Socher-SST}
\bibfield{author}{\bibinfo{person}{Richard Socher}, \bibinfo{person}{Alex
  Perelygin}, \bibinfo{person}{Jean Wu}, \bibinfo{person}{Jason Chuang},
  \bibinfo{person}{Christopher~D. Manning}, \bibinfo{person}{Andrew Ng}, {and}
  \bibinfo{person}{Christopher Potts}.} \bibinfo{year}{2013}\natexlab{}.
\newblock \showarticletitle{Recursive deep models for semantic compositionality
  over a sentiment treebank}. In \bibinfo{booktitle}{\emph{Proceedings of the
  2013 Conference on Empirical Methods in Natural Language Processing}}.
  \bibinfo{pages}{1631--1642}.
\newblock


\bibitem[\protect\citeauthoryear{Solakidis, Vavliakis, and Mitkas}{Solakidis
  et~al\mbox{.}}{2014}]%
        {IAT'14-Solakidis-keyword}
\bibfield{author}{\bibinfo{person}{Georgios~S. Solakidis},
  \bibinfo{person}{Konstantinos~N. Vavliakis}, {and}
  \bibinfo{person}{Pericles~A. Mitkas}.} \bibinfo{year}{2014}\natexlab{}.
\newblock \showarticletitle{Multilingual sentiment analysis using emoticons and
  keywords}. In \bibinfo{booktitle}{\emph{Proceedings of the IEEE/WIC/ACM
  International Joint Conferences on Web Intelligence and Intelligent Agent
  Technologies}}. \bibinfo{pages}{102--109}.
\newblock


\bibitem[\protect\citeauthoryear{SPDX}{SPDX}{2018a}]%
        {url-Apache-2.0}
\bibfield{author}{\bibinfo{person}{SPDX}.} \bibinfo{year}{2018}\natexlab{a}.
\newblock \bibinfo{title}{Apache License 2.0}.
\newblock \bibinfo{howpublished}{https://spdx.org/licenses/Apache-2.0.html}.
\newblock


\bibitem[\protect\citeauthoryear{SPDX}{SPDX}{2018b}]%
        {url-BSD-3-Clause}
\bibfield{author}{\bibinfo{person}{SPDX}.} \bibinfo{year}{2018}\natexlab{b}.
\newblock \bibinfo{title}{BSD 3-Clause "New" or "Revised" License}.
\newblock \bibinfo{howpublished}{https://spdx.org/licenses/BSD-3-Clause.html}.
\newblock


\bibitem[\protect\citeauthoryear{SPDX}{SPDX}{2018c}]%
        {url-CCBYSA}
\bibfield{author}{\bibinfo{person}{SPDX}.} \bibinfo{year}{2018}\natexlab{c}.
\newblock \bibinfo{title}{Creative Commons Attribution Share Alike 4.0
  International}.
\newblock \bibinfo{howpublished}{https://spdx.org/licenses/CC-BY-SA-4.0.html}.
\newblock


\bibitem[\protect\citeauthoryear{SPDX}{SPDX}{2018d}]%
        {url-LGPL-3.0-only}
\bibfield{author}{\bibinfo{person}{SPDX}.} \bibinfo{year}{2018}\natexlab{d}.
\newblock \bibinfo{title}{GNU Lesser General Public License v3.0 only}.
\newblock \bibinfo{howpublished}{https://spdx.org/licenses/LGPL-3.0-only.html}.
\newblock


\bibitem[\protect\citeauthoryear{SPDX}{SPDX}{2018e}]%
        {url-MIT}
\bibfield{author}{\bibinfo{person}{SPDX}.} \bibinfo{year}{2018}\natexlab{e}.
\newblock \bibinfo{title}{The MIT License}.
\newblock \bibinfo{howpublished}{https://spdx.org/licenses/MIT.html}.
\newblock


\bibitem[\protect\citeauthoryear{SPDX}{SPDX}{2018f}]%
        {url-ZPL-2.1}
\bibfield{author}{\bibinfo{person}{SPDX}.} \bibinfo{year}{2018}\natexlab{f}.
\newblock \bibinfo{title}{Zope Public License 2.1}.
\newblock \bibinfo{howpublished}{https://spdx.org/licenses/ZPL-2.1.html}.
\newblock


\bibitem[\protect\citeauthoryear{SPDX}{SPDX}{2021a}]%
        {url-CC3.0}
\bibfield{author}{\bibinfo{person}{SPDX}.} \bibinfo{year}{2021}\natexlab{a}.
\newblock \bibinfo{title}{Creative Commons Attribution 3.0 Unported}.
\newblock \bibinfo{howpublished}{https://spdx.org/licenses/CC-BY-3.0.html}.
\newblock


\bibitem[\protect\citeauthoryear{SPDX}{SPDX}{2021b}]%
        {url-spdx-licenses}
\bibfield{author}{\bibinfo{person}{SPDX}.} \bibinfo{year}{2021}\natexlab{b}.
\newblock \bibinfo{title}{SPDX License List}.
\newblock \bibinfo{howpublished}{https://spdx.org/licenses/}.
\newblock


\bibitem[\protect\citeauthoryear{Statsite}{Statsite}{2021}]%
        {url-statsite}
\bibfield{author}{\bibinfo{person}{Statsite}.} \bibinfo{year}{2021}\natexlab{}.
\newblock \bibinfo{title}{Statsite}.
\newblock \bibinfo{howpublished}{https://github.com/statsite/statsite}.
\newblock


\bibitem[\protect\citeauthoryear{Timo, Jussi, and Tommi}{Timo
  et~al\mbox{.}}{2009}]%
        {ase`09-Timo-ASLA}
\bibfield{author}{\bibinfo{person}{Tuunanen Timo}, \bibinfo{person}{Koskinen
  Jussi}, {and} \bibinfo{person}{Kärkkäinen Tommi}.}
  \bibinfo{year}{2009}\natexlab{}.
\newblock \showarticletitle{Automated software license analysis}.
\newblock \bibinfo{journal}{\emph{Automated Software Engineering}}
  \bibinfo{volume}{16} (\bibinfo{year}{2009}), \bibinfo{pages}{455–490}.
\newblock


\bibitem[\protect\citeauthoryear{Wheeler}{Wheeler}{2007}]%
        {Wheeler`07-Wheeler}
\bibfield{author}{\bibinfo{person}{David~A. Wheeler}.}
  \bibinfo{year}{2007}\natexlab{}.
\newblock \bibinfo{title}{The free-libre / open source software (FLOSS) license
  slide}.
\newblock \bibinfo{howpublished}{http://www.dwheeler.
  com/essays/floss-license-slide.pdf}.
\newblock


\bibitem[\protect\citeauthoryear{Xia, Liu, and Zhang}{Xia
  et~al\mbox{.}}{2019}]%
        {CSAI`19-Xia-BiLSTM}
\bibfield{author}{\bibinfo{person}{Linzhong Xia}, \bibinfo{person}{Jun Liu},
  {and} \bibinfo{person}{Zhenjiu Zhang}.} \bibinfo{year}{2019}\natexlab{}.
\newblock \showarticletitle{Automatic essay scoring model based on two-layer
  bi-directional long-short term memory network}. In
  \bibinfo{booktitle}{\emph{Proceedings of the 2019 3rd International
  Conference on Computer Science and Artificial Intelligence}}.
  \bibinfo{pages}{133–137}.
\newblock


\bibitem[\protect\citeauthoryear{Xu, Yang, Wan, and Wan}{Xu
  et~al\mbox{.}}{2010}]%
        {cise`10-Xu-OSLC}
\bibfield{author}{\bibinfo{person}{HongBo Xu}, \bibinfo{person}{HuiHui Yang},
  \bibinfo{person}{Dan Wan}, {and} \bibinfo{person}{JiangPing Wan}.}
  \bibinfo{year}{2010}\natexlab{}.
\newblock \showarticletitle{The design and implement of open source license
  tracking system}. In \bibinfo{booktitle}{\emph{Proceddings of the 2010
  International Conference on Computational Intelligence and Software
  Engineering}}. \bibinfo{pages}{1--4}.
\newblock


\bibitem[\protect\citeauthoryear{Xu, Gao, Fan, Liu, Liu, and Ji}{Xu
  et~al\mbox{.}}{2021a}]%
        {lidetector}
\bibfield{author}{\bibinfo{person}{Sihan Xu}, \bibinfo{person}{Ya Gao},
  \bibinfo{person}{Lingling Fan}, \bibinfo{person}{Zheli Liu},
  \bibinfo{person}{Yang Liu}, {and} \bibinfo{person}{Hua Ji}.}
  \bibinfo{year}{2021}\natexlab{a}.
\newblock \bibinfo{title}{LiDetector: License Incompatiblity Detection for Open
  Source Software}.
\newblock \bibinfo{howpublished}{https://sites.google.com/view/lidetector}.
\newblock


\bibitem[\protect\citeauthoryear{Xu, Gao, Fan, Liu, Liu, and Ji}{Xu
  et~al\mbox{.}}{2021b}]%
        {lidetector-github}
\bibfield{author}{\bibinfo{person}{Sihan Xu}, \bibinfo{person}{Ya Gao},
  \bibinfo{person}{Lingling Fan}, \bibinfo{person}{Zheli Liu},
  \bibinfo{person}{Yang Liu}, {and} \bibinfo{person}{Hua Ji}.}
  \bibinfo{year}{2021}\natexlab{b}.
\newblock \bibinfo{title}{LiDetector: License Incompatiblity Detection for Open
  Source Software}.
\newblock \bibinfo{howpublished}{https://github.com/XuSihan/LiDetector}.
\newblock


\bibitem[\protect\citeauthoryear{Zhan, Fan, Chen, Wu, Liu, Luo, and Liu}{Zhan
  et~al\mbox{.}}{2021a}]%
        {zhan2021atvhunter}
\bibfield{author}{\bibinfo{person}{Xian Zhan}, \bibinfo{person}{Lingling Fan},
  \bibinfo{person}{Sen Chen}, \bibinfo{person}{Feng Wu},
  \bibinfo{person}{Tianming Liu}, \bibinfo{person}{Xiapu Luo}, {and}
  \bibinfo{person}{Yang Liu}.} \bibinfo{year}{2021}\natexlab{a}.
\newblock \showarticletitle{Atvhunter: Reliable version detection of
  third-party libraries for vulnerability identification in android
  applications}. In \bibinfo{booktitle}{\emph{2021 IEEE/ACM 43rd International
  Conference on Software Engineering (ICSE)}}. IEEE,
  \bibinfo{pages}{1695--1707}.
\newblock


\bibitem[\protect\citeauthoryear{Zhan, Fan, Liu, Chen, Li, Wang, Xu, Luo, and
  Liu}{Zhan et~al\mbox{.}}{2020}]%
        {zhan2020automated}
\bibfield{author}{\bibinfo{person}{Xian Zhan}, \bibinfo{person}{Lingling Fan},
  \bibinfo{person}{Tianming Liu}, \bibinfo{person}{Sen Chen},
  \bibinfo{person}{Li Li}, \bibinfo{person}{Haoyu Wang}, \bibinfo{person}{Yifei
  Xu}, \bibinfo{person}{Xiapu Luo}, {and} \bibinfo{person}{Yang Liu}.}
  \bibinfo{year}{2020}\natexlab{}.
\newblock \showarticletitle{Automated third-party library detection for android
  applications: Are we there yet?}. In \bibinfo{booktitle}{\emph{2020 35th
  IEEE/ACM International Conference on Automated Software Engineering (ASE)}}.
  IEEE, \bibinfo{pages}{919--930}.
\newblock


\bibitem[\protect\citeauthoryear{Zhan, Liu, Fan, Li, Chen, Luo, and Liu}{Zhan
  et~al\mbox{.}}{2021b}]%
        {zhan2021research}
\bibfield{author}{\bibinfo{person}{Xian Zhan}, \bibinfo{person}{Tianming Liu},
  \bibinfo{person}{Lingling Fan}, \bibinfo{person}{Li Li}, \bibinfo{person}{Sen
  Chen}, \bibinfo{person}{Xiapu Luo}, {and} \bibinfo{person}{Yang Liu}.}
  \bibinfo{year}{2021}\natexlab{b}.
\newblock \showarticletitle{Research on Third-Party Libraries in Android Apps:
  A Taxonomy and Systematic Literature Review}.
\newblock \bibinfo{journal}{\emph{IEEE Transactions on Software Engineering}}
  (\bibinfo{year}{2021}).
\newblock


\end{thebibliography}

\end{document}